\documentclass[prc,letterpaper,twocolumn,showpacs,showkeys,lengthcheck,
               nofootinbib,preprintnumbers,superscriptaddress]{revtex4-1}

\usepackage[utf8]{inputenc}
\usepackage{newtxtext}
\usepackage[libertine]{newtxmath}
\usepackage[main=english]{babel}
\usepackage{microtype}

\usepackage{graphicx}
\usepackage[usenames,dvipsnames]{xcolor}
\usepackage{color}
\usepackage{tikz}

\usepackage[colorlinks=true,linkcolor=blue,urlcolor=blue,citecolor=blue]{hyperref}

\usepackage{amsmath,amsfonts}
\usepackage{slashed}
\usepackage{bm}

\usepackage{titlesec}
\usepackage{dcolumn}

\usepackage{blindtext}

\titlespacing{\section}{4pt}{12pt plus 4pt minus 2pt}{8pt plus 2pt minus 2pt}
\titlespacing{\subsection}{0pt}{12pt plus 4pt minus 2pt}{8pt plus 2pt minus 2pt}
\titlespacing{\subsubsection}{0pt}{12pt plus 4pt minus 2pt}{0pt plus 2pt minus 2pt}

\graphicspath{{.}{figures/}}

\def\lf{\left}
\def\rg{\right}

\newcommand{\vect}[1]{\boldsymbol{#1}}

\def\widebar{\bar}

\newcommand{\be}{\begin{equation}}
\newcommand{\ee}{\end{equation}}

\def\emdash{\nobreak\hspace{0.1em}\textemdash\nobreak\hspace{0.1em}}
\newcommand*\dd{\mathop{}\!\mathrm{d}}

\begin{document}
%\title{Composite octet baryons in nuclear matter and neutron stars}
\title{Composite octet baryons in a relativistic mean field description of nuclear and neutron star matter}
%Equation of State of High Density Matter in the Two-Flavor Nambu--Jona-Lasinio Model}

\author{Kaito Noro}
\email[Corresponding author:~]{1ctad005@mail.u-tokai.ac.jp}
\affiliation{Graduate School of Science and Technology, Tokai University,
4-1-1 Kitakaname, Hiratsuka-shi, Kanagawa 259-1292, Japan}
\affiliation{Micro/Nano Technology Center, Tokai University,
4-1-1 Kitakaname, Hiratsuka-shi, Kanagawa 259-1292, Japan}

\author{Wolfgang Bentz}
\email[]{bentz@keyaki.cc.u-tokai.ac.jp}
\affiliation{Department of Physics, School of Science, Tokai University,
4-1-1 Kitakaname, Hiratsuka-shi, Kanagawa 259-1292, Japan}

\author{Ian C. Clo\"et}
\email[]{icloet@anl.gov}
\affiliation{Physics Division, Argonne National Laboratory, Argonne, Illinois 60439, USA}

\author{Teruyuki Kitabayashi}
\email[]{teruyuki@keyaki.cc.u-tokai.ac.jp}
\affiliation{Department of Physics, School of Science, Tokai University,
4-1-1 Kitakaname, Hiratsuka-shi, Kanagawa 259-1292, Japan}

\begin{abstract}
We examine the properties of composite octet baryons in the nuclear medium and neutron star matter.
The internal quark-diquark structure of the octet baryons and the equations of state of nuclear matter and neutron star matter 
in the mean field approximation are described
by using the three-flavor Nambu--Jona-Lasinio (NJL) model as an effective quark theory of QCD. After introducing our model, 
we first discuss the
properties of single baryons and their effective meson exchange interactions in symmetric nuclear matter by using concepts
of Fermi liquid theory. Several model independent implications of this description are derived, and illustrated by numerical 
results obtained in our model. 
%Special emphasis is put on the role of the internal quark-diquark structure of hadrons in the nuclear medium. 
Second, we extend the model description to high baryon densities, and investigate the equation of state of neutron
star matter and the resulting star masses. We find that the so called hyperon puzzle persists also 
for the case of composite hadrons. To get more information on this point, we also investigate 
the role of 6-fermi and 8-fermi interactions, in addition to the standard 4-fermi interactions.
The strengths of those higher order fermi interactions is determined so as not to spoil the saturation properties of nuclear matter.
Among them, an interaction characterized by a product of four quark current operators plays a special role
to stabilize the stars over a large region of central baryon densities, although it has little effect
on the maximum star masses.   
%star, with a color superconducting quark matter core, exceeds $2.01 \pm 0.04\,M_\odot$ which is the value of the recently observed %massive neutron star PSR J0348+0432. The mass-radius relation is also consistent with gravitational wave observations (GW170817).
%
\vskip 1.0em
\noindent \textit{PhySH}: {
        Quark model;
        Asymmetric nuclear matter;
        Nuclear matter in neutron stars.
    }
\end{abstract}

\maketitle
%===============================================================================
%===============================================================================
\section{INTRODUCTION\label{sec:Intro}}

Systems of strongly interacting baryons are fascinating objects of current research, because their properties
reflect the basic interaction between baryons which is intimately related to their quark substructure, and they
connect microscopic nuclear systems to macroscopic astrophysical objects like supernovae and neutron stars.
Besides the familiar building blocks of nuclear systems\emdash protons and neutrons made of up ($u$) and down ($d$) quarks\emdash
baryons which carry strangeness are receiving much attention now because experimental and theoretical tools
have become available to study their interactions and their role in nuclear and neutron star matter.

On the theoretical side, the baryon-baryon interactions have been extensively studied by using the meson exchange picture~\cite{Haidenbauer:2005zh,Rijken:2006ep}, effective field theories~\cite{Bogner:2009bt,Petschauer:2020urh},
and quantum Monte Carlo calculations~\cite{Lonardoni:2013gta}.
The parameters characterizing the two-body and possible three-body interactions are usually adjusted to 
scattering data, quark model predictions, or experimental data on nuclei and hypernuclei.
Another line of approach, based on nonrelativistic constituent quark models, has been pursued vigorously~\cite{Faessler:1982ik,Oka:1986fr,Fujiwara:2006yh}, mainly
to understand the origin of the short-range repulsion on the basis of the Pauli principle on the quark level~\cite{Oka:2023hdc}. More recent investigations are based on  
first principles derived from QCD~\cite{Inoue:2010hs,Hyodo:2020czb}. These approaches, together, provide vital information
to understand the properties of baryonic systems, in particular hypernuclei~\cite{Gibson:2010zz,Gal:2016boi,
Hiyama:2018lgs,Tamura:2022jet}, and are useful tools to analyze new data on hyperon-nucleon scattering~\cite{Miwa:2022coz,J-PARCE40:2022nvq}. 

A test stone for theoretical models was provided by the observation of heavy neutron stars with about 2 solar masses~\cite{Demorest:2010bx,Antoniadis:2013pzd,Riley:2021pdl,Fonseca:2021wxt}. Because the presence of hyperons 
usually leads to a softening of the equation of state of neutron star matter~\cite{Glendenning:1997wn}, many
models were and are still unable to reproduce such heavy stars, and this problem is commonly called the
``hyperon puzzle''~\cite{Bombaci:2016xzl}. For extensive reviews on this subject and possible solutions, see
for example Refs.~\cite{Chatterjee:2015pua,Burgio:2021vgk}. Most of the proposed solutions require additional
repulsion between the baryons in the system, either via the exchange of vector mesons with particular forms of their couplings to baryons~\cite{Weissenborn:2011ut,Spinella:2018dab}, pomeron exchange~\cite{Yamamoto:2013ada}, 
or new kinds of three-body interactions~\cite{Haidenbauer:2016vfq,Kohno:2018gby,Logoteta:2019utx,Gerstung:2020ktv}. 
Another possible solution~\cite{Contrera:2022tqh} is based on the idea of a phase transition from
nuclear matter to color superconducting quark matter~\cite{Buballa:2003qv,Alford:2007xm} at densities
below or near the hyperon threshold.      
 
As we mentioned at the beginning of this section, the properties of baryons and their interactions reflect their quark substructure, which
changes in the nuclear medium. In order to study this aspect of the problem over a wide range of densities, relativistic
quark models based on QCD are very useful tools. Two models of this kind, which have been used to describe nuclear
phenomena in terms of quark degrees of freedom, are the quark-meson-coupling (QMC) model~\cite{Guichon:1987jp}, 
which is based on the MIT bag model~\cite{Chodos:1974je}, 
and the Nambu--Jona-Lasinio (NJL) model~\cite{Nambu:1961tp,Nambu:1961fr,Vogl:1991qt,Hatsuda:1994pi}, for which a full Faddeev
approach~\cite{Ishii:1995bu} and a closely related but much simpler quark-diquark approach~\cite{Bentz:2001vc} to baryons have been developed.
The degrees of freedom in the QMC model are quarks coupled to elementary mesons via Yukawa couplings, while the NJL model
in its original form uses 4-fermi interactions between quarks to generate mesons as quark-antiquark bound states.
Both models have been used extensively to explore the effects of medium modification on the quark level to
nuclear observables~\cite{Geesaman:1995yd,Lu:1997mu,Stone:2016qmi,Cloet:2005rt,Cloet:2006bq,Cloet:2015tha}.

The QMC model has also been applied to a wide range of hypernuclei~\cite{Saito:2005rv}. Because
meson exchange interactions usually tend to overbind the $\Lambda$ and $\Sigma$ baryons in 
nuclei~\cite{Sammarruca:2008hy,Petschauer:2020urh}, in these earlier calculations a phenomenological repulsive interaction
was introduced in order to reproduce the data. In a later version~\cite{Guichon:2008zz}, the observation was made
that the effect of spin-spin correlations between quarks, associated with the hyperfine interaction from gluon exchange, 
become enhanced in the nuclear medium if the $u$ and $d$ quark masses decrease as functions of the density 
but the strange ($s$) quark mass remains constant.
(For a simplified argument, see also Ref.~\cite{Close:1979bt}.) 
Because of the different spin-flavor structures of the $\Lambda$ and $\Sigma$ baryons, this leads to the expectation that
the $\Sigma - \Lambda$ mass difference increases with the nuclear density. 
%A simple estimate was presented to show that the $\Sigma$ 
%will be less bound than the $\Lambda$ in the nuclear medium by about 20 MeV, which was confirmed by the model calculations. 
This kind of mechanism relies on the assumption of a constant $s$ quark mass, which is well satisfied
in hypernuclei where the density of $s$ quarks is essentially zero, but may become less effective in neutron star matter
as soon as a finite density of strange baryons appears.   

In the present work, we will use the NJL model to describe the internal quark-diquark structure of the octet baryons,
the equation of state of nuclear and neutron star matter in the mean field approximation, 
the corresponding in-medium effective meson exchange interactions
between the baryons, and the resulting neutron star masses. The purposes of our work are as follows: 
First, we wish to explore the role of the quark-diquark substructure of baryons in the nuclear medium. 
For this purpose, we extend our previous work~\cite{Carrillo-Serrano:2016igi} on the properties of octet baryons in free space. 
Our model is well suited to examine the above expectation about the in-medium $\Sigma - \Lambda$ mass difference, 
because the spin-spin correlations in the scalar ($0^+$) and axial vector ($1^+$) diquark channels are built in from 
the outset. Second, in close connection to this, we wish to
introduce ideas of the successful theory of Fermi liquids due to Landau~\cite{Landau:1956aa,Landau:1957aa,Landau:1959aa} and Migdal~\cite{Migdal:1967aa,Migdal:1990vm}, and its relativistic 
extensions~\cite{Baym:1975va}, to hyperons in the nuclear medium. Because the power of the Fermi liquid theory
to respect symmetries, conservation laws, and the renormalization group in many-fermion systems is well
known~\cite{Nozieres:1964aa,Negele:1988aa,Shankar:1993pf}, we find it desirable and timely to provide such a connection.
Third, we wish to present a consistent formulation of isospin asymmetric baryonic systems on the background
of the three independent Lorentz scalar and Lorentz vector mean fields, which are defined in Eq.~(\ref{eq:fields})
of the following section.  
%This naturally implies that, for example, we must treat the constituent masses of $u$, $d$, and $s$ quarks in the baryons 
%as independent parameters which have to be determined by minimization of the energy density of the whole system.
Finally, we wish to investigate the status of the hyperon puzzle in the NJL model, and investigate the roles of 
6-fermi~\cite{tHooft:1976snw} and 8-fermi~\cite{Osipov:2006ns} interactions on the equation of state and star masses in the
mean field approximation. In order to achieve these aims as clearly as possible, we will make no attempt to reproduce
any empirical data related to octet baryons, their mutual interactions, or properties of neutron stars.
Rather than this, we wish to explain problems which arise from chiral
symmetry restrictions on the form of the interaction Lagrangian, which were not encountered in our previous work on the flavor SU(2) case~\cite{Tanimoto:2019tsl}.

The outline of the paper is as follows: Sec.~\ref{sec:NJL} discusses our effective quark model for octet baryons and baryonic matter; Sec.~\ref{sec:nuclearmatter} discusses the properties of baryons and their effective meson exchange interactions in symmetric nuclear matter using concepts of Fermi liquid theory; Sec.~\ref{sec:neutronstar} presents our results for neutron star matter and the resulting star masses; Sec.~\ref{sec:sixeight} discusses the roles of 6-fermi and 8-fermi interactions; and Sec.~\ref{sec:summary} gives a summary of our results. 
%App.~\ref{app:baryons} explains our construction of quark-diquark bound states of baryons based on the Faddeev
%approach.  

%===============================================================================
%===============================================================================\
\section{MODEL FOR BARYONS AND BARYONIC MATTER\label{sec:NJL}}
The three-flavor NJL Lagrangian with 4-fermi interactions in the $\bar{q}  q$ channels relevant for this study reads~\cite{Vogl:1991qt,Hatsuda:1994pi}:
\begin{align}
\mathcal{L} &= \widebar{q} \lf(i \slashed{\partial} - \hat{m} \rg) q
+ G_{\pi} \left[ \big(\widebar{q} \lambda_a q\big)^{2} - \big(\widebar{q} \lambda_a \gamma_{5}  q\big)^{2} \right]  \allowdisplaybreaks \nonumber \\
&- G_{v} \left[ \big(\widebar{q} \lambda_a \gamma^{\mu} q \big)^{2} 
 + \big(\widebar{q} \lambda_a \gamma^{\mu} \gamma_5 q \big)^{2} \right] \,,  
%& + G_{s} \left(\widebar{\psi} \gamma_{5} C \tau_{2} \beta^{A} \widebar{\psi}^{T}\right) 
%          \left(\psi^{T} C^{- 1} \gamma_{5} \tau_{2} \beta^{A} \psi\right),
\label{eq:lagrangian}
\end{align}
where $q = (q_1, q_2, q_3)$ with $1 \equiv u$, $2 \equiv d$, $3 \equiv s$ is the quark field, $\hat{m}$ the current quark mass matrix with diagonal elements $(m_u, m_d, m_s)$, and $\lambda_a$ ($a=0,1,2, \dots 8$) are the Gell-Mann flavor matrices plus $\lambda_0 = \sqrt{\frac{2}{3}} \mathbf{1}$. The 4-fermi coupling constants in the scalar--pseudoscalar and the vector--axial-vector channels are denoted by $G_{\pi}$ and $G_v$, respectively.
The Lagrangian~(\ref{eq:lagrangian}) has the  $SU(3)_L \otimes SU(3)_R \otimes U(1)_V \otimes U(1)_A$
symmetry of QCD, which contains the familiar flavor $SU(3)$ as a subgroup. 
The explicit breaking of the $U(1)_A$ symmetry, which is known as the axial anomaly in QCD, can be realized in the
NJL model by the 6-fermi (determinant) interaction~\cite{tHooft:1976snw}, which will be investigated together with possible 8-fermi interactions in Sec.~\ref{sec:sixeight}. It is important to note that in this work we will follow the successful path established by various low energy theorems and octet mass formulas, that current quark masses are the only
sources of explicit breaking of the flavor and the chiral symmetries, and all other symmetry breakings are dynamical.
%due the the presence of symmetry breaking ground states (vacuum, isospin symmetric nuclear matter, and neutron star matter). 
As we will see, this leads to very strong, sometimes unwelcome, restrictions on the model parameters in the mean field approximation.

In order to construct the octet baryons as quark-diquark bound states, we also need the interaction Lagrangian in the $qq$ channels with the same symmetries, which is specified in App.~\ref{app:baryons}.  
Our model description of the octet baryons is a straight forward extension of the quark-diquark model based on the Faddeev framework, 
%taking into account scalar ($0^+$) and axial vector ($1^+$) diquark correlations, 
as described in Refs.~\cite{Carrillo-Serrano:2014zta,Carrillo-Serrano:2016igi}, to the case where the isospin symmetry is broken, like in neutron star matter.   
In the vacuum isospin symmetry is assumed to be intact, i.e., we use $m_u = m_d \equiv m$ throughout this work.

\subsection{Mean field approximation\label{sec:mf}}

In order to construct the equation of state of nuclear matter and neutron star matter in the mean field approximation, we will take into account three scalar fields $\sigma_{\alpha}$ and three
4-vector fields $\omega_{\alpha}^{\mu}$, where $\alpha= u, d, s$. We use the following definitions:
\begin{align}
\sigma_{\alpha} &= 4 G_{\pi} \langle \widebar{q}_{\alpha} \, q_{\alpha} \rangle \,,  &
\omega_{\alpha}^{\mu} &= 4 G_v \langle \widebar{q}_{\alpha} \gamma^{\mu} \, q_{\alpha} \rangle \,,
\label{eq:fields}
\end{align}
where $\langle \dots \rangle$ denotes the expectation value in the ground state of the medium under consideration
(vacuum, nuclear matter, or neutron star matter). 
The presence of the scalar fields leads to spontaneous breaking of the chiral symmetry, 
and gives rise to the effective quark masses
\begin{align}
M_{\alpha} = m_{\alpha} - \sigma_{\alpha} \,, 
\label{eq:quarkmass}
\end{align}
which must be treated independently if the isospin symmetry is broken in the medium. 
The presence of the vector fields leads to shifts in the 4-momenta of the particles in the system. As a result, the energy of a baryon with flavor $b$ and 3-momentum $\vect{k}$ is obtained from the pole of the quark-diquark equation in the variable
$k_0$ as\footnote{Here and in the following, a summation over multiple flavor indices ($\alpha, \beta, \dots$ for quarks, $b, b'$ for octet baryons, $\tau$ for the special case of nucleons, and $i$ for baryons and leptons) in a product,
including squares like $\omega_{\alpha}^2$, is implied if those indices appear only
on one side of an equation. (As usual, the same convention is used for the Lorentz indices $\mu, \nu, \dots$.)
The Fermi momentum of particle $i$ will be denoted as $p_i$.}  
\begin{align}
\varepsilon_b(k) = \sqrt{\vect{k}_b^2 + M_b^2}  + n_{\alpha/b} \, \omega^0_{\alpha} 
\equiv E_b(k_b) +  n_{\alpha/b} \, \omega^0_{\alpha} \,,
\label{eq:baryonenergy}
\end{align}
where $n_{\alpha/b}$ is the number of quarks with flavor $\alpha$ in the baryon $b$, and
$\vect{k}_b = \vect{k} - n_{\alpha/b} \, \boldsymbol{\omega}_{\alpha}$. The effective mass of the baryon, $M_b$, is a function of the effective quark masses $M_u, M_d, M_s$, as described in App.~\ref{app:baryons}.  

The mean field approximation is implemented into the Lagrangian~(\ref{eq:lagrangian}) in the standard way by decomposing the various quark bilinears into classical (c-number) parts and quantum (normal ordered) parts. We will assume that the only non-vanishing classical parts are the mean fields given in Eq.~(\ref{eq:fields}).  The normal ordered parts, together with the $q q$ interaction parts given in App.~\ref{app:baryons}, are used to calculate bound state masses of pseudoscalar mesons and octet baryons, as well as the pion decay constant. 

The quantity of central interest in our work is the energy density (${\cal E}$) of baryonic matter
%which we will determine directly as a function of the densities of the particles without recourse to chemical potentials. In the following we wish to specify the form of ${\cal E}$ 
in the mean field approximation. The basic physical picture can be visualized by composite baryons moving in scalar and vector mean fields on the background of the constituent quark vacuum~\cite{Bentz:2001vc}. 
Except for the vacuum contributions, this is similar in spirit to the QMC model~\cite{Guichon:1987jp,Saito:1994ki}, although the mesons in our approach are composite objects.
The term which describes the Fermi motion of the baryons is given by (note our
summation convention for multiple flavor indices)
\begin{align}
&2 \int \frac{{\rm d}^3 k}{(2 \pi)^3} \, \varepsilon_b(k) \, n_b(k)  \nonumber \\ 
&= 2 \int \frac{{\rm d}^3 k}{(2 \pi)^3} E_b(k) \, n_b(k) + \rho_{\alpha} \, \omega^0_{\alpha}
\equiv {\cal E}_B + \rho_{\alpha} \, \omega^0_{\alpha} \,,
\label{eq:fermimotion}
\end{align}
where $n_b(k)$ is the Fermi distribution function of baryon $b$, and we defined the quark number densities $\rho_{\alpha}$ in terms of the baryon number densities $\rho_b$ by $\rho_{\alpha} = n_{\alpha/b} \rho_{b}$. For the case of neutron star matter we also include the contributions from the Fermi gas of leptons ($\ell = e^-, \mu^-$) in chemical equilibrium with the baryons. The total energy density in the mean field approximation is then expressed as 
\begin{align}
{\cal E} &= {\cal E}_{\rm vac} - \frac{\omega_{\alpha}^2}{8 G_v} + \rho_{\alpha} \omega^0_{\alpha}  
+ {\cal E}_B + {\cal E}_{\ell} \, .
\label{eq:energydensity}
\end{align}
Here the unregularized form of the vacuum (Mexican hat shaped) contribution is
\begin{align}
{\cal E}_{\rm vac}  = 6 i \int  \frac{{\rm d}^4 k}{(2 \pi)^4} \, 
\ln \frac{k^2 - M_{\alpha}^2}{k^2 - M_{\alpha 0}^2} 
+ \frac{\sigma_{\alpha}^2 - \sigma_{\alpha 0}^2}{8 G_{\pi}} \,,
\label{eq:vacuum}
\end{align}
where a sum over the quark flavors $\alpha$ is implied, and the sub-index $0$ refers to the vacuum with zero baryon density.

The scalar and vector fields are determined for given baryon density $\rho_B$ by the conditions 
\begin{align}
\partial {\cal E} / (\partial \sigma_{\alpha}) = \partial {\cal E} / (\partial \omega^{\mu}_{\alpha}) =0 \,.
\label{eq:minimization}
\end{align}
For the scalar fields, the minimizations~(\ref{eq:minimization}) have to be done numerically. It is, however, easy to confirm that they are equivalent to the relation 
\begin{align}
\sigma_{\alpha} = 4 \, G_{\pi} \frac{\partial {\cal E}}{\partial m_{\alpha}} = 4 \,G_{\pi} \, 
\frac{\partial {\cal E}}{\partial M_{\alpha}}  \,,
\label{eq:fhtheorem}
\end{align}
where the first equality is the general Feynman-Hellman theorem, while the second equality holds if the 
energy density is expressed in such a way that the constituent quark masses $M_{\alpha}$ always appear together
with the current quark masses $m_{\alpha}$, i.e., in the first term of the vacuum energy~(\ref{eq:vacuum}) and in the term ${\cal E}_B$ of~(\ref{eq:fermimotion}). 
For the vector fields, Eq.~(\ref{eq:minimization}) leads to 
\begin{align}
\omega_{\alpha}^{\mu} = 4 \, G_v \, j_{\alpha}^{\mu} = 4 \, G_v \, n_{\alpha/b} \, j_b^{\mu} \,,   
\label{eq:vectorfields}
\end{align}
where $j_{\alpha}^{\mu} = \left(\rho_{\alpha}, \vect{j}_{\alpha}\right)$ is the contribution to the baryon current carried by the quark of flavor $\alpha$, and $j_b^{\mu}$ is the corresponding quantity for the baryon $b$. Eq.~(\ref{eq:vectorfields}) is in accordance with the definition 
given in Eq.~\eqref{eq:fields}. 
 
For neutron star matter, the minimization w.r.t. the scalar fields\emdash or equivalently the solution to Eq.~(\ref{eq:fhtheorem})\emdash has to be done under the requirements of chemical equilibrium and charge neutrality~\cite{Glendenning:1997wn}
\begin{align}
\mu_b - \mu_n + q_b \, \mu_e  = \mu_{\mu} - \mu_e = q_i \, \rho_i = 0 \,,  
\label{eq:constraints}
\end{align}
where the chemical potentials for baryons and leptons are given by $\mu_b = \varepsilon_b(k=p_b)$ and
$\mu_{\ell} = \sqrt{p_{\ell}^2 + m_{\ell}^2}$. The Fermi momenta $p_i$ of baryons ($i=b$) and
leptons ($i=\ell$) are related to their number densities $\rho_i$ by $\rho_i = \frac{p_i^3}{3 \pi^2}$. 
In Eq.~(\ref{eq:constraints}), $q_i$ are the electric charges of baryons and leptons. 
In the general case, for given baryon density, the nine 
independent relations in Eq.~\eqref{eq:constraints} determine the densities of 10 particles in the system (8 baryons and 2 leptons). The pressure of the system can then be obtained as a function of baryon density from the relation $P =  \rho_i \, \mu_i \, - {\cal E}$.

%We close this subsection by a comment on the patterns of dynamical symmetry breakings which are caused by the presence of the mean fields
%~(\ref{eq:fields}), assuming for the moment that the symmetries are not explicitly broken by the current quark masses 
%or interactions other than the strong ones.
%On the mean field level, the dynamically broken generators of the chiral  transformations are $\lambda_i \gamma_5$, where $i=1,2,3,\dots 8$.
%If the residual interaction in those channels is turned on, the symmetry is restored in the Goldstone mode by the appearance of massless pseudoscalar excitations.
%Similarly, on the mean field level the dynamically broken generators of flavor transformations are 
%$\lambda_1$, $\lambda_2$, $\lambda_4$, $\lambda_5$, $\lambda_6$, $\lambda_7$, or equivalently 
%the raising and lowering operators
%of isospin ($T_{\pm}$), $U$-spin ($U_{\pm}$) and $V$-spin ($V_{\pm}$). By the residual interactions in those channels, the symmetry is restored 
%by the appearance of massless excitations, which may correspond to the the isobaric analog states in the case of isospin~\cite{Engelbrecht:1970zz,Danchev:1994zz}. 
%Simple examples of relations of this kind are given by Eq.~(\ref{eq:cs}) in 
%App.~\ref{app:baryons}, which will be useful when we describe neutron star matter in Sec.~\ref{sec:neutronstar}. 
%Any kind of ad-hoc explicit symmetry breaking of the
%interaction part of the Lagrangian~(\ref{eq:lagrangian}) would inevitably destroy these features.        

\subsection{Effective baryon-baryon interaction\label{sec:bbi}}

\begin{figure}
\centering\includegraphics[width=0.7\columnwidth]{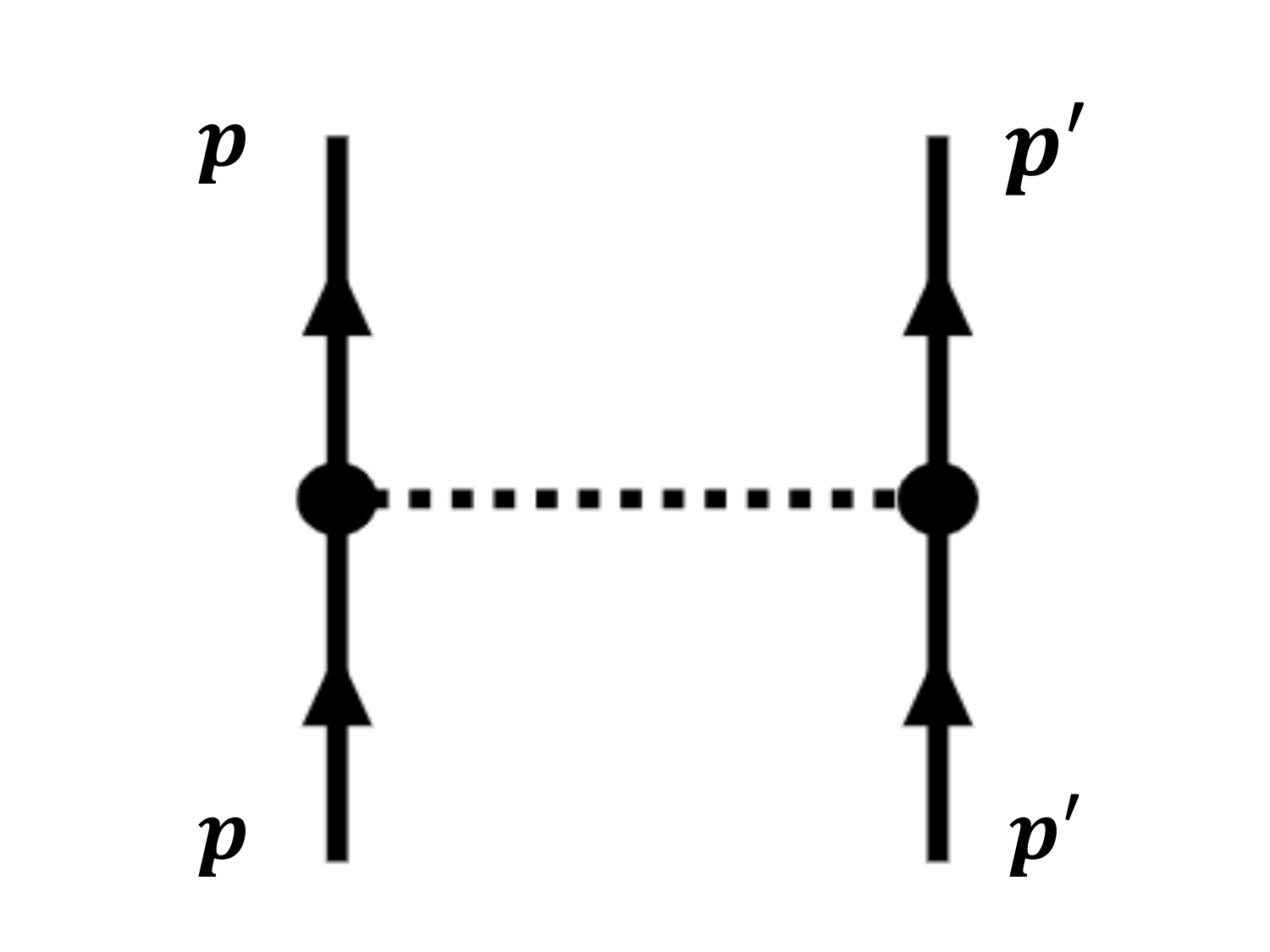}
%\vspace{-0.5cm}
\caption{Graphical representation of the effective baryon-baryon interaction~(\ref{eq:int1})
as a meson exchange potential. For explanation of symbols, see the text.}
    \label{fig:interaction}
\end{figure}

For the purpose of discussions, it will be useful to know the form of the effective baryon-baryon interaction which underlies the mean field approximation described above. For this purpose, we follow the ideas of the Fermi liquid theory~\cite{Nozieres:1964aa,Migdal:1967aa} 
and its relativistic extensions~\cite{Baym:1975va}, and define the spin averaged effective baryon-baryon interaction $F_{b b'}(p,p')$ by the variation of the energy of one of the baryons, $\varepsilon_b(p)$, w.r.t. the distribution function of the other baryon, $n_{b'}(p')$. We wish to express this interaction as a generalized meson-exchange potential. 
%with 3 types of scalar and vector mesons, which correspond to the mean fields given by Eq.(\ref{eq:fields}). 
Because our baryon energies in Eq.~(\ref{eq:baryonenergy}) do no depend explicitly on the distribution functions, we have
\begin{align}
F_{b b'} = \frac{\delta \varepsilon_b}{\delta n_{b'}} = 
\frac{\partial \varepsilon_b}{\partial \sigma_{\alpha}} \frac{\delta \sigma_{\alpha}}{\delta n_{b'}}
+ \frac{\partial \varepsilon_b}{\partial \omega^{\mu}_{\alpha}} \frac{\delta \omega^{\mu}_{\alpha}}{\delta n_{b'}}  \,,
\label{eq:interaction}
\end{align}
where we omitted the dependence on the momenta $p$ and $p'$ to simplify the notations.
Because the conditions given in Eq.~(\ref{eq:minimization}) hold for any fixed set of distribution functions, we can make use of the relations
\begin{align}
\frac{\delta}{\delta n_{b'}} \left( \frac{\partial {\cal E}}{\partial \sigma_{\alpha}} \right)
&= 0 = \frac{\partial {\varepsilon_{b'}}}{\partial \sigma_{\alpha}} 
+ \frac{\partial^2 {\cal E}}{\partial \sigma_{\alpha} \partial \sigma_{\beta}} \, 
\frac{\delta \sigma_{\beta}}{\delta n_{b'}} \,, 
\nonumber \\
\frac{\delta}{\delta n_{b'}} \left( \frac{\partial {\cal E}}{\partial \omega^{\mu}_{\alpha}} \right)
&= 0 = \frac{\partial {\varepsilon_{b'}}}{\partial \omega^{\mu}_{\alpha}} 
+ \frac{\partial^2 {\cal E}}{\partial \omega^{\mu}_{\alpha} \partial \omega^{\nu}_{\beta}} \, 
\frac{\delta \omega^{\nu}_{\beta}}{\delta n_{b'}} \,,
\label{eq:relations}
\end{align}
where the second equalities hold in our model when the whole system is at rest, in which case there are no mixings between scalar and vector mean fields. Using~(\ref{eq:relations}) in~(\ref{eq:interaction}) we obtain
\begin{align}
F_{b b'} &= - \frac{M_b}{E_b}  \, \frac{\partial M_b}{\partial \sigma_{\alpha}}
\left( S^{-1} \right)_{\alpha \beta}  \frac{\partial M_{b'}}{\partial \sigma_{\beta}} \, \frac{M_{b'}}{E_{b'}}
- n_{\alpha/b} \left(V^{-1}\right)^{00}_{\alpha \beta} \, n_{\beta/b'}  
\nonumber \\
& - \frac{p_i}{ E_b} \, n_{\alpha/b} \left(V^{-1} \right)^{ij}_{\alpha \beta} n_{\beta/b} \, \frac{p'_j}{E_{b'}} \,. 
\label{eq:int1}
\end{align}
Here $E_b \equiv E_b(p)$, $E_{b'} \equiv E_{b'}(p')$, and we defined the $3 \times 3$ flavor matrices $S$ and $V$ by
\begin{align}
S_{\alpha \beta} &\equiv \frac{\partial^2 {\cal E}}{\partial \sigma_{\alpha} \partial \sigma_{\beta}}
\,,  &  V^{\mu \nu}_{\alpha \beta} &\equiv \frac{\partial^2 {\cal E}}{\partial \omega_{\alpha \mu} \partial \omega_{\beta \nu}} \,.
\label{eq:propagators}
\end{align}
We illustrate the effective interaction of Eq.~(\ref{eq:int1}) by Fig.~\ref{fig:interaction}, where the solid lines express the baryons, the dashed line expresses the generalized propagators of neutral scalar mesons ($S^{-1}$) and vector meson ($V^{-1}$) for zero momenta, and the vertices stand for the factors to the left and the right of the meson propagators in Eq.~(\ref{eq:int1}). 

In isospin asymmetric baryonic matter, like neutron star matter, the 
$\bar{u} u$, $\bar{d} d$ and $\bar{s} s$ components of the exchanged mesons are mixed by the baryon loop term ${\cal E}_B$. To disentangle them, one could make an orthogonal transformation to diagonalize $S$ and $V$ at fixed baryon density, and express the couplings of each exchanged flavor to the baryon by a linear combination of vertices.
In the present work we will not carry out such an analysis for the case of neutron star matter.   
In the case of isospin symmetric nuclear matter, on the other hand, the matrices $S$ and $V$ become diagonal automatically by taking isoscalar and isovector combinations of the interacting baryons in the particle-hole channel ($t$-channel), and we will show the explicit forms in the next section.

%In this simplified case, two of the three scalar mesons correspond the familiar isoscalar $\sigma$ and the isovector $\delta_0$ meson, while the third one is an effective scalar meson with the quark content $\bar{s} s$. For the case of the vector mesons, one can identify them with the isoscalar $\omega$, the isovector $\rho_0$, and the $\phi$ meson with the quark content $\bar{s} s$.
%===============================================================================
%===============================================================================

\section{BARYONS IN SYMMETRIC NUCLEAR MATTER\label{sec:nuclearmatter}}

In this section we wish to discuss our results for the properties of baryons and their mutual effective
meson exchange interactions in isospin symmetric nuclear matter. In this case, the mean fields~(\ref{eq:fields}) with $\alpha = u$ and
$\alpha=d$ are the same because of the isospin symmetry, and $\omega_s^{\mu}=0$ because the density of strange quarks is zero. Because the $s$-quark mass enters only in the vacuum energy~(\ref{eq:vacuum}), the minimization condition $\partial {\cal E} / \partial \sigma_s = 0$ gives
$\sigma_s = \sigma_{s 0}$ and therefore $M_s = M_{s 0}$. (Note that this holds only in the present case
of 4-fermi interactions. The 6-fermi and 8-fermi interactions considered in Sec.~\ref{sec:sixeight} 
lead to a slight density dependence of $M_s$ even in symmetric nuclear matter.)
The energy density of the system is given by Eq.~(\ref{eq:energydensity}) without the leptonic 
term ${\cal E}_{\ell}$. 

\subsection{Effective meson exchange interaction\label{sec:effbb}}

In the case of isospin symmetric nuclear matter, the flavor matrices of Eq.~(\ref{eq:propagators}), which characterize the effective interaction~(\ref{eq:int1}), become diagonal automatically by taking appropriate combinations, e.g., for the case of the baryon-nucleon interaction we define
\begin{align}
f_{bN} &\equiv \frac{1}{2} \left( F_{bp} + F_{bn} \right), &  
f'_{bN} &\equiv \frac{1}{2} \left( F_{bp} - F_{bn} \right).
\label{eq:int2}
\end{align}
Within the isospin multiplet to which the baryon $b$ belongs, $f_{bN}$ is an isoscalar and
the same for all members of the multiplet, while $f'_{bN}$ is an isovector proportional to the
isospin 3-component of the baryon $b$. We find the explicit forms
\begin{align}
&f_{bN}(p,p') = - \frac{M_b}{E_b} \frac{M_N}{E_N} \left(\frac{\partial M_b}{\partial M} \right)
\left(\frac{\partial M_N}{\partial M} \right) \nonumber \\
&\times \left( \frac{1}{2 G_{\pi}} + 2 g(M) 
+ \rho_N^{(s)} \frac{\partial^2 M_N}{\partial M^2} 
+ \phi_N \left(\frac{\partial M_N}{\partial M}\right)^2 \right)^{-1}  \nonumber \\
&+ \frac{ 6 \left(1 + \frac{y_b}{2} \right)}{ 1/(2 G_v)}  
- 6 \left(1 + \frac{y_b}{2} \right) 
\frac{\vect{p} \cdot \vect{p}'}{ E_b \, E_N} \,
\left(\frac{1}{2 G_v}  + \frac{9 \rho_B}{E_N} \right)^{-1},  
\label{eq:isoscalar} 
\end{align}
\begin{align}
&f'_{bN}(p,p') = - \frac{M_b}{E_b} \frac{M_N}{E_N} \left(\frac{\partial M_b}{\partial \Delta M} \right)
\left(\frac{\partial M_p}{\partial \Delta M} \right) \nonumber \\
&\times  \left( \frac{1}{2 G_{\pi}}  + 2 g(M) 
+ \rho_N^{(s)} \frac{\partial^2 M_p}{\partial (\Delta M)^2} 
+ \phi_N \left(\frac{\partial M_p}{\partial \Delta M}\right)^2 \right)^{-1}   \nonumber \\
&+ \frac{ 2 \, t_b}{ 1/(2 G_v)} - 2 t_b \,  \frac{\vect{p} \cdot \vect{p}'}{ E_b \, E_N} \,
\left(\frac{1}{2 G_v} + \frac{\rho_B}{E_N} \right)^{-1}. 
\label{eq:isovector} 
\end{align}
Here $M$ is the effective mass of $u,d$ quarks, $M_N$ the nucleon mass in symmetric nuclear matter, and
the derivatives w.r.t. $\Delta M \equiv M_u - M_d$ should be evaluated at $M_u = M_d$.
The unregularized form of the function $g(M)$ is 
\begin{align}
g(M) = - 12 i \int \frac{{\rm d}^4 k}{(2\pi)^4} \frac{k^2 + M^2}{\left(k^2 - M^2\right)^2} \,,
\label{eq:g}
\end{align}
and the scalar density of the nucleon ($\rho_N^{(s)}$) and the function $\phi_N$ are defined by
\begin{align}
\rho_N^{(s)} &= 2 \int \frac{{\rm d}^3 k}{(2 \pi)^3} \, n_N(k) \, \frac{M_N}{E_N(k)}  \,, 
\label{eq:scalardensity} \\ 
\phi_N &= 2 \int \frac{{\rm d}^3 k}{(2 \pi)^3} \, n_N(k) \, \frac{k^2}{E_N(k)^3} \,. 
\label{eq:phibaryon}
\end{align}
In Eqs.~(\ref{eq:isoscalar}) and (\ref{eq:isovector}), $E_b \equiv E_b(p)$, $E_N \equiv E_N(p')$, and we used the relations 
$n_{u/b} + n_{d/b} = 2 \left(1 + \frac{y_b}{2} \right)$ and $n_{u/b} - n_{d/b} = 2 t_b$ where $y_b$ and $t_b$ are the hypercharge and the isospin
3-component of the baryon $b$. 
%where the momenta of nucleons are always defined to be the Fermi momentum, while the momenta of
%hyperons are considered free variables.

The interpretation of~(\ref{eq:isoscalar}) for the case $b=N$ in terms of the meson exchange
processes of Fig.~\ref{fig:interaction} has been discussed in detail in Ref.~\cite{Bentz:2001vc}, 
and the generalization is almost self evident:\footnote{To make the connection to the baryon-meson coupling constants and meson masses, the numerator functions and
denominator functions (i.e., those parts which involve $\frac{1}{2 G_{\pi}}$ or $\frac{1}{2 G_v}$)
must be multiplied by the squares of the relevant quark-meson coupling constants, i.e., by $g^{(q)2}_{\sigma} = g^{(q)2}_{\delta}$ 
for the first 2 lines of~(\ref{eq:isoscalar}) and~(\ref{eq:isovector}), and by $g^{(q)2}_{\omega} =
g^{(q)2}_{\rho}$ for the third lines. See App.~\ref{app:mesonexchange} for details.}
The first two lines (third line) in~(\ref{eq:isoscalar})
correspond to $\sigma$ ($\omega$) meson exchange, while the first two lines (third line) in~(\ref{eq:isovector})
correspond to neutral $\delta$ ($\rho$) meson exchange. The coupling constants of a baryon $b$ to the $\sigma$
($\delta$) meson are proportional to the derivative of $M_b$ w.r.t. $M$ ($\Delta M$), while the couplings
to the vector mesons ($\omega$ and $\rho$) are determined by the isoscalar and isovector combination
of the quark numbers in the baryon. The function $g(M)$ in the denominators of~(\ref{eq:isoscalar}) and~(\ref{eq:isovector}) is the one quark-loop self energy of the scalar meson in the vacuum, the terms
involving the scalar density are the Fermi averages over effective $\sigma \sigma NN$ ($\delta \delta NN$)
contact interactions which are induced by the scalar-isoscalar polarizability $\frac{\partial^2 M_N}{\partial M^2}$ (the scalar-isovector polarizability $\frac{\partial^2 M_p}{\partial (\Delta M)^2}$) of the nucleon~\cite{Birse:1994eg,Saito:1995up,Wallace:1994ux}, 
and the terms proportional to $\phi_N$ are the Fermi averages over the ``Z-graph'' contributions, which also
appear in hadronic theories~\cite{Brown:1985gt,Wallace:1998mx}. 
In our numerical calculations, discussed in Sec.~\ref{sec:numericalnm}, 
we find that the numerators of the $\sigma$ and $\delta$ exchange parts in~(\ref{eq:isoscalar}) and~(\ref{eq:isovector}) substantially decrease, while their denominators
slightly increase as the baryon density increases. The increase of the denominators is partially related to the fact that 
the scalar-isoscalar and scalar-isovector polarizabilities of nucleons\emdash both being positive\emdash increase as the baryon density increases. 
The density dependence of the coupling constants and mesons self energies in~(\ref{eq:isoscalar}) and~(\ref{eq:isovector})
then suppresses the attractive effects of scalar meson exchanges, and at higher densities the vector meson exchanges
become dominant. The terms $\propto \vect{p} \cdot \vect{p}'$ correspond to the contributions from the exchange of
$\vect{\omega}$ and neutral $\vect{\rho}$ mesons, and their self energies arise only from the Fermi averages over the
corresponding Z-graphs. More details will be given in Sec.~\ref{sec:fermiliquid1} and App.~\ref{app:mesonexchange}.

\subsection{Physical implications of the interaction\label{sec:fermiliquid}}

In order to explain some physical implications of the effective meson exchange interactions $f_{bN}$ and $f'_{bN}$
of Eqs.~(\ref{eq:isoscalar}) and~(\ref{eq:isovector}),  
we extend a few basic points of Fermi liquid theory to octet baryons in the nuclear medium. 
%The general relations given below are well kown for the case of nucleons, but - to best of our knowledgle - they have not yet been discussed in the literature for the case of hyperons.
%In the following, we take up three general relations which follow from an application of Fermi liquid theory, and give more details in App.~\ref{app:fermiliquid}.
In the following discussions, we will use the following notations:\footnote{We remind that ``isovector'' ($T=1$) and ``isoscalar'' ($T=0$) refers to the particle-hole channel (t-channel) of the interacting baryons, not to the isospin of the two incoming baryons.}
\begin{itemize}
\item $B = N, \Sigma, \Lambda, \Xi$ stands for the isospin multiplets (including the isospin singlet $\Lambda$), 
while $b$ continues to stand for a member of the baryon octet;
\item $f_{BN}$ denotes the 4 independent isoscalar
baryon-nucleon interactions, defined by Eq.~(\ref{eq:isoscalar}) with $b$ a member of $B$;
\item $f'_{BN}$ denotes the 3 independent isovector baryon-nucleon interactions ($B=N, \Sigma, \Xi$),    
defined by Eq.~(\ref{eq:isovector}) with $b$ a member of $B$ with the largest value of the isospin 3-component $t_b$ (i.e., $p$, $\Sigma^+$, and $\Xi^0$).
\end{itemize}
We also separate the terms $\propto \vect{p} \cdot \vect{p}'$, which 
involve the transfer of one unit of orbital angular momentum ($\ell = 1$) between the baryons, from the other terms
which involve no angular momentum transfer ($\ell=0$):
\begin{align}
f_{bN} &= f_{0, bN} + \cos \theta \, f_{1, bN} \,, &
f'_{bN} &= f'_{0, bN} + \cos \theta \, f'_{1, bN} \,,
\label{eq:separation}
\end{align}
and similarly for $f_{BN}$, $f'_{BN}$, $F_{bp}$, and $F_{bn}$, where $\theta$ is the angle between $\vect{p}$ and $\vect{p}'$.
The parameters $f_{\ell, NN}$ and $f'_{\ell, NN}$ defined in this way agree with the familiar Landau-Migdal parameters,
usually denoted by $f_{\ell}$ and $f'_{\ell}$. 

\subsubsection{\bf{\em{Nucleon density variations}}\label{sec:densityvar}}

The $\ell=0$ baryon-nucleon interactions $f_{0,bN}$ and $f'_{0,bN}$ express the change of the baryon
energies, $\varepsilon_b$, caused by variations of the Fermi momenta of the background nucleons.   
If we denote the corresponding variations of nucleon densities by $\delta \rho_{\tau}$ ($\tau = p, n$), the change of the distribution functions to first order is given by $\delta n_{\tau}(k) = \frac{\pi^2}{p^2_{\tau}} \, (\delta \rho_{\tau}) \, \delta( k - p_{\tau})$. Then, according to the general definition given by the first equality in
Eq.~(\ref{eq:interaction}), the energy of a baryon $b$ in nuclear matter changes by an amount
\begin{align}
\delta \varepsilon_b(k) &= 2 \int \frac{{\rm d}^3 p}{(2 \pi)^3}  \, F_{b \tau}(k, p) \, \delta n_{\tau}(p)
=  \, F_{0, b \tau} \, \delta \rho_{\tau} \,.
\label{eq:densvar}
\end{align}
Separating the isoscalar from the isovector contributions then gives the general relations
\begin{align}
\frac{\partial \varepsilon_b(k)}{\partial \rho_B} &= f_{0, bN}(k, p_N) \,,  
& 
\frac{\partial \varepsilon_b(k)}{\partial \rho_{(3)}} &= f'_{0, bN}(k, p_N) \,,
\label{eq:energychanges}
\end{align}
where $\rho_{(3)} = \rho_p - \rho_n$, and the limit of isospin symmetric nuclear matter
($\rho_{(3)} \rightarrow 0$) is understood. For the case where $b$ is a nucleon, the parameters $f_{0, NN}$ and $f'_{0, NN}$ for $k=p_N$ are related to the incompressibility ($K$) and the symmetry energy ($a_s$) as follows:
\begin{align}
K &= 9 \rho_B \left( \frac{\pi^2}{2 p_N E_N} + f_{0, NN} \right), 
\label{eq:incomp}  \\
a_s &= \frac{\rho_B}{2} \left(  \frac{\pi^2}{2 p_N E_N} + f'_{0, NN} \right).
\label{eq:symmetry}
\end{align} 

In our model the baryon energy is given by Eq.~(\ref{eq:baryonenergy}), and by using Eq.~(\ref{eq:vectorfields}) in nuclear matter at rest, we have
\begin{align}
\varepsilon_b(k) = E_b(k) + 12 G_v \rho_B \, \left( 1 + \frac{y_b}{2} \right)
+ 4 G_v \, \rho_{(3)} \, t_b \,.  
\label{eq:baryonenergy1}
\end{align}
Using this in Eq.~(\ref{eq:energychanges}), we see that in our model the $\ell=0$ baryon-nucleon interaction
reflects the density dependence of the baryon effective masses:
\begin{align}
f_{0, bN}(k, p_N) &= \frac{M_b}{E_b(k)} \, \frac{\partial M_b}{\partial \rho_B} + 12 G_v \left(1 + \frac{y_b}{2} \right) \,,
\label{eq:f0bn} \\
f'_{0, bN}(k, p_N) &= \frac{M_b}{E_b(k)} \, \frac{\partial M_b}{\partial \rho_{(3)}} + 4 \, G_v \, t_b \,.
\label{eq:f0pbn}
\end{align}
%In terms of the meson exchange picture, the first term in Eq.(\ref{eq:f0bn}) corresponds to $\sigma$ meson exchange and the second term to $\omega$ meson eachange, while in Eq.(\ref{eq:f0pbn}) the first term corresponds to $\delta_0$ meson exchange and the second term to $\rho_0$ meson exchange.    
For the case where $b$ is a nucleon, the two terms in $f_{0,NN}$, when multiplied
by $9 \rho_B$, give the contributions of $\sigma$ meson and $\omega$ meson
exchange to the incompressibility. Similarly, the two terms in $f'_{0,NN}$, 
when multiplied by $\frac{\rho_B}{2}$, give the contributions of $\delta$ meson and $\rho$ meson exchange to the symmetry energy.

\subsubsection{\bf{\em{Lorentz invariance}}\label{sec:lorentz}}

There are two basic requirements from Lorentz invariance in the present context: First, the distribution function
of the nucleons is Lorentz invariant: $n'_{\tau}(k') = n_{\tau}(k)$, where we use a prime to denote a system which moves with
velocity $\vect{u}$ relative to the reference system which we assume to be at rest, and $k' = \Lambda_{\vect{u}} \, k$ where $\Lambda_{\vect{u}}$ is the Lorentz matrix. A Lorentz transformation then leads to a variation of the distribution
function for fixed momentum according to~\cite{Baym:1975va} $\delta n_{\tau}(k) \equiv n_{\tau}'(k) - n_{\tau}(k) = - \varepsilon_{\tau}(k) \vect{u} \cdot \hat{\vect{k}} \delta(k - p_{\tau})$ to first order in $\vect{u}$. 
Second, the change of the energy of a baryon in symmetric nuclear matter, induced by this density variation,
\begin{align}
\delta \varepsilon_b(k) &= 2 \int \frac{{\rm d}^3 p}{(2 \pi)^3}  \, F_{b \tau}(k,p) \, \delta n_{\tau}(p)   \nonumber \\
&= - \vect{u} \cdot \hat{\vect{k}} \, \frac{ 2 p_N^2 \varepsilon_N}{3 \pi^2} \, f_{1, bN}(k,p_N) \,,    
\label{eq:deltaepsilonb}
\end{align}
must be equivalent to a Lorentz transformation applied directly to the baryon energy, $\delta \varepsilon_b(k) \equiv
\varepsilon_b'(k) - \varepsilon_b(k) = - \vect{u} \cdot \vect{k} + \varepsilon_b(k) \, \vect{u} \cdot \vect{v}_b(k)$ to first
order in $\vect{u}$, where $\vect{v}_b(k) = \vect{\nabla}_k \varepsilon_b(k)$ is the velocity of the baryon. 
This requirement leads to the relation
\begin{align}
\frac{k}{\varepsilon_b(k)} = v_b(k) + \frac{\varepsilon_N}{\varepsilon_b(k)} \frac{2 p_N^2}{3 \pi^2} \, f_{1, bN}(k, p_N) \,.
\label{eq:landau}
\end{align}
In Eqs.~(\ref{eq:deltaepsilonb}) and (\ref{eq:landau}), and in all following relations, $\varepsilon_N \equiv \varepsilon_N(p_N)$ is the Fermi energy of the nucleon in symmetric nuclear matter, while the momentum $k$ is arbitrary. For the case where $b$ is a nucleon, Eq.~(\ref{eq:landau})
agrees with the relativistic form of the Landau effective mass relation for variable momentum $k$~\cite{Baym:1975va}.
%***************************************REVISION(begin)*************************************************** 
%The new and important point here is the factor $\frac{\varepsilon_N}{\varepsilon_b(k)}$. 
%***************************************REVISION(end)*******************************************************
%In a nonrelativistic framework, where one requires the invariance under Galilei transformations, this becomes a factor $\frac{M_{N0}}{M_{b0}}$, i.e., the ratio of the free masses of a nucleon and a baryon $b$,
The velocity of the baryon is usually expressed in terms of the Landau effective mass ($M_b^*(k)$) by $v_b(k) \equiv \frac{k}{M_b^*(k)}$.
By taking the limits $k \rightarrow 0$ on both sides of Eq.~(\ref{eq:landau}), we then obtain a simple relation of the form
\begin{align}
\frac{1}{\varepsilon_b(0)} = \frac{1}{M_b^*(0)}+ \frac{\varepsilon_N}{\varepsilon_b(0)} \, \rho_B \, \hat{f}_{1, bN}(0, p_N) \,,
\label{eq:landau0}
\end{align}
where we defined $\hat{f}_{1, bN}(p,p')$ such that the ``full'' $\ell=1$ interaction (like for example the last term
in Eq.~(\ref{eq:isoscalar})) is expressed in the form $\left(\vect{p} \cdot \vect{p}'\right) \, \hat{f}_{1, bN}(p,p')$. 

It is easy to check that our model satisfies the requirement~(\ref{eq:landau}): The energy of a baryon with momentum $k$ and the
Fermi energy $\varepsilon_N$ of a nucleon in symmetric nuclear matter are obtained from~(\ref{eq:baryonenergy1}) by setting $\rho_{(3)}=0$:  
\begin{align}
\varepsilon_b(k) &= E_b(k) + 12 \, G_v \rho_B \left( 1 + \frac{y_b}{2} \right) \,, \nonumber \\
& \varepsilon_N =  E_N + 18 \, G_v \rho_B \,, 
\label{eq:checklandau}
\end{align}
while $f_{1, bN}$, which corresponds to $\vect{\omega}$ exchange, is given by the last term in Eq.~(\ref{eq:isoscalar}) without
the factor $\cos \theta$, see Eq.~(\ref{eq:separation}).
It is then clear that the general relation~(\ref{eq:landau}) is valid in our model.

\subsubsection{\bf{\em{Currents carried by baryons}}\label{sec:currents}}

The Lorentz invariance requirement of Eq.~(\ref{eq:landau}) is related to the isoscalar $\ell=1$ Fermi liquid parameter $f_{1, bN}$. To give an example where also the isovector part enters, let us consider the currents carried by a baryon $b$ moving with momentum $k$ in nuclear matter, for the case where no momentum is transferred by the external field to the baryon. From gauge invariance and the integral equations for the vertex functions, the Fermi liquid theory leads to the following result~\cite{Bentz:1985qh} 
(see also, for example, Eqs.~(1)-(33) of Ref.~\cite{Nozieres:1964aa} or Eq.~(2.16) of Ref.~\cite{Migdal:1967aa}):  
\begin{align}
\vect{j}_b^{(X)}(k) &=  \vect{v}_b(k) \, Q_b^{(X)} + 2  \,p_{\tau}^2 \, Q^{(X)}_{\tau} \int  \frac{{\rm d} \Omega_p}{(2 \pi)^3} \, 
\hat{\boldsymbol p} \, F_{b \tau}(k, p) \,, 
\label{eq:current}
\end{align}
where $X$ characterizes the type of current, e.g., $X = B$ for the baryon current, $X=I$ for the isospin current, and $X=E$ for the electric current, and $Q_b^{(X)}$ are the corresponding bare charges of the baryon $b$, i.e., $Q_b^{(B)}=1$, $Q_b^{(I)} = t_b$, and $Q_b^{(E)}=q_b$. 
The second term in Eq.~(\ref{eq:current}) is the backflow due to the nuclear medium. 

The magnitude of the baryon current (case $X=B$ in Eq.~(\ref{eq:current})) can be expressed in a model independent way by using the Lorentz invariance relation~(\ref{eq:landau}):
\begin{align}
j_b^{(B)}(k) &= \frac{k}{\varepsilon_N} - v_b(k) \left( \frac{\varepsilon_b(k)}{\varepsilon_N} - 1 \right) \,.
\label{eq:baryoncurrent}
\end{align}
%In a nonrelativistic theory, the corresponding result would be expressed as 
%$j_b^{(B)}(k) = \frac{k}{M_N} - \frac{k}{M^*_b(k)} \left(\frac{M_b}{M_N} - 1 \right)$, where the effective mass of the baryon is defined as usual by $v_b(k) = \frac{k}{M^*_b(k)}$. 
For the case of a nucleon at the Fermi surface, Eq.~(\ref{eq:baryoncurrent}) gives the well known result $j_N^{(B)}(k=p_N) = \frac{p_N}{\varepsilon_N}$, which reduces to the free current $\frac{p_N}{M_{N0}}$ in the nonrelativistic limit. 
%The new point here is the mass dependent correction factor in (\ref{eq:baryoncurrent}). 
%the baryon current carried by a hyperon in the nuclear medium is smaller than this result.  

For the electric current (case $X=E$ in Eq.~(\ref{eq:current})) we obtain generally
\begin{align}
j_b^{(E)}(k) &= 
v_b(k) \, q_b + \frac{1}{2} \left(\frac{k}{\varepsilon_N} - 
\frac{\varepsilon_b(k)}{\varepsilon_N} \, v_b(k) \right) \nonumber \\
&+ \frac{p_N^2}{3 \pi^2} \, f'_{1, bN}(k,p_N) \,.  \label{eq:electric}
\end{align} 
Here we can insert our model result for $f'_{1, bN}$, given by the last term in Eq.~(\ref{eq:isovector}) without the factor $\cos \theta$ (see~(\ref{eq:separation})). 
We can express the result in terms of an effective angular momentum $g$-factor of the baryon ($g_{\ell, b}$), which we define here\emdash in a naive way\emdash so that it becomes unity for a free proton, i.e.,  
$j_b^{(E)}( k \rightarrow 0) \equiv \frac{k}{M_{N0}} \, g_{\ell, b}$. This gives 
%where $M_{N0}$ is the mass of a free nucleon. We obtain
\begin{align}
g_{\ell, b} = \frac{M_{N0}}{M_b} \left( q_b - \frac{3 x}{ 1 + 9 x} \left( 1 + \frac{y_b}{2} \right)
- \frac{x}{1 + x} \, t_b \right) \,,
\label{eq:gellbaryon}
\end{align}
where $x = 2 G_v \rho_B/ E_N = \frac{1}{9} \left(\frac{\varepsilon_N}{E_N} - 1 \right)$ characterizes the strength of the vector interaction. 
The quantities which depend on the baryon density in~(\ref{eq:gellbaryon}) are the baryon effective mass $M_b$ and $x$.

\subsection{Numerical results\label{sec:numericalnm}}

In order to illustrate several physics points of our above discussions, in this subsection we present numerical results 
for symmetric nuclear matter.

\subsubsection{\bf{\em{Model parameters}}\label{sec:parameters}}

First we explain the choice of our model parameters. 
The Lagrangian of Eq.~(\ref{eq:lagrangian}) contains the coupling constants $G_{\pi}$ and $G_v$, and the
current quark masses $m$ and $m_s$, which are related to the constituent quark masses in the vacuum,
$M_0$ and $M_{s0}$, by the gap equations~(\ref{eq:quarkmass}). The other parameters, which are necessary to
define the model, are     
the infrared (IR) and ultraviolet (UV) cut-offs $\Lambda_{\mathrm{IR}}$ and $\Lambda_{\mathrm{UV}}$, which are used with the proper-time regularization scheme~\cite{Schwinger:1951nm,Hellstern:1997nv}, see App.~\ref{app:regularization}. 
In this scheme, the UV cut-off is necessary to give finite integrals, while the IR cut-off is necessary to avoid unphysical decay thresholds of hadrons into quarks, thereby
simulating one important aspect of confinement.    
%\footnote{We mention that chiral symmetry of the interaction Lagrangian requires an additional term
%$G_{\rho} \big(\widebar{\psi} \gamma^{\mu} \gamma_5 \vec{\tau} \psi\big)^{2}$, however this is not directly related to
%our calculation.}
These parameters are determined as follows: The IR cut-off should be similar to $\Lambda_{\rm QCD}$, and we  
choose $\Lambda_{\rm IR} = 0.24\,$GeV. $\Lambda_{\rm UV}$, $m$, and $G_{\pi}$ are determined so as to give a constituent quark mass in vacuum of $M_0 = 0.4\,$GeV, the pion decay constant $f_{\pi} = 0.93\,$GeV, and the pion mass $m_{\pi} = 0.14\,$GeV, using the standard methods based on the Bethe-Salpeter equation in the pionic $\overline{q} q$ channel~\cite{Vogl:1991qt,Hatsuda:1994pi}. $m_s$ is determined so as to give a constituent
$s$-quark mass in vacuum of $M_{s0} = 0.562$ GeV, which reproduces the observed mass of the $\Omega$ baryon
$M_{\Omega} = 1.67$ GeV by using the quark-diquark bound state equations explained in App.~\ref{app:baryons}.
The vector coupling $G_{v}$ is determined from the binding energy per-nucleon in symmetric nuclear matter ($E_B/A = -16 \,$MeV) at the saturation density, which becomes $\rho_0 = 0.15\,$fm$^{-3}$. In the present flavor SU(3) NJL model, the
vector couplings in the isoscalar and isovector $\overline{q} q$ channels are the same because of constraints from chiral symmetry,
and we do not have an independent parameter (like the coupling $G_{\rho}$ in the flavor SU(2) model used in Ref.~\cite{Tanimoto:2019tsl}) 
to fit the symmetry energy.\footnote{Chiral symmetry would allow different vector couplings in the flavor singlet and octet terms
of Eq.~(\ref{eq:lagrangian}), but in the mean field approximation used here it is easy to check that there remains only one 
independent vector coupling in any case. This follows from  
the identity $\sum_{a=0,3,8} \left(\overline{q} \lambda_a \Gamma q \right)^2 = 
2 \left[ \left(\overline{q}_1 \Gamma q_1 \right)^2 + \left(\overline{q}_2 \Gamma q_2 \right)^2
+ \left(\overline{q}_3 \Gamma q_3 \right)^2 \right]$ for any Dirac matrix $\Gamma$.}
The resulting values of the cut-offs, coupling constants in the $\bar{q} q$ channels, and quark masses are shown in Tab.~\ref{tab:parameters}. They are identical to those used in Ref.~\cite{Tanimoto:2019tsl} except for the $s$-quark masses 
which were not needed there. 
Two additional model parameters are the coupling constants in the scalar and axial vector $qq$ channels, $G_S$ and $G_A$
of Eq.~(\ref{eq:lagqq}). As explained in App.~\ref{app:baryons}, they are fixed to the free nucleon and delta
masses ($M_{N0} = 0.94$ GeV, $M_{\Delta 0} = 1.23$ GeV). The resulting free masses of octet baryons are then predictions of the model, and are summarized in 
Tab.~\ref{tab:octetmasses} together with the observed values.

%===============================================================================
\begin{table}[tbp]
\addtolength{\extrarowheight}{2.2pt}
\centering
\caption{Values for the model parameters which are determined in the vacuum, single hadron, and nuclear matter sectors. The regularization parameters, constituent quark masses in the vacuum (sub-index $0$) and current quark masses are given in units of GeV, and the coupling constants in units of GeV$^{-2}$.}
\begin{tabular*}{\columnwidth}{@{\extracolsep{\fill}}cccccccc}
\hline\hline
$\Lambda_{\rm IR}$  & $\Lambda_{\rm UV}$  &  $G_{\pi}$  &  $G_v$  &  $M_0$  &  $M_{s0}$ &  $m$ & $m_s$ \\
\hline
0.240 & 0.645 & 19.04 & 6.03 & 0.40 & 0.562 & 0.016 & 0.273  \\
\hline\hline
\end{tabular*}
\label{tab:parameters}
\end{table}
%===============================================================================

%===============================================================================
\begin{table}
\addtolength{\extrarowheight}{2.2pt}
    \centering
    \caption{Masses of octet baryons (in units of GeV) calculated in the vacuum (sub-index $0$)
    by using the coupling constants $G_S = 8.76$ GeV$^{-2}$ and $G_A = 7.36$ GeV$^{-2}$ in the scalar
    and axial vector diquark channels fitted to the vacuum masses of the
    nucleon and the $\Delta$ baryon, 
    in comparison to the observed values.\footnote{The fact that our calculated masses agree slightly better with the Gell-Mann Okubo
octet mass relation~\cite{Gell-Mann:1962yej,Okubo:1961jc} 
$M_{N0} + M_{\Xi0} = \frac{1}{2} \left(M_{\Sigma0} + 3 M_{\Lambda0}\right)$ than the
experimental values (using either neutral or isospin averaged masses) may be a mere coincidence.}
}
    \begin{tabular*}{\columnwidth}{@{\extracolsep{\fill}}ccccc}
        \hline\hline
            & $M_{N0}$ & $M_{\Lambda 0}$ & $M_{\Sigma 0}$
        & $M_{\Xi 0}$ 
        \\ \hline
        calc. & 0.94 & 1.12 & 1.17 & 1.32 \\
        obs. & 0.94 & 1.12 & 1.19 & 1.32 \\
        \hline\hline
    \end{tabular*}
    \label{tab:octetmasses}
\end{table}
%===============================================================================

\subsubsection{\bf{\em{Energies per nucleon and single baryon energies}}\label{sec:energies}}

\begin{figure}
\centering\includegraphics[width=\columnwidth]{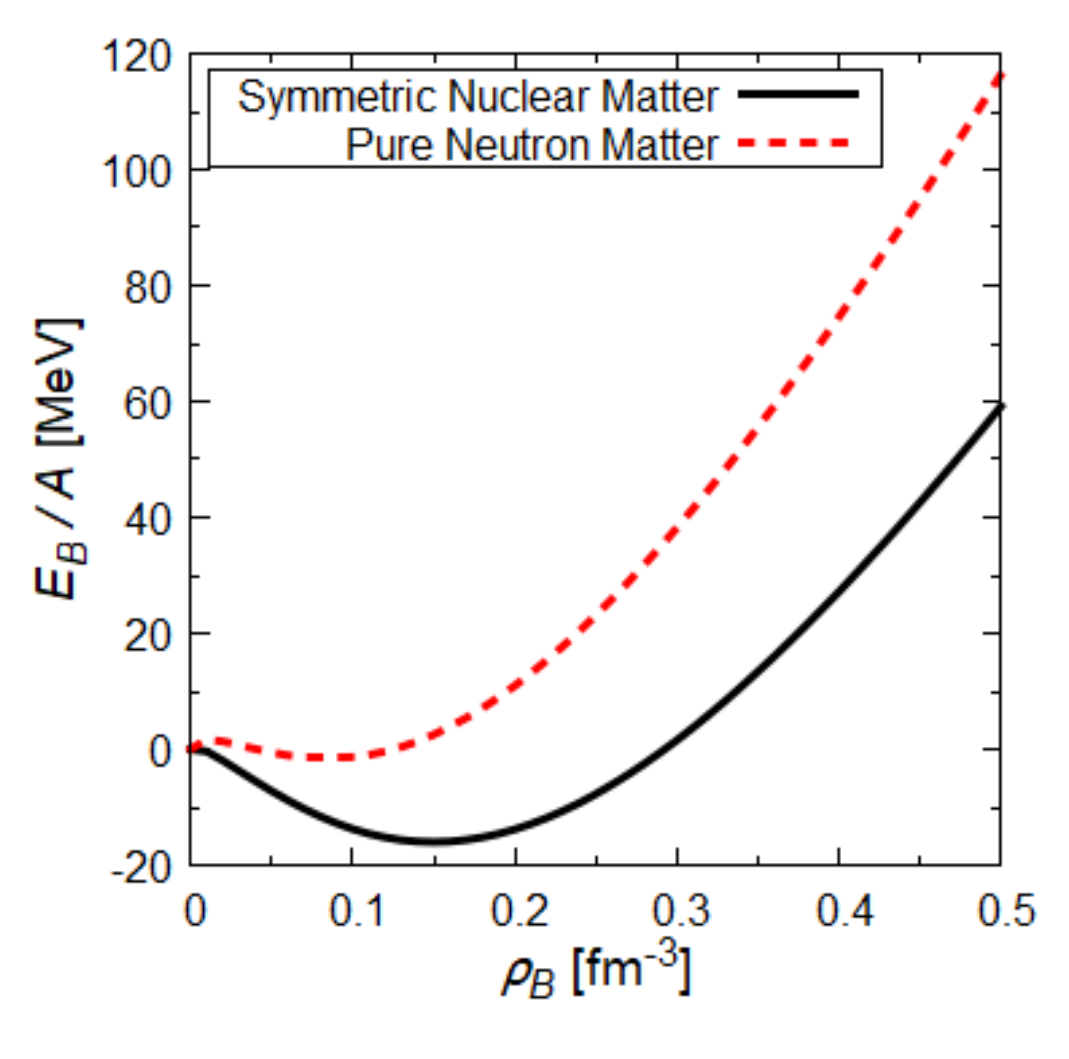} 
\centering\includegraphics[width=\columnwidth]{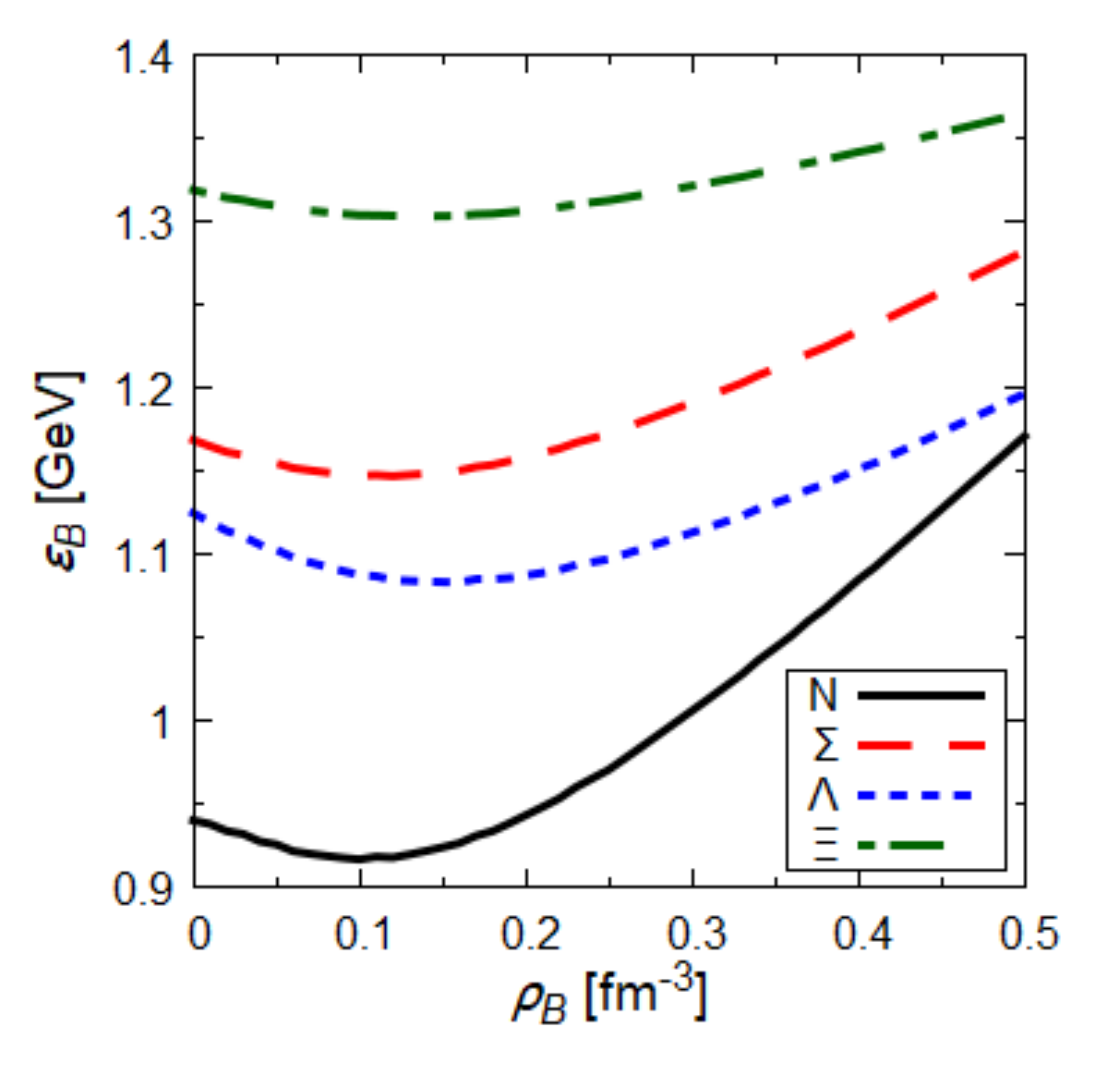}
\caption{(Color online) Binding energy per nucleon in symmetric nuclear matter and pure neutron matter ({\it top panel}), and baryon
energies $\varepsilon_B$ of Eq.~(\ref{eq:baryonenergy}) in symmetric nuclear matter ({\it bottom panel}) as functions of the baryon density.
In the bottom panel, the nucleon Fermi momentum is used in $\varepsilon_N$, while the momentum is set to zero for the other
baryons.}
\label{fig:energies_snm}
\end{figure}

\begin{figure}
  \centering\includegraphics[width=\columnwidth]{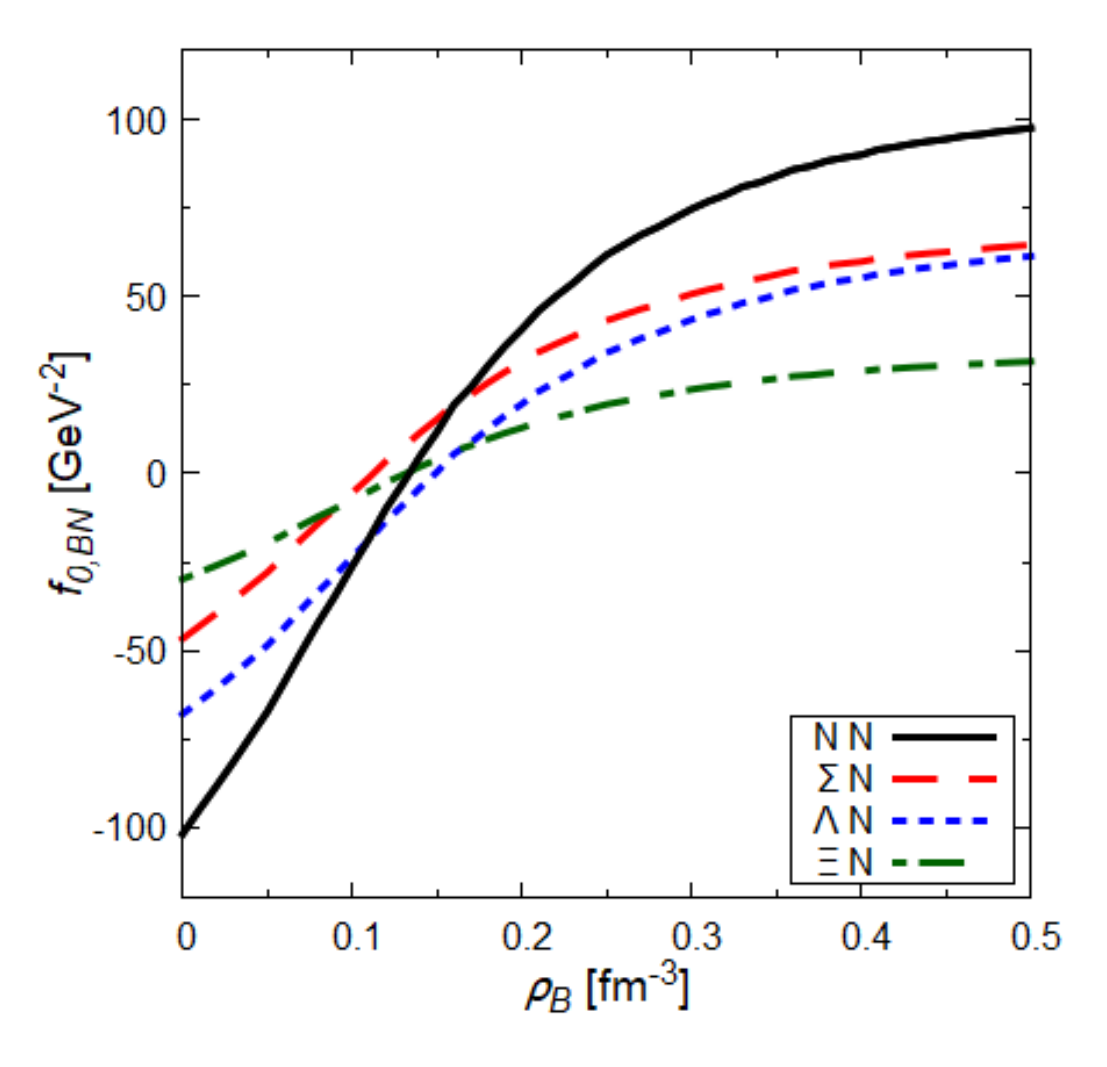} \\ 
  \centering\includegraphics[width=\columnwidth]{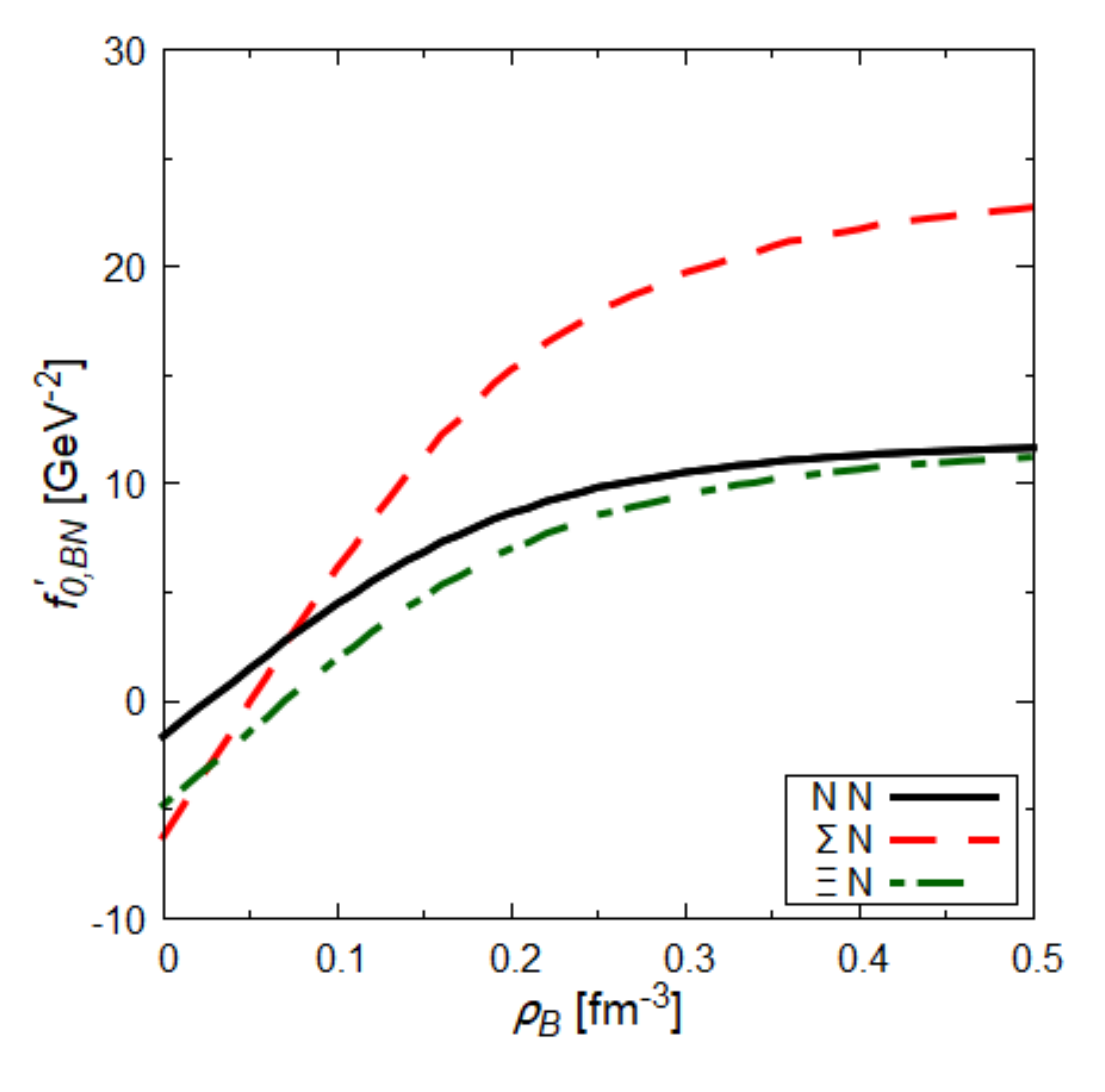}
\caption{(Color online) The $\ell=0$ part of the isoscalar baryon-nucleon interaction $f_{0,BN}$ ({\it top panel}), and
the corresponding isovector interaction $f'_{0,BN}$ ({\it bottom panel}) in symmetric nuclear matter 
as functions of the baryon density.} 
\label{fig:interactions_snm}
\end{figure}

In the top panel of Fig.~\ref{fig:energies_snm} we show the binding energies per nucleon 
(${\cal E}/\rho_B - M_{N0}$) in symmetric nuclear matter (SNM)
in comparison to pure neutron matter (PNM). Although we have only one parameter ($G_v$) to fit the binding energy at saturation
in SNM, the
result for the saturation density agrees with the empirical value. On the other hand, as we do not have any further free 
parameters, our results for the incompressibility (symmetry energy) in SNM are too large (too small) compared to the
empirical values, as will be discussed in more detail in connection to Fig.~\ref{fig:interactions_snm} later. 
Because of the small symmetry energy, our PNM is slightly bound around densities of $\rho_B = 0.1$ fm$^{-3}$.

%===============================================================================
%===============================================================================

In order to show the effects of isospin breaking in PNM on the effective quark and nucleon masses, we list in
Tab.~\ref{tab:masses} the masses in SNM and PNM for four values of the baryon density. Here we can see several points:
First, as can be shown from the gap equation~(\ref{eq:quarkmass}), for systems with an excess of $d$-quarks, 
the magnitude of the mean scalar field $\sigma_d$ decreases more
rapidly with density than the magnitude of $\sigma_u$. Therefore $M_u > M_d$, and one can expect that in an isospin multiplet the 
baryons with more $u$-quarks will be heavier. Second, as will be explained in detail later, the isospin splittings for the 
baryons are generally smaller than for the quarks, because of the scalar isovector polarizabilities of the baryons.

%===============================================================================
\begin{table}[tbp]
\addtolength{\extrarowheight}{2.2pt}
    \centering
    \caption{Effective masses of quarks and nucleons (in units of GeV) in symmetric nuclear matter (SNM) and pure neutron matter
    (PNM) for four values of the baryon density (in units of fm$^{-3}$).}
    \begin{tabular*}{\columnwidth}{@{\extracolsep{\fill}}cccccc}
        \hline\hline
         case   & $\rho_B$ & $M_u$ & $M_d$  &  $M_p$  &   $M_n$    
         \\ \hline
        SNM & 0     &  0.4      & 0.4      &   0.94   &  0.94  \\
            & 0.15  &  0.325    & 0.325    &   0.765  &  0.756  \\
            & 0.3   &  0.284    & 0.284    &   0.683  &  0.683  \\
            & 0.5   &  0.257    & 0.257    &   0.648  &  0.648  \\
         \hline
        PNM & 0     &  0.4      & 0.4      &   0.94   &  0.94   \\
            & 0.15  &  0.340    & 0.314    &   0.768  &  0.755   \\ 
            & 0.3   &  0.301    & 0.275    &   0.695  &  0.686   \\
            & 0.5   &  0.271    & 0.251    &   0.656  &  0.651  \\
        \hline\hline
    \end{tabular*}
    \label{tab:masses}
\end{table}
%===============================================================================

In the bottom panel of Fig.~\ref{fig:energies_snm} we show the Fermi energies (chemical potentials) of the baryons
in symmetric nuclear matter. (As the Fermi momenta of hyperons immersed in nuclear matter are zero, the corresponding lines show the energies 
of hyperons at rest.) The line $\varepsilon_N$ shows that nuclear matter is unstable for densities below
$0.1$ fm$^{-3}$, and at the saturation density it takes the value $940 - 16 = 924$ MeV. It is seen that the $\Lambda$
is bound stronger than the nucleon around the saturation density, although its effective mass (not shown
here) drops more slowly than $M_N$ with increasing density. The reasons are, first, that the curve for $\varepsilon_{\Lambda}$ 
refers to zero momentum, corresponding to low energy orbitals in finite nuclei, while $\varepsilon_N$
refers to the Fermi surface. Second, as shown by Eq.~(\ref{eq:baryonenergy}), the vector repulsion for the 
$\Lambda$ in symmetric nuclear matter ($12 \, G_v \, \rho_B$)  is smaller than for the nucleon ($18 \, G_v \, \rho_B$), 
because $\omega^0_s$ vanishes here.
%This makes the curve $\varepsilon_{\Lambda}$ flatter than $\varepsilon_N$. 

The curves $\varepsilon_{\Lambda}$ and $\varepsilon_{\Sigma}$ in Fig.~\ref{fig:energies_snm} 
show a quite different behavior. Because in this case 
the vector repulsion is the same, the increase of the difference between the two lines with increasing baryon density 
reflects the different dependence of their effective masses on $\rho_B$. As a result, around the saturation density the $\Sigma$ is bound by only 
half of the amount of the $\Lambda$, i.e., by about 22 MeV less than the $\Lambda$ in our model, which is consistent 
with the estimate of about 20 MeV presented in Ref.~\cite{Guichon:2008zz}. The reason for this lies in the different quark substructure: 
The scalar diquark made of $(u,d)$
quarks, which is the main source of attraction in the $\Lambda$ as well as the nucleon, is absent in the $\Sigma$ as well as in the $\Delta$ baryon. 
%Eqs.(\ref{eq:sigma}) and (\ref{eq:lambda}) in App.~\ref{app:baryons}. 
This difference in quark structure, which is well known from the constituent quark model~\cite{Close:1979bt}, generates the mass
difference between the free $\Sigma$ and $\Lambda$ baryons shown in Tab.~\ref{tab:octetmasses}, and increases with increasing
baryon density because the mass of the scalar diquark decreases more rapidly than the mass of the axial vector diquark.
The strong $(u,d)$ correlations in the scalar channel, as compared to the axial vector channel, play a role similar to the color magnetic 
spin-spin interaction from gluon exchange. 
%which plays an important role in the QMC model ~\cite{Guichon:2008zz,Leong:2022aam}, 
In our model we adjusted this strength to reproduce the $\Delta - N$ mass difference in free space.

%Because the mass of the scalar diquark decreases more rapidly with density than the 
%mass of the axial vector diqaurk, the in-medium mass difference between the $\Sigma$ and the $\Lambda$ increases.
The flattening of the curves $\varepsilon_B$ with increasing energy, shown in the lower panel of Fig.~\ref{fig:energies_snm}, 
continues further to the $\Xi$, because in the present model with 4-fermi
interactions the $s$-quark does not participate in the nuclear interactions in symmetric nuclear matter.

In spite of the increasing $\Sigma - \Lambda$ mass difference due to their different quark substructures, the $\Sigma$ baryon
is still bound in our mean field model. It is now believed that the $\Sigma$ is unbound in the nuclear medium~\cite{Gal:2016boi}, and
recent experiments support this view~\cite{J-PARCE40:2022nvq}. 
%Because we cannot increase our vector repulsion acting on strange quarks by hand because of the flavor and chiral symmetry restrictions on the form of the
%underlying Lagrangian (\ref{eq:lagrangian}), 
It would be natural as a next step to include the effects of antisymmetrization (exchange terms), 
both on the level of baryons and the level of quarks. It is, in fact, well known that quark exchange effects appear naturally
in the hadronization of the NJL model in the path integral formalism~\cite{Bentz:2002um,Nagata:2003gg}. The effects of the Pauli exclusion principle on the level of
quarks to produce the $\Sigma N$ repulsion have been emphasized very much recently~\cite{Oka:2023hdc,J-PARCE40:2022nvq}. 
Since the aim of the present work is to explore the effects of the quark
substructure of baryons in a mean field approximation for many baryon systems, we will leave this interesting subject for future studies.

%===============================================================================
%===============================================================================

\subsubsection{\bf{\em{Baryon-nucleon Fermi liquid parameters}}\label{sec:fermiliquid1}}

The top panel of Fig.~\ref{fig:interactions_snm} shows the $\ell=0$ part of the isoscalar baryon-nucleon interaction, given by Eq.~(\ref{eq:isoscalar}) without the last term $\propto \vect{p} \cdot \vect{p}'$, 
and the bottom panel shows the corresponding isovector one, Eq.~(\ref{eq:isovector}). As in the figure for the
baryon energies, the momentum of the nucleons is set to the Fermi momentum ($p_N$), and for the
hyperons it is set to zero. The behavior of all curves in this figure
reflects the change from attraction due to scalar meson exchange at low densities to repulsion from vector
meson exchange at higher densities. We find that the third and fourth factors in the first
lines of~(\ref{eq:isoscalar}) and~(\ref{eq:isovector}), which reflect the couplings of the scalar mesons to the baryons, 
decrease substantially in magnitude as the density increases,
while the denominators given in the second lines of~(\ref{eq:isoscalar}) and~(\ref{eq:isovector}) become slightly enhanced 
because of cancellations between
the attractive quark loop and repulsive baryon loop contributions. Therefore    
the attraction from scalar meson exchange decreases much faster with increasing density than for the case of elementary hadrons.
In order to illustrate this point more quantitatively, we show in Tab.~\ref{tab:factors} the various factors which characterize
the meson-baryon couplings and meson masses in Eqs.~(\ref{eq:isoscalar}) and~(\ref{eq:isovector}). (The full results for the couplings and meson masses, including the effects of the quark-meson couplings, are given in App.~\ref{app:mesonexchange}.)

%===============================================================================
\begin{table}[tbp]
\addtolength{\extrarowheight}{2.2pt}
\centering
\caption{Values of various factors in the effective $\ell=0$ interactions $f_{0,BN}$ and $f'_{0,BN}$ of Eq.~(\ref{eq:isoscalar}) and~(\ref{eq:isovector}) which characterize the meson-baryon couplings and meson masses,
for four values of the baryon density (in units of fm$^{-3}$). The scalar isoscalar polarizability 
($\frac{\partial^2 M_N}{\partial M^2}$) and the scalar isovector polarizability ($\frac{\partial^2 M_p}{\partial (\Delta M)^2}$)
of the nucleon are given in units of GeV$^{-1}$, and the denominators $D$ given in the second lines of~(\ref{eq:isoscalar}) and~(\ref{eq:isovector}) are expressed in units of GeV$^{-2}$.}  
\begin{tabular*}{\columnwidth}{@{\extracolsep{\fill}}c|cccc|cc}
\hline\hline
$\rho_B$   &    $\frac{\partial M_N}{\partial M}$   & $\frac{\partial M_{\Lambda}}{\partial M}$    &  
$\frac{\partial M_{\Sigma}}{\partial M}$   &  $\frac{\partial M_{\Xi}}{\partial M}$   & $\frac{\partial^2 M_N}{\partial M^2}$  & D   \\
\hline
0      &   2.74  &   1.83  &  1.55  &  0.86 &   5.90  &   0.0357  \\
0.15   &   2.07  &   1.45  &  1.15  &  0.65 &   12.6  &   0.0398  \\
0.3    &   1.49  &   1.14  &  0.86  &  0.49 &   15.9  &   0.0532  \\
0.5    &   1.04  &   0.89  &  0.63  &  0.37 &   17.2  &   0.0737   \\
\hline   \hline  
$\rho_B$   &    $\frac{\partial M_p}{\partial (\Delta M)}$   &     &     
$\frac{\partial M_{\Sigma^+}}{\partial (\Delta M)}$   &  $\frac{\partial M_{\Xi^0}}{\partial (\Delta M)}$   & 
$\frac{\partial^2 M_p}{\partial (\Delta M)^2}$  & D   \\
\hline
0      & 0.70  &      &    1.55    &  0.86     &   9.35  &   0.0357  \\
0.15   & 0.49  &      &    1.15    &  0.65     &   14.7  &   0.0418  \\
0.3    & 0.33  &      &    0.86    &  0.49     &   18.1  &   0.0572  \\
0.5    & 0.21  &      &    0.63    &  0.37     &   19.8  &   0.0819  \\
\hline\hline
\end{tabular*}
\label{tab:factors}
\end{table}
%===============================================================================

%is the scalar polarizability of the hadrons.
%That is, the hadron masses or their isospin splittings develop a positive curvature as $M$ or $\Delta M$ decreases, respectively.
%These scalar-isoscalar and scalar-isovector polarizabilities, $\frac{\partial M}{\partial M^2}$ and 
%$\frac{\partial M_p}{\partial (\Delta M)^2}$, appear explicitly as repulsive contributions to the $\sigma$ and $\delta$ masses
%in (\ref{eq:isoscalar}) and (\ref{eq:isovector}), and lead to a reduction of the meson-baryon couplings.

The curves in the top panel of Fig.~\ref{fig:interactions_snm} are related to the baryon energies
of Fig.~\ref{fig:energies_snm} by the first of the two general relations given in Eq.~(\ref{eq:energychanges}).\footnote{For the case of the nucleon, however, the momentum $k$ is set to the Fermi momentum $p_N$ {\em after} the
differentiation in~(\ref{eq:energychanges}).} It is thus natural that the average values of
$f_{0, BN}$ become smaller in the sequence $NN \rightarrow \Lambda N \rightarrow \Sigma N \rightarrow \Xi N$.
In particular, as explained above, $M_{\Lambda}$ decreases faster with density than $M_{\Sigma}$, and therefore  
the first term in~(\ref{eq:f0bn}) shows that the $\Lambda N$ attraction at low densities is stronger than the
$\Sigma N$ attraction.
%$\Sigma \, N$ attraction at low densities is smaller than the $\Lambda \, N$ attraction due to
%the smaller coupling constant ($|\partial M_{\Sigma}/\partial M| <  |\partial M_{\Lambda}/\partial M|$). 
%In particular,
%the $\Sigma N$ interaction is much more repulsive than the $\Lambda N$ interaction in the low density region, for the same
%reason as discussed above in connection to the baryon energies. 
Around the saturation density, the $\Lambda N$ and
the isoscalar $\Xi N$ interactions are similar and very small, while the isoscalar $NN$ and $\Sigma N$ interactions 
are both repulsive.\footnote{We remind again that $f_{0, BN}$ refers to the spin averaged interaction characterized by $\ell=0$ and $T=0$ in the particle-hole channel.}
For the $NN$ case, we can use Eqs.~(\ref{eq:incomp}) and~(\ref{eq:f0bn}) to split the incompressibility
as $K = \frac{3 p_N^2}{E_N} + 9 \rho_B \frac{M_N}{E_N} \frac{\partial M_N}{\partial \rho_B} + 162 \, G_v \, \rho_B
= (0.253 - 1.014 + 1.124)$ GeV = $0.363$ GeV, where the first term refers to noninteracting quasiparticles with $E_N = 0.8$ GeV, 
the second term corresponds to $\sigma$ meson exchange and the third term to $\omega$ meson exchange. In order to reproduce the 
empirical value $K \simeq 0.25$ GeV, we would need $f_{0, NN} \simeq 0$ at saturation density, instead of the positive value indicated in Fig.~\ref{fig:interactions_snm}.

The 3 curves in the lower panel of Fig.~\ref{fig:interactions_snm} similarly result from the attraction due to
$\delta$ meson exchange at low densities and the repulsion from $\rho$ meson exchange at higher densities. 
The fact that the isovector $\Sigma N$ repulsion is stronger than the others is simply because of the
isospin factor $t_b$ in Eq.~(\ref{eq:f0pbn}), which indicates that the energy of $\Sigma^{\pm}$ is most sensitive to changes of the
isovector nucleon density. The overall size of the isovector interactions is
small compared to the isoscalar ones. For the $NN$ case, we can use Eqs.~(\ref{eq:symmetry}) and~(\ref{eq:f0pbn}) 
to split the symmetry energy
as $a_s = \frac{p_N^2}{6 \, E_N} + \frac{\rho_B}{2} \frac{M_N}{E_N} \frac{\partial M_p}{\partial \rho_{(3)}} + G_v \, \rho_B
= (14 - 3 + 7)$ MeV = $18$ MeV, 
where the first term refers to noninteracting quasiparticles, the second
term corresponds to $\delta$ meson exchange and the third term to $\rho$ meson exchange. 
It is known from the case of elementary nucleons~\cite{Ulrych:1997es} that 
the mechanism of $\delta$ meson exchange gives a negative
contribution to the symmetry energy, and in our model this effect is small. 
Our value of $a_s$ is considerably smaller than the empirical value $a_s \simeq 32$ MeV, which reflects the fact that
our 3-flavor Lagrangian~(\ref{eq:lagrangian}) does not allow for an independent vector coupling in the
isovector channel because of the assumed flavor and chiral symmetry, in contrast to the 2-flavor case~\cite{Tanimoto:2019tsl}.

Finally in this subsection, we add two more comments. The first concerns the isospin splittings which can be expected for 
isospin asymmetric matter.  
Because our $f'_{0,BN}$ is negative at small densities, the first term in  
Eq.~(\ref{eq:f0pbn}) is negative for $b=p, \Sigma^+, \Xi^0$. For systems with neutron excess ($\rho_{(3)}<0$) 
we can then expect that the in-medium isospin splittings will be ordered such that 
the particles with more $u$-quarks become heavier, which is consistent with our finding that $M_p > M_n$ and $M_u > M_d$ in
neutron rich matter, see Tab.~\ref{tab:masses}.
The reason why the mass splittings for baryons are smaller than for quarks is now clear from Tab.~\ref{tab:factors}, 
which shows that the isovector couplings $\partial M_b / \partial (\Delta M)$ strongly
decrease with increasing baryon density. Expressed in a different way, the scalar isovector polarizability
of the nucleon ($\frac{\partial^2 M_p}{\partial (\Delta M)^2}$) strongly increases with the density.  

Second, it is well known that any two-body interaction with non-explicit density dependence, for example through
masses and couplings, contains effects from an effective three-body interaction. Taking the $\ell=0$ part of Eq.~(\ref{eq:isoscalar})
as an example, in the case of point nucleons the only density dependence of this kind resides in the factor 
$\frac{M_N}{E_N}$ and in the function $\phi_N$ in the denominator. 
The decrease of our couplings and the increase of meson masses due to the scalar isoscalar polarizability of the nucleons
reflect the presence of additional repulsive three-body interactions.\footnote{The variation of $f_{\,0, bN}$ with density can be expressed as an effective three-body interaction:
\begin{align}
\frac{\delta f_{0, bN}}{\delta \rho_B} = \frac{1}{4} \sum_{\tau=p,n} 
\left(h_{0, b p \tau} + h_{0, b n \tau} \right)  \,,
\nonumber
\end{align}
where the three-particle forward scattering amplitudes ($h_{0}$) are defined as averages over the angles 
between the momenta of the three interacting particles~\cite{Bentz:2020mdk}.}
The rapid decrease of the $bN$ attraction with increasing
density, expressed by Fig.~\ref{fig:interactions_snm}, shows that our effective three-particle interaction
is strongly repulsive, but\emdash as Fig.~\ref{fig:energies_snm} (lower panel) has shown\emdash not sufficient to generate an overall 
repulsion between the $\Sigma$ baryon and the nucleon. 
%To generate such a repulsion, effects of the Pauli
%principle, both on the level of baryons and on the level of quarks from different baryons, have to be taken into account.  

%===============================================================================
\begin{figure}
  \centering\includegraphics[width=\columnwidth]{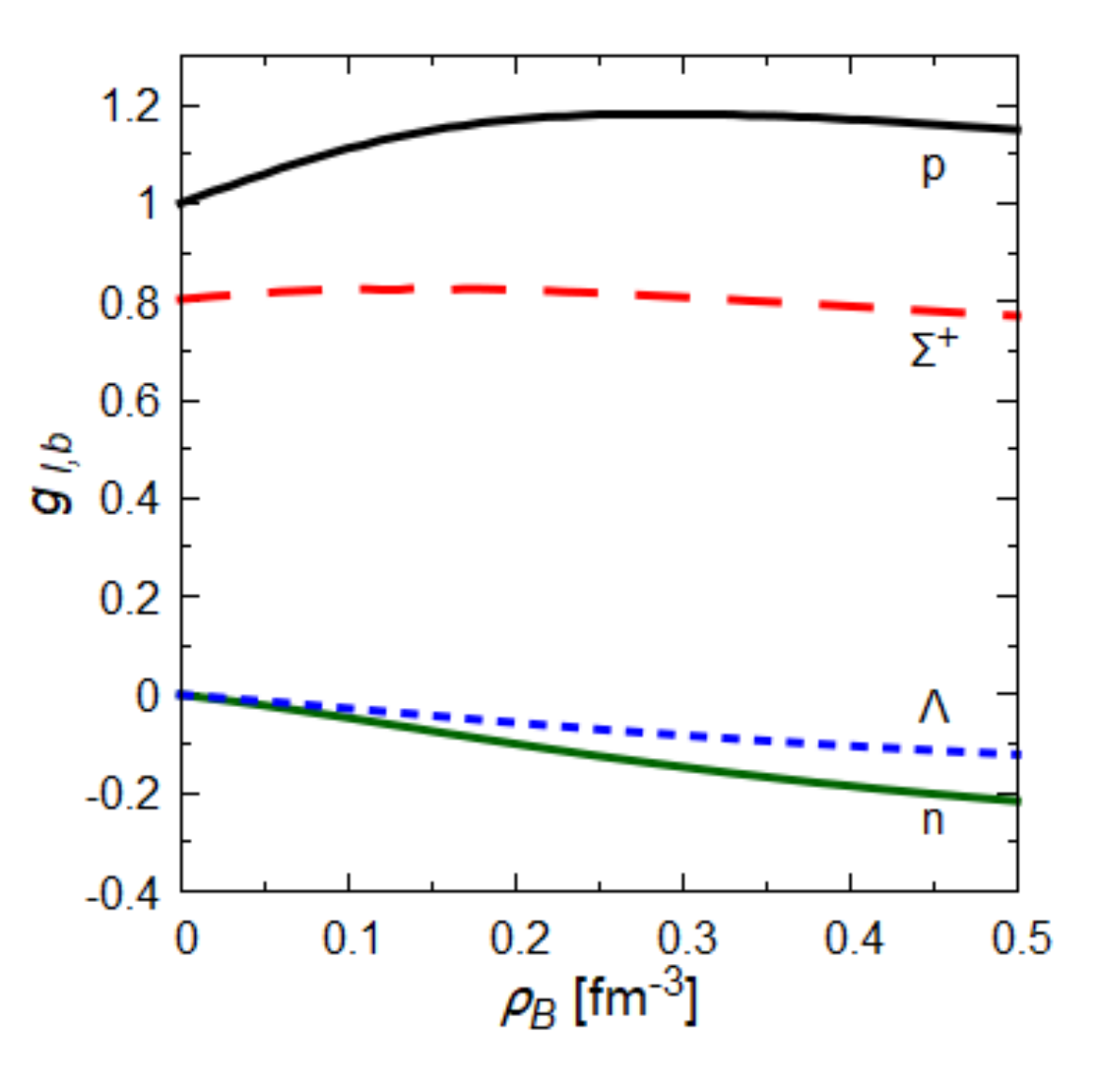} 
\caption{(Color online) The angular momentum $g$-factors (see Eq.~(\ref{eq:gellbaryon})) of the proton in 
comparison to the $\Sigma^+$, and
of the neutron in comparison to the $\Lambda$, in symmetric nuclear matter as functions 
of the baryon density. } 
\label{fig:gell_snm}
\end{figure}
%===============================================================================

\subsubsection{\bf{\em{In-medium orbital $g$-factors of baryons}}\label{sec:gell}}

%*********************************REVISION (begin)*********************************************
Here we wish to illustrate the renormalization of the orbital angular momentum
$g$-factors, given by Eq.~(\ref{eq:gellbaryon}), for a few cases. 
%*******************************REVISION (end)************************************************
In Sec.~\ref{sec:currents}
we used the concept of the backflow, which is central to the Fermi liquid theory, but the same
results can be obtained in relativistic meson-nucleon theories by using the response of
the core (filled Fermi sea of nucleons) to the addition of one nucleon
~\cite{McNeil:1986zz,Ichii:1988jn} or one hyperon~\cite{Cohen:1986tt,Cohen:1993zr}. In such
a description, the backflow arises from RPA-type vertex corrections due to 
virtual $N \overline{N}$ excitations of the core,\footnote{The ``antinucleons'' which show up in those vertex corrections, or in the Z-graph
contributions to the scalar meson propagators mentioned in Sec.~\ref{sec:effbb}, are highly virtual objects,
mathematically necessary to form a complete set of spinors, 
and have little to do with real observable antinucleons.} 
%Unfortunately this point was overlooked in many papers which aimed to show a conceptual problem
%with relativistic meson-nucleon theories.}, 
and the importance of these contributions
to give reasonable magnetic moments in relativistic theories is well known~\cite{Arima:1987hib,Furnstahl:1987rd}. 
As examples for baryons with positive charge, we illustrate the relation~(\ref{eq:gellbaryon})
for the proton and the $\Sigma^+$, and as examples for neutral baryons we show the
cases of the neutron and the $\Lambda$ in Fig.~\ref{fig:gell_snm}.
For the isoscalar combination $g_{\ell, p} + g_{\ell, n}$, the backflow reduces the enhancement
($\frac{M_{N0}}{M_N} \simeq 1.24$ near the saturation density) by a factor of 
$\frac{E_N}{\varepsilon_N} \simeq 0.87$, while for the isovector combination
$g_{\ell, p} - g_{\ell, n}$ there is almost no reduction, because the last term in~(\ref{eq:gellbaryon})
is very small. As a result, the isovector combination remains enhanced, i.e., $g_{\ell}$ of the proton 
(neutron) is larger (smaller) than its free value. 
%These results are consistent with previous findings. 
For the $\Sigma^+$, the
enhancement due to its reduced mass is only about half of the case of the proton,
and the reduction from the backflow gives results which change only mildly with density. 
For the $\Lambda$, the backflow corrections are similar in
magnitude to the case of the neutron, but its effective mass, and therefore also $g_{\ell}$, 
decreases more slowly with density. For more expensive discussions on backflow effects for the
magnetic moments of hypernuclei, we refer to Ref.~\cite{Cohen:1993zr}.

%***********************************REVISION(begin)*************************************************

\subsubsection{\bf{\em{Comments on sizes of quark cores of in-medium nucleons}}\label{sec:sizes}}

Finally in this section, we wish address the question whether the size of in-medium nucleons invalidates the basic physical picture of the mean field approximation. 
The relevance of this question is underlined by the fact that the NJL model is known 
to predict a moderate swelling of nucleons in the medium at normal densities, a feature which 
is important for
the EMC effect~\cite{Cloet:2006bq} or the Coulomb sum rule~\cite{Cloet:2015tha}.
If the nucleons swell considerably at higher densities, the Pauli principle would become inapplicable
at the nucleon level.  

\begin{figure}
  \centering\includegraphics[width=\columnwidth]{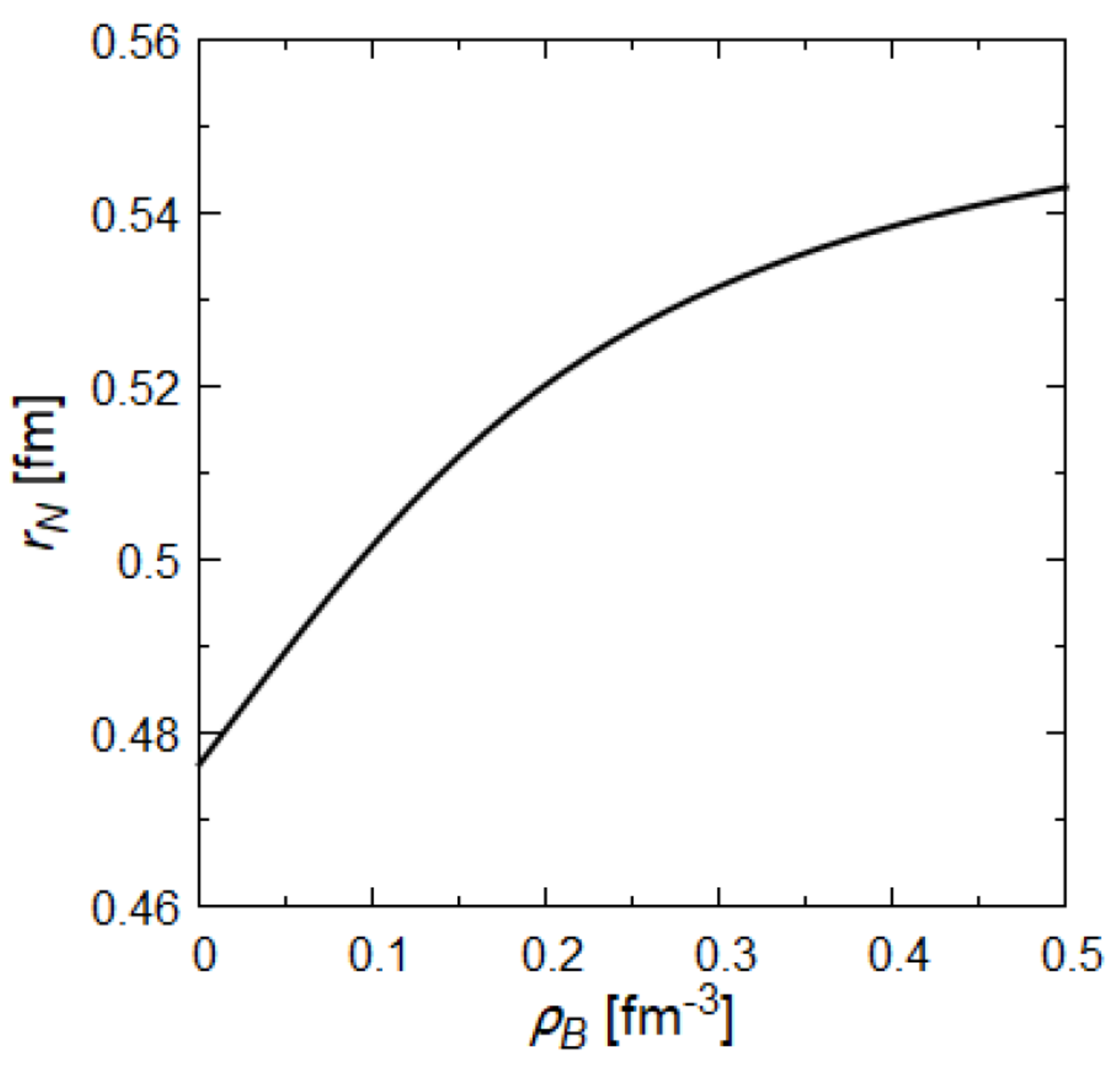} 
\caption{Rms radius of the baryon density distribution of the quark cores in SNM as 
function of the baryon density.} 
\label{fig:sizes_snm}
\end{figure}

Rather than the physical size of nucleons including their meson clouds,
the quantity which seems more relevant for role of the Pauli principle
is the size of the quark cores of the nucleons in the medium. Here we consider the rms radius
of the baryon density distribution of the quark cores, denoted as $r_N(\rho_B)$, which is an 
isoscalar quantity 
and therefore the same for protons and neutrons. The definitions and further details
are given in App.~\ref{app:sizes}, and the results are shown in 
Fig.~\ref{fig:sizes_snm}. Our free nucleon (zero density) value is $r_N(0) = 0.475$ fm, which   
increases by $7\%$ at saturation density ($0.15\,$fm$^{-3}$), and by $13\%$ at $\rho_B=0.5\,$fm$^{-3}$.
Even at very large densities ($\rho_B \simeq 1.0$ fm$^{-3}$) the baryon radius of the quark core increases only by $16\%$ of its free value. 
%and stays almost constant. 
This behavior reflects our phenomenological implementation of confinement effects via the infrared cut-off ($\Lambda_{\rm IR}$). It is interesting to note that our values of $r_N$ are similar to the
radii which have been assumed in the excluded volume framework in QMC model calculations~\cite{Panda:2002iu,Leong:2023lmw}, 
although we do not go into further details here.

%===============================================================================
%===============================================================================

By using the values of $r_N$ shown in 
Fig.~\ref{fig:sizes_snm}, we can estimate the volume fractions occupied by the quark cores in SNM (see App.~\ref{app:sizes}).
We obtain $9\%$ at saturation density, and $36\%$ at the highest density 
shown in the previous figures of this section ($\rho_B = 0.5$ fm$^{-3}$). 
Although these numbers may give us some confidence in the overall physical picture of the 
mean field approximation, they leave room for corrections and improvements of the model.
We also remind that the Pauli principle at the quark level has been predicted to play an important 
role in producing the $\Sigma N$ repulsion even at normal densities, as mentioned at the end of Sec.~\ref{sec:energies}. 
Further investigations on these points are necessary.
 
%*********************************REVISION(end)*************************************************************            

\section{NEUTRON STAR MATTER\label{sec:neutronstar}}

In this section we wish to discuss our results for neutron star matter and the resulting star masses,
based on the expression~(\ref{eq:energydensity}) for the energy density and the equilibrium and charge
neutrality conditions~(\ref{eq:constraints}). Our parameters are the same as used in symmetric nuclear matter,
see Sec.~\ref{sec:parameters}. As mention at the end of Sec.~\ref{sec:bbi}, we will not analyze the effective
baryon-baryon interactions in neutron star matter as exhaustively as we have done it for nuclear matter, in order to keep the length of the paper within reasonable limits. 

\subsection{Single particle properties in-medium\label{sec:singleparticle}}

\begin{figure}
\hspace{-1cm}
  \centering\includegraphics[width=\columnwidth]{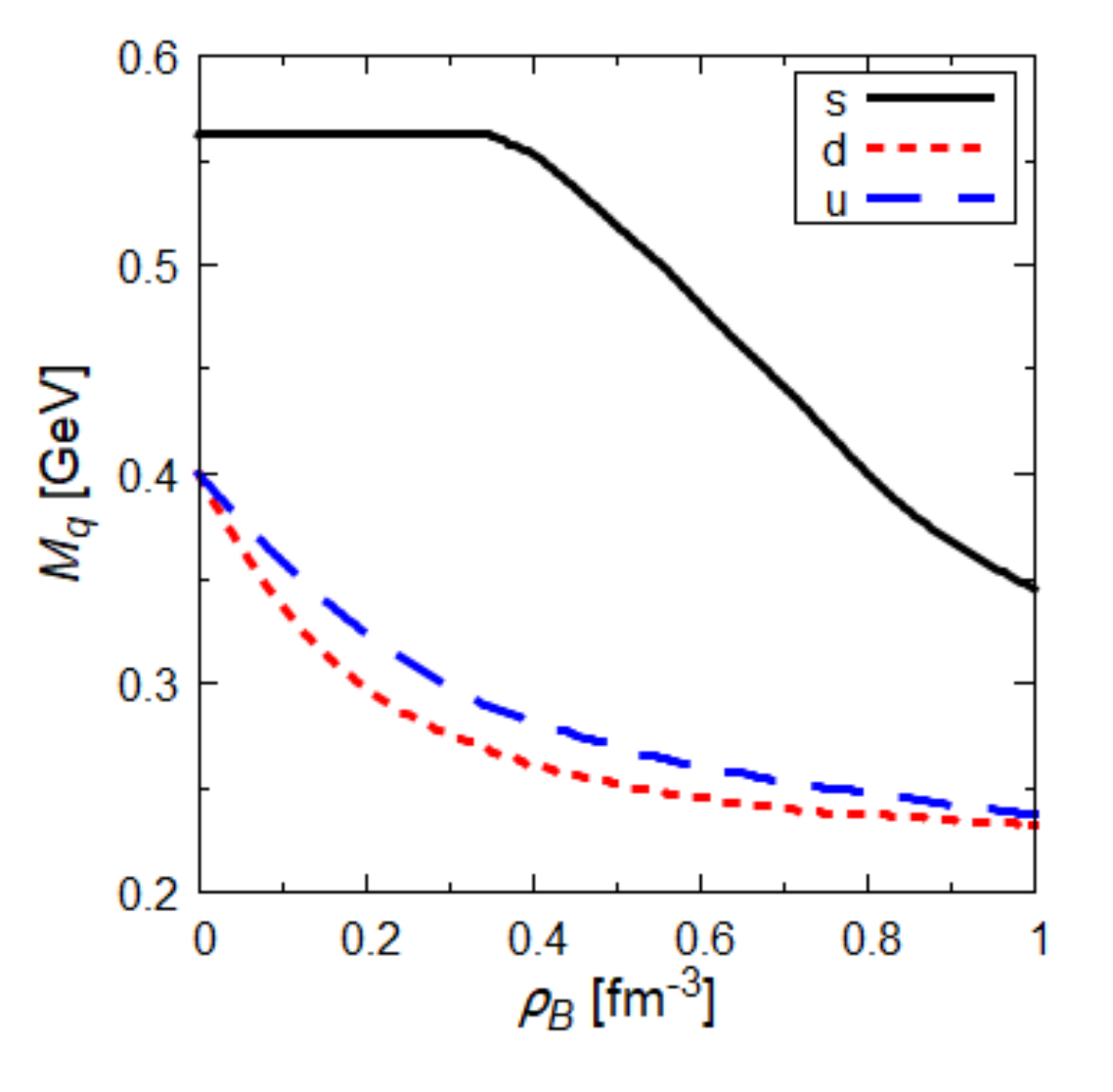} 
\caption{(Color online) The effective quark masses in neutron star matter as functions of the baryon density.} 
\label{fig:effq_nsm}
\end{figure}

First we show our results for the quark effective masses in Fig.~\ref{fig:effq_nsm} as functions of the baryon
density. Because of the isospin asymmetry (excess of $d$-quarks) in neutron star matter, 
the $u$-quark becomes heavier than the $d$-quark by  25 MeV at baryon densities around $0.3$ fm$^{-3}$. 
As discussed already in Sec.~\ref{sec:energies},
this is expected from $|\sigma_u| > |\sigma_d|$ in neutron rich matter, or equivalently from the 
effective $\delta$-meson
exchange mechanism~\cite{Ulrych:1997es} in hadronic theories. The $s$-quark mass, on the other hand, starts to
decrease as soon as hyperons appear in the
system, i.e., as soon as the condition $\partial {\cal E}/\partial \sigma_s = 0$ receives contributions 
from hyperons in the baryon loop term ${\cal E}_B$ of Eq.~(\ref{eq:energydensity}). In this case, the 
$\widebar{s} s$ exchange between hyperons can proceed without violating the OZI rule~\cite{Okubo:1963fa,Zweig:1964jf,Iizuka:1966fk}, and, as anticipated in
Sec.~\ref{sec:Intro}, this gives rise to an appreciable attraction in neutron star matter. 
%which are opposite in sign to the vacuum contributions from the negative energy Dirac sea of quarks. 
We will explain later how this decrease of $M_s$ influences the masses of neutron stars.

%===============================================================================
%===============================================================================

%===============================================================================
\begin{figure}
\hspace{-1cm}
 \centering\includegraphics[width=\columnwidth]{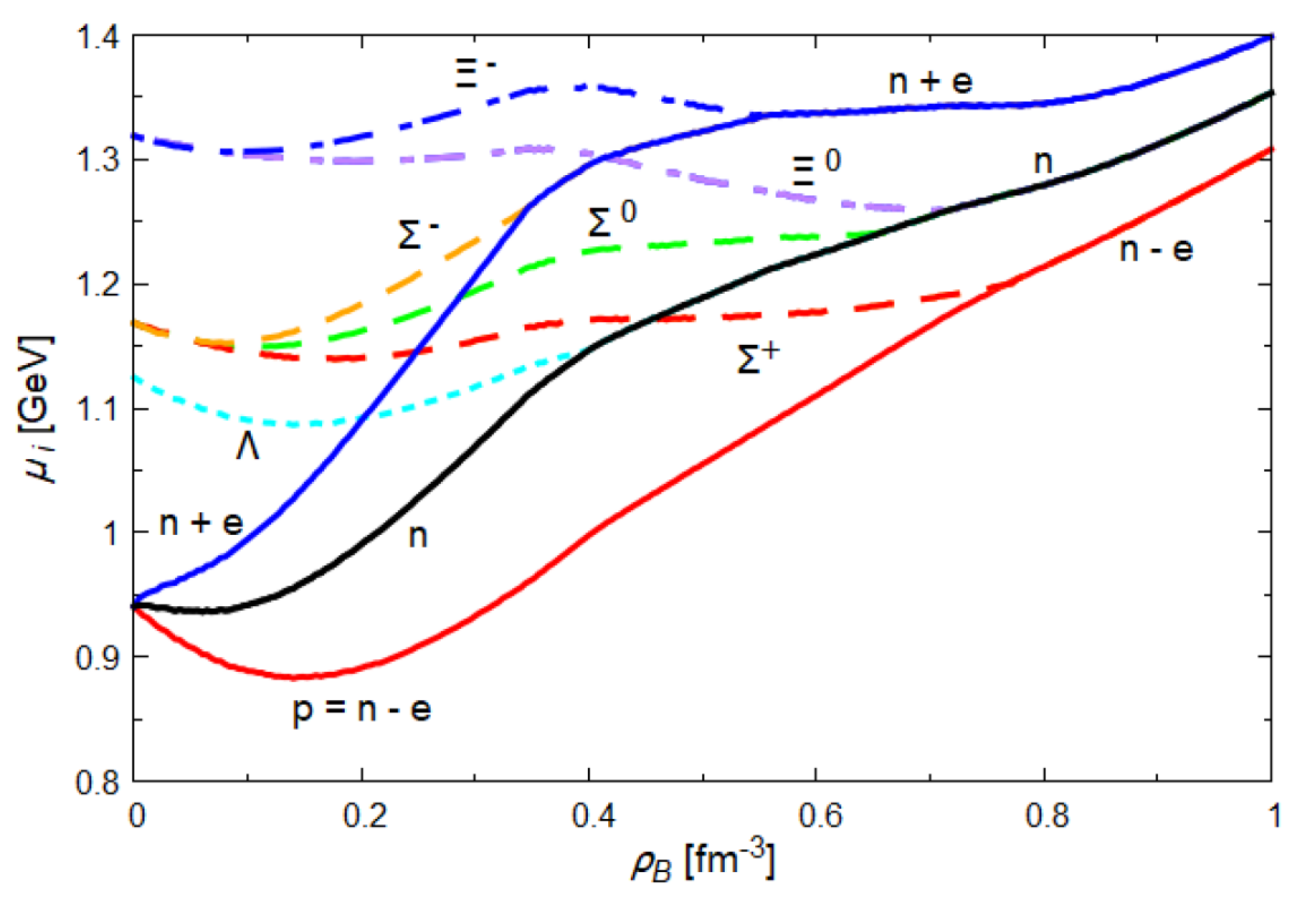} \\
\hspace{-1cm}
  \centering\includegraphics[width=\columnwidth]{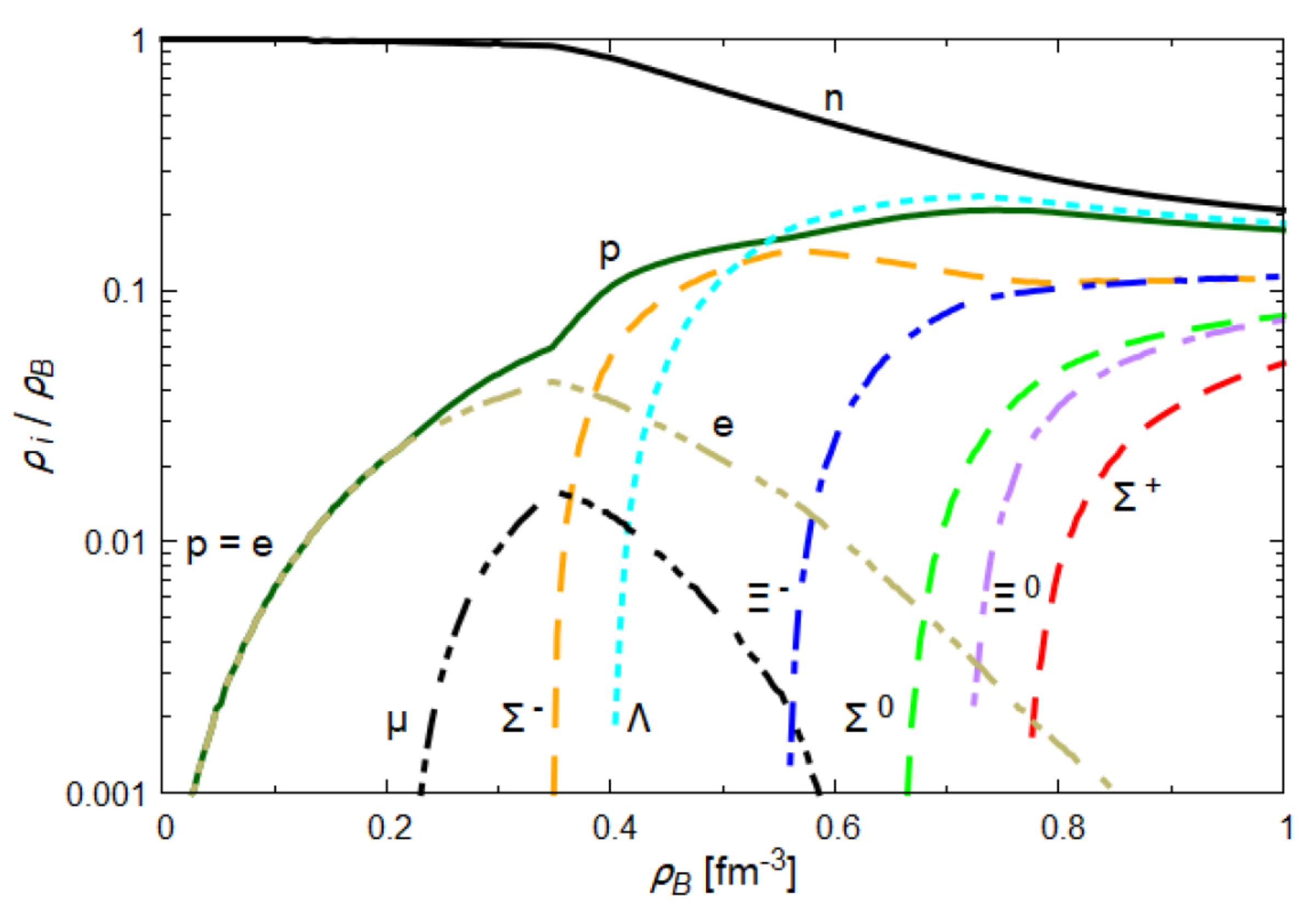}
\caption{(Color online) The chemical potentials of baryons ({\it top panel}), and the density fractions of
baryons and leptons ({\it bottom panel}) in neutron star matter as functions of the baryon density.}  
\label{fig:chemdens_nsm}
\end{figure}
%===============================================================================

The upper panel of Fig.~\ref{fig:chemdens_nsm} shows our results for the chemical potentials, and the lower panel
shows the density fractions of the particles, as functions of the baryon density.
The upper panel is the analogue of the lower panel of Fig.~\ref{fig:energies_snm}, discussed in the previous section for symmetric nuclear matter. 
The three solid lines in the upper panel of Fig.~\ref{fig:chemdens_nsm} (from bottom to top) show $\mu_p = \mu_n - \mu_e$, $\mu_n$, and $\mu_n + \mu_e$. Because of the conditions~(\ref{eq:constraints}), the density where the chemical potential of a hyperon with electric charge $q_b$ touches the solid line
$\mu_n - q_b \mu_e$ from above is the threshold density for this hyperon. Below the threshold
densities, the chemical potentials are simply the energies of hyperons at rest (zero Fermi momentum).
Compared to the symmetric nuclear matter case of Fig.~\ref{fig:energies_snm}, the lines show a considerable isospin 
splitting, which mainly comes
from the vector potential term in Eq.~(\ref{eq:baryonenergy}). For example, the vector potential for 
the $\Sigma^-$ is $4 G_v \left(2 \rho_d + \rho_s \right)$, which is larger than the vector potential
for the $\Sigma^+$, which is $4 G_v \left( 2 \rho_u + \rho_s \right)$. The mass splittings are in the opposite order, 
e.g., the $\Sigma^+$ is heavier than
the $\Sigma^-$, as can be expected also from Fig.~\ref{fig:effq_nsm}. The mass splittings in baryon isospin multiplets are, 
however, small compared
to the splittings from the vector potential. For example, at baryon densities around $0.3$ fm$^{-3}$ the mass splitting
between the $\Sigma^+$ and the $\Sigma^-$ is only about $20$ MeV, and the proton-neutron mass difference is 
only about $10$ MeV, both being smaller than the naive expectation from the quark mass difference shown in 
Fig.~\ref{fig:effq_nsm} for the reasons explained in the previous section.
In the low density region, where $\rho_s=0$ and $\rho_u+\rho_d = 3 \rho_B$, the vector potentials
for the $\Sigma^0$ and the $\Lambda$ are the same ($4 G_v (\rho_u + \rho_d)$), and we see again the different
behaviors of their energies with increasing baryon density, which is caused by their different quark substructures as discussed in Sec.~\ref{sec:energies}. 

As we can see from Fig.~\ref{fig:chemdens_nsm}, the threshold density for the $\Sigma^-$ is $\rho_B = 0.35$ fm$^{-3}$
in our calculation.
Although it has been conjectured for long on energetic reasons that the $\Sigma^-$ will appear as the first hyperon in neutron star
matter~\cite{Heiselberg:1999mq}, this point is controversial nowadays~\cite{Stone:2019blq,Motta:2022nlj,Leong:2023yma}, mainly because the
$\Sigma N$ interaction is believed to be repulsive (see the related discussions at the end of Sec.~\ref{sec:energies}). 
However, we wish to note that also in our present mean field model the onset of the $\Sigma^-$ depends on several details:
First, we are underestimating the free $\Sigma$ mass by about 20 MeV (see Tab.~\ref{tab:octetmasses}); second, the
in-medium mass of the $\Sigma^-$ is shifted down by a similar amount relative to the $\Sigma^+$ as explained above; and third, 
our electron chemical potential is rather large in this density region. Therefore, apart from the more fundamental problem
on the $\Sigma N$ repulsion,    
the question whether $\mu_{\Sigma^-}$ touches $\mu_n + \mu_e$ or not, and if it does at which baryon density, 
depends on several details of the model. (We will return to this point in a different context in Sec.~\ref{sec:summary}.)

%===============================================================================
\begin{figure}
  \centering\includegraphics[width=\columnwidth]{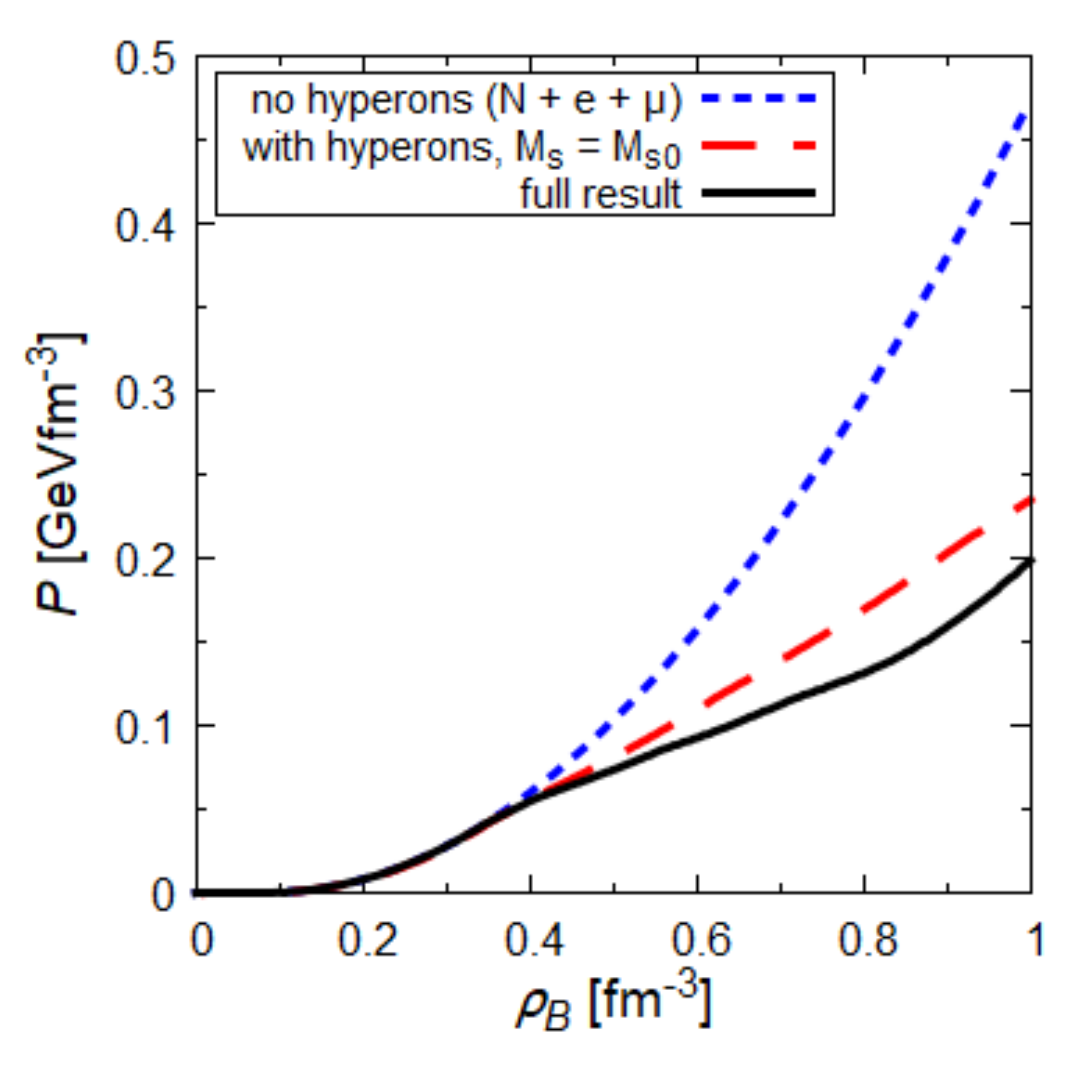} \\ 
  \centering\includegraphics[width=1.05\columnwidth]{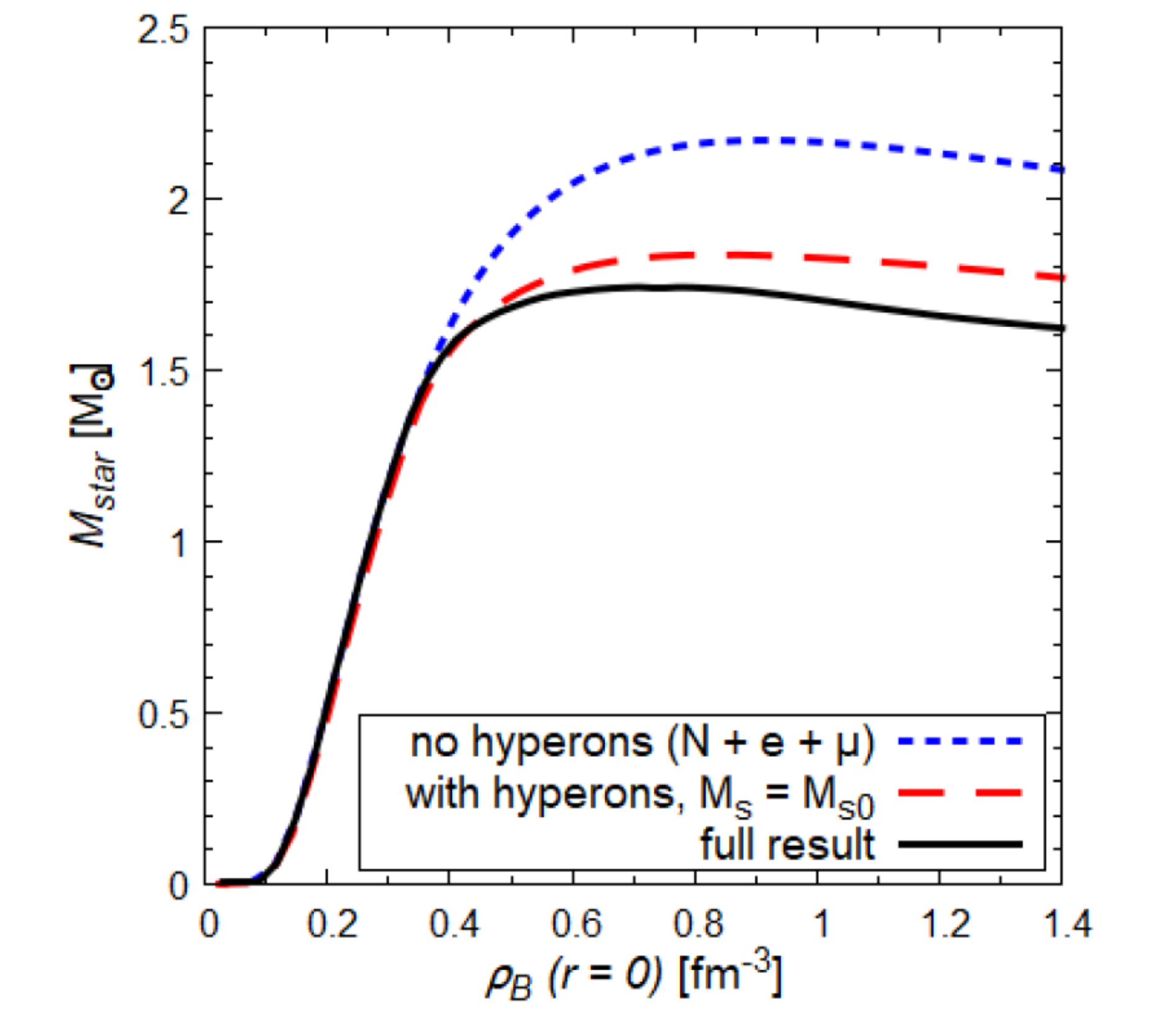}
\caption{(Color online) The pressure in neutron star matter as function of the baryon density ({\it top panel}) and
the resulting neutron star masses as functions of the central baryon density ({\it bottom panel}).
The three lines in each panel show the cases of nucleons and leptons only, the case including hyperons but fixing the 
$s$-quark mass to its free value, and the full result obtained with by treating all three quark masses as independent
variational parameters.} 
\label{fig:Press_4fermi}
\end{figure}
%===============================================================================

\subsection{Equation of state and neutron star masses\label{sec:eos_star}}

The upper panel of Fig.~\ref{fig:Press_4fermi} shows our results for the pressure in neutron star matter as function of the
baryon density, and the lower panel shows the neutron star masses, as obtained from the solution of the
Tolman-Oppenheimer-Volkoff (TOV) equations~\cite{Tolman:1939jz,Oppenheimer:1939ne}, with the constraints of Eq.~(\ref{eq:constraints}) imposed, as functions of the central baron
density. We show the cases of nucleons and leptons only, the case including hyperons but fixing artificially the effective $s$-quark mass 
to its free value ($M_{s0}$), and the
full result with the $s$-quark mass determined by minimization of the energy density. 
The results for nucleons and leptons only are very similar to the results obtained in Ref.~\cite{Tanimoto:2019tsl}
for the flavor $SU(2)$ case, although there it was possible to reproduce the symmetry energy without explicit breaking
of chiral symmetry of the interaction Lagrangian. (See App.~\ref{app:baryons} for a more detailed comparison.) 

It is well known that
the presence of hyperons can lead to a sizable reduction of the pressure in neutron star matter and a decrease of the maximum mass of 
neutron stars~\cite{Glendenning:1997wn,Bombaci:2016xzl}, and Fig.~\ref{fig:Press_4fermi} shows that the same situation is encountered 
in a relativistic mean field calculation which takes into account the internal quark-diquark structure
of the octet baryons. Our results suggest that most of the reduction of the pressure
arises simply because nucleons and leptons with high Fermi momenta can be converted 
to hyperons with low Fermi momenta by weak processes. The reduction of the $s$-quark mass in the medium is not so
important for the overall size of the pressure and the maximum star mass, but it works towards destabilization of the
star as the central baryon density increases. 
The values of the maximum central baryon densities which gives stable stars, the maximum star masses, 
and the radii of the stars with maximum mass for the three cases shown in Fig.~\ref{fig:Press_4fermi} are as follows: 
\begin{align}
(\rho_{B}^{\rm max}(r=0), \, M_{\rm star}^{\rm max}, \, R) 
= (0.9 \, {\rm fm}^{-3}, \,\, 2.17 \, M_{\odot}, \,\,  11.5 \, {\rm km})  
\nonumber
\end{align}
for the case of no hyperons;
\begin{align}
(\rho_{B}^{\rm max}(r=0), \, M_{\rm star}^{\rm max}, \, R) 
= ( 0.85 \, {\rm fm}^{-3}, \,\, 1.83 \, M_{\odot}, \,\, 11.8 \, {\rm km}) 
\nonumber
\end{align}
for the case with hyperons but $M_s$ fixed to $M_{s0}$; and
\begin{align}
(\rho_{B}^{\rm max}(r=0), \, M_{\rm star}^{\rm max}, \, R) 
= ( 0.72 \, {\rm fm}^{-3},  \,\, 1.73 \, M_{\odot}, \,\,  12.3 \, {\rm km})
\nonumber
\end{align}
for the case with hyperons and $M_s$ determined from minimization of the energy density.

\section{ROLE OF 6-FERMI AND 8-FERMI INTERACTIONS\label{sec:sixeight}}

Because the maximum mass of neutron stars is sensitive to the high density behavior of the 
equation of state, it is natural to investigate the role of higher order Fermi interactions,
i.e., the 6-fermi~\cite{tHooft:1976snw} and 8-fermi~\cite{Osipov:2006ns} interactions. 
%******************************************REVISION(begin)*****************************************
While there is little doubt about the importance of the 6-fermi
(flavor determinant) interaction to break the $U_A(1)$ symmetry of the 4-fermi Lagrangian
of Eq.~(\ref{eq:lagrangian}) and to split the masses of the otherwise degenerate pseudoscalar mesons $\pi$ and $\eta$,
%******************************************REVISION(end)*******************************************
the situation is not so clear for the 8-fermi interactions, because many possible flavor structures are allowed by chiral symmetry. In this work we will limit ourselves to three types of chiral invariant 8-fermi interactions with the simplest structure, namely the square of the scalar-pseudoscalar term in Eq.~(\ref{eq:lagrangian}), the product of this term with the vector-axial vector term, and the square of the vector-axial vector term. We wish to investigate whether those higher order Fermi interactions in the $\overline{q} q$ channels, with coupling constants restricted by the basic properties of symmetric nuclear matter around the saturation point, can lead to appreciable changes in high density neutron star matter or not. We will not include higher order interactions in the $qq$ channels used to construct the baryons as quark-diquark bound states, i.e., the Lagrangian of Eq.~(\ref{eq:lagqq}) is left unchanged.

\subsection{Basic formulas and new parameters\label{sec:basic}}  

To the basic NJL Lagrangian of Eq.~(\ref{eq:lagrangian}), we add the 6-fermi (flavor determinant) interaction~\cite{Buballa:2003qv}
\begin{align}
{\cal L}_6 &= G_6 \, {\rm det} \left[ \widebar{q}_{\alpha} \left(1 - \gamma_5 \right) q_{\beta} 
+  \widebar{q}_{\alpha} \left(1 + \gamma_5 \right) q_{\beta} \right] \,,   
\label{eq:sixfermilag} 
\end{align}
and the following 8-fermi interactions:
\begin{align}
{\cal L}_8 &= G^{(ss)}_8 \, \left({\cal L}_s \, {\cal L}_s \right)  
-  G^{(sv)}_8 \, \left({\cal L}_s \, {\cal L}_v \right)  
- G^{(vv)}_8 \, \left({\cal L}_v \, {\cal L}_v \right)   \nonumber \\
& \equiv {\cal L}_8^{(ss)} + {\cal L}_8^{(sv)} + {\cal L}_8^{(vv)} \,. 
\label{eq:eightfermilag}
\end{align}
Here ${\cal L}_s$ means the 4-fermi interaction in the scalar-pseudoscalar
channel of Eq.~(\ref{eq:lagrangian}) without the factor $G_{\pi}$,
and ${\cal L}_v$ means the one in the vector-axial vector channel of
Eq.~(\ref{eq:lagrangian}) without the factor $(- G_v)$.
In this simplest possible form, each factor ${\cal L}_s$ or ${\cal L}_v$
is closed under the summations over Dirac, flavor, and color indices.
Altogether 4 new coupling constants are involved in~(\ref{eq:sixfermilag}) and~(\ref{eq:eightfermilag}).

The mean field approximation is implemented in the same way as 
for the 4-fermi interactions in Sec.~\ref{sec:mf}. 
The gap equation~(\ref{eq:quarkmass}) is now replaced by the more complicated form
\begin{align}
M_{\alpha} = m_{\alpha} - \sigma_{\alpha} \left(1+  \frac{G_8^{(ss)}}{4 G_{\pi}^3} \, \sigma_{\beta}^2 
- \frac{G_8^{(sv)}}{8 G_{\pi} G_v^2} \, \omega_{\beta}^2 \right)
+ \frac{G_6}{8 G_{\pi}^2} \, \sigma_{\beta} \, \sigma_{\gamma} \,, 
\label{eq:masstot}
\end{align}
where in the 6-fermi term $(\alpha, \beta, \gamma)$ is any set of three different quark flavors, and in the other terms a sum over the quark
flavors $\beta$ is implied. The baryon energies~(\ref{eq:baryonenergy}) are replaced by
\begin{align}
\varepsilon_b(k) = E_b(k_b) + n_{\alpha/b} V_{\alpha}^0  \,,
\label{eq:epsilontot}
\end{align}
where $\vect{k}_b = \vect{k} - n_{\alpha/b} \vect{V}_{\alpha}$, with the vector fields $V_{\alpha}^{\mu}$ defined by
\begin{align}
V_{\alpha}^{\mu} = \omega_{\alpha}^{\mu} \left( 1 + \frac{G_8^{(sv)}}{8 G_{\pi}^2 G_v} \, \sigma_{\beta}^2
+ \frac{G_8^{(vv)}}{4 G_{v}^3} \, \omega_{\beta}^2 \right) \,.
\label{eq:vecttot}
\end{align}
The new contributions from the 6-fermi and 8-fermi interactions to the energy density are 
\begin{align}
{\cal E}_6 &= - \frac{G_6}{16 G_{\pi}^3} \, 
\left(\sigma_u \, \sigma_d \, \sigma_s - \sigma_{u0} \, \sigma_{d0} \, 
\sigma_{s0} \right) \,,
\label{eq:esix}  \\
{\cal E}_8 &=  \frac{3 G_8^{(ss)}}{64 G_{\pi}^4} \, \left(\sigma_{\alpha}^2 \, \sigma_{\beta}^2 - \sigma_{\alpha 0}^2 \, \sigma_{\beta 0}^2 \right)  
- \frac{3 G_8^{(sv)}}{64 G_{\pi}^2 G_v^2} \, \sigma_{\alpha}^2 \, \omega_{\beta}^2  \nonumber \\
& - \frac{3 G_8^{(vv)}}{64 G_{v}^4} \, \omega_{\alpha}^2 \, \omega_{\beta}^2  \,, 
\label{eq:eeight}
\end{align} 
%
%**************************REVISION(begin)************************************************
which are added to~Eq.~\eqref{eq:energydensity}, after replacing $\omega_\alpha^{\mu}$ in \eqref{eq:energydensity} by the expression given in Eq.~\eqref{eq:vecttot}. 
%*************************REVSION(end)***************************************************
It is easy to check that the basic conditions~(\ref{eq:minimization}), which determine the three scalar and three vector mean fields $\sigma_{\alpha}$ and $\omega_{\alpha}^{\mu}$,
lead to the same expressions (\ref{eq:fhtheorem}) and (\ref{eq:vectorfields}) as before, because those expressions simply reflect the definitions given by Eq.~(\ref{eq:fields}). 
If we eliminate the vector fields by using~(\ref{eq:vectorfields}), it becomes clear
that $G_8^{(ss)}$ and $G_8^{(vv)}$ must be positive in order that the energy density is bounded from below, 
while the sign of $G_8^{(sv)}$ is not determined generally.  
For the case of neutron star matter, the conditions of chemical equilibrium and charge neutrality
are given by Eq.~(\ref{eq:constraints}) with the modified baryon chemical potentials $\mu_b = \varepsilon_b(k=p_b)$.

We also note that the 6-fermi and 8-fermi interactions lead to a renormalization of the residual 4-fermi interactions.
The only physical quantities, for which we
use the residual 4-fermi interactions in the $\overline{q} q$ channel to fix model parameters in this work, 
are the mass of the pion, the pion decay constant,  
and the $\eta$ - $\eta'$ mass difference, where the pseudoscalar mesons $\eta$ and $\eta'$ arise 
from mixing~\cite{Hatsuda:1991jn,Rehberg:1995kh} between the
$\eta_0$ and $\eta_8$. The effective 4-fermi coupling constants in the vacuum, relevant for those quantities, are
given by (see, for example, Refs.~\cite{Kato:1993np,Rehberg:1995kh} for the 6-fermi case)  
\begin{align}
\tilde{G}_{\pi} &= \left(G_{\pi}+ \frac{G_8^{(ss)}}{4 G_{\pi}^2} \,\sigma_{\alpha 0}^2 \right)  
- \frac{G_6}{8 G_{\pi}} \, \sigma_{s0}  \equiv 19.04 \, {\rm GeV}^{-2} \,,   
\label{eq:gpitilde} \\
\tilde{G}_{00} &=  \left(G_{\pi}+ \frac{G_8^{(ss)}}{4 G_{\pi}^2} \,\sigma_{\alpha 0}^2 \right) 
+ \frac{G_6}{12 G_{\pi}} \left(2 \sigma_{0} + \sigma_{s0}\right) \,,  
\label{eq:g00}  \\
 \tilde{G}_{08} &= - \frac{\sqrt{2} \, G_6}{12 G_{\pi}} \left(\sigma_{0} - \sigma_{s0}\right) \,,
\label{eq:g08}  \\
\tilde{G}_{88} &=  \left(G_{\pi}+ \frac{G_8^{(ss)}}{4 G_{\pi}^2} \,\sigma_{\alpha 0}^2 \right) 
- \frac{G_6}{24 G_{\pi}} \left(4 \sigma_{0} - \sigma_{s0}\right) \,.  \label{eq:g88}  
\end{align}
We require that $\tilde{G}_{\pi}$ has the same value as $G_{\pi}$ in the 4-fermi calculation in order to reproduce
the observed pion mass (see
Tab.~\ref{tab:parameters}), and that $G_6$ reproduces the observed mass difference 
$m_{\eta'} - m_{\eta} = 0.41$ GeV. 
One can use Eq.~(\ref{eq:gpitilde}) to express the quantity 
$\left(G_{\pi}+ \frac{G_8^{(ss)}}{4 G_{\pi}^2} \,\sigma_{\alpha 0}^2  \right)$ in the form 
$\tilde{G}_{\pi} + \frac{G_6}{8 G_{\pi}} \, \sigma_{s0}$. By inserting this into (\ref{eq:g00}) and (\ref{eq:g88}), we see
that the three coupling constants
(\ref{eq:g00}), (\ref{eq:g08}) and (\ref{eq:g88}), which are used to calculate the $\eta-\eta'$ mass difference,
can be expressed in terms of $\tilde{G}_{\pi}$, $G_6$, and the quark condensates in the vacuum,  
$\langle \overline{u} u \rangle_0 = \langle \overline{d} d \rangle_0 = \sigma_0/(4 G_{\pi})$ and 
$\langle \overline{s} s \rangle_0 = \sigma_{s0}/(4 G_{\pi})$, which are fixed by the constituent quark masses
in the vacuum and the cut-offs given in Tab.~\ref{tab:parameters}.
Therefore $G_6$ can be adjusted to the $\eta - \eta'$ mass difference in the standard way~\cite{Kato:1993np},
without recourse to the value assumed for $G_8^{(ss)}$. 
It is also easy to see that the gap equation~(\ref{eq:masstot}) for the
$u, d$ quarks in the vacuum remains numerically the same as in the pure 4-fermi case, because it can be expressed as
$M_0 = m - 4 \tilde{G}_{\pi} \langle \overline{u} u \rangle_0$. Therefore the value of $m$, given in
Tab.~\ref{tab:parameters}, is unchanged.\footnote{The value of the current $s$-quark mass depends 
slightly on the values assumed for $G_6$ and $G_8^{(ss)}$. Also the original 4-fermi coupling constant $G_{\pi}$
changes according to Eq.~(\ref{eq:gpitilde}), although this has no effect on any physical quantity.}
By the standard calculations, we find that $G_6 = 1260$ GeV$^{-5}$ reproduces the observed 
$\eta - \eta'$ mass difference. 
 
Next we comment on the role of the 8-fermi coupling constants. 
As one can expect from the gap equation~(\ref{eq:masstot}), $G_8^{(ss)}$ works into the same direction
as the original 4-fermi coupling $G_{\pi}$, i.e., it gives attraction, while a positive coupling $G_8^{(sv)}$
gives repulsion. The coupling $G_8^{(vv)}$, on the other hand, is not related to the gap equation,
but after eliminating the vector fields according to~(\ref{eq:vectorfields}) it is easily seen to give    
a repulsive contribution of $4 G_8^{(vv)} \rho_{\alpha}^2 \rho_{\beta}^2 = 81 G_8^{(vv)} \rho_B^4$ to the energy density, 
and $12 \, G_8^{(vv)} \rho_{\alpha}^2 \rho_{\beta}^2 = 243 \, G_8^{(vv)} \rho_B^4$ to the pressure in symmetric
nuclear matter. Although the 8-fermi coupling constants can be treated as free parameters, their choice is strongly limited 
by the requirements that the saturation point of isospin symmetric nuclear matter is unchanged, and the
discrepancies of the calculated incompressibility and the symmetry energy to the empirical values
do not increase. In the present calculation, we achieved this by making use of the balance between the 
attractive $(ss)$-type interaction and the repulsive $(vv)$-type interaction. Concerning the $(sv)$-type
interaction, which can work as an attraction ($G_8^{(sv)}<0$) or a repulsion ($G_8^{(sv)}>0$), we found that
the case of attraction leads to conflicts with the nuclear matter equation of state, and the case of
repulsion gives a much smaller effect in neutron star matter than the repulsive $(vv)$-type interaction.
We therefore consider only the case $G_8^{(sv)}=0$ in the calculations described below. We also
note that changes in the original 4-fermi vector coupling constant $G_v$, under the
constraints imposed by symmetric nuclear matter, do not lead to any noteworthy improvements of the equation of state 
of neutron star matter, so we keep the same value as given in Tab.~\ref{tab:parameters}.

In Tab.~\ref{tab:sixeight}, we list as case 1 the pure 4-fermi case, where the 6-fermi and 8-fermi
coupling constants are zero, and in case 2 the 6-fermi interaction with the value
of $G_6$ as determined above is added. Case 3 gives one possible choice
for the 8-fermi coupling constants, where the balance between the attractive $(ss)$-type interaction
and the repulsive $(vv)$-type interaction is used to keep the nuclear matter properties around the
saturation point unchanged, while the $(sv)$-type interaction is assumed to vanish.

%===============================================================================
\begin{table}
\addtolength{\extrarowheight}{2.2pt}
    \centering
    \caption{Values of the 6-fermi coupling constant $G_6$ in units of GeV$^{-5}$, and the 8-fermi
     coupling constants $G_8^{(ss)}$ and $G_8^{(vv)}$ in units of GeV$^{-8}$, for the
     three cases discussed in this section. The coupling $G_8^{(sv)}$ is set to zero in all three cases.} 
    \begin{tabular*}{\columnwidth}{@{\extracolsep{\fill}}cccc}
        \hline\hline
      case &   $G_6$   & $G_8^{(ss)}$  & $G_8^{(vv)}$  
        \\ \hline
       1  &    0    &  0    &    0 \\
       2  &  1260   &  0    &    0 \\
       3  &  1260   & 2330  &   1220   \\
        \hline\hline
    \end{tabular*}
    \label{tab:sixeight}
\end{table}
%===============================================================================

\subsection{Numerical results including 6-fermi and 8-fermi interactions\label{sec:results468}}

%===============================================================================
\begin{figure}
  \centering\includegraphics[width=\columnwidth]{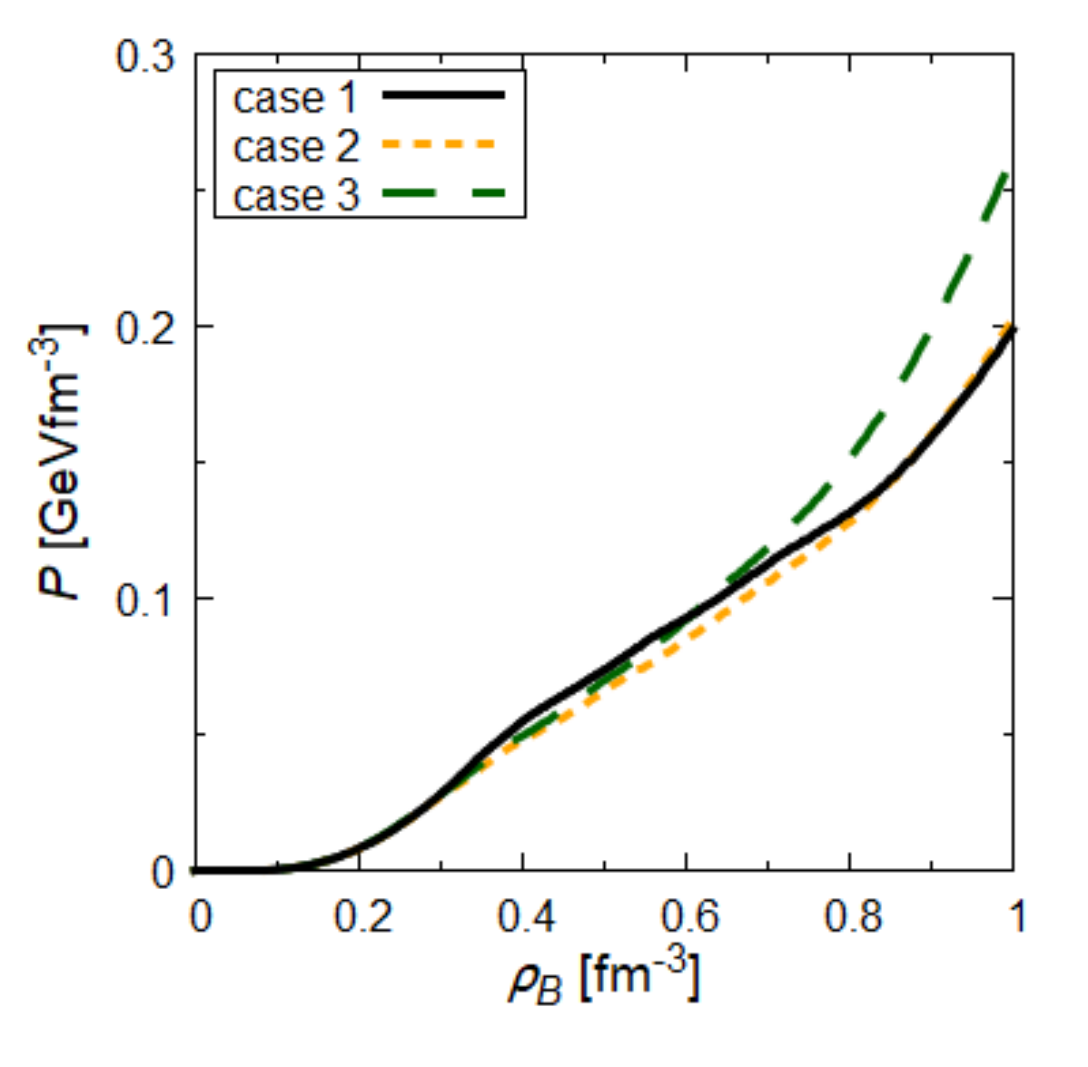} \\ 
  \centering\includegraphics[width=1.05\columnwidth]{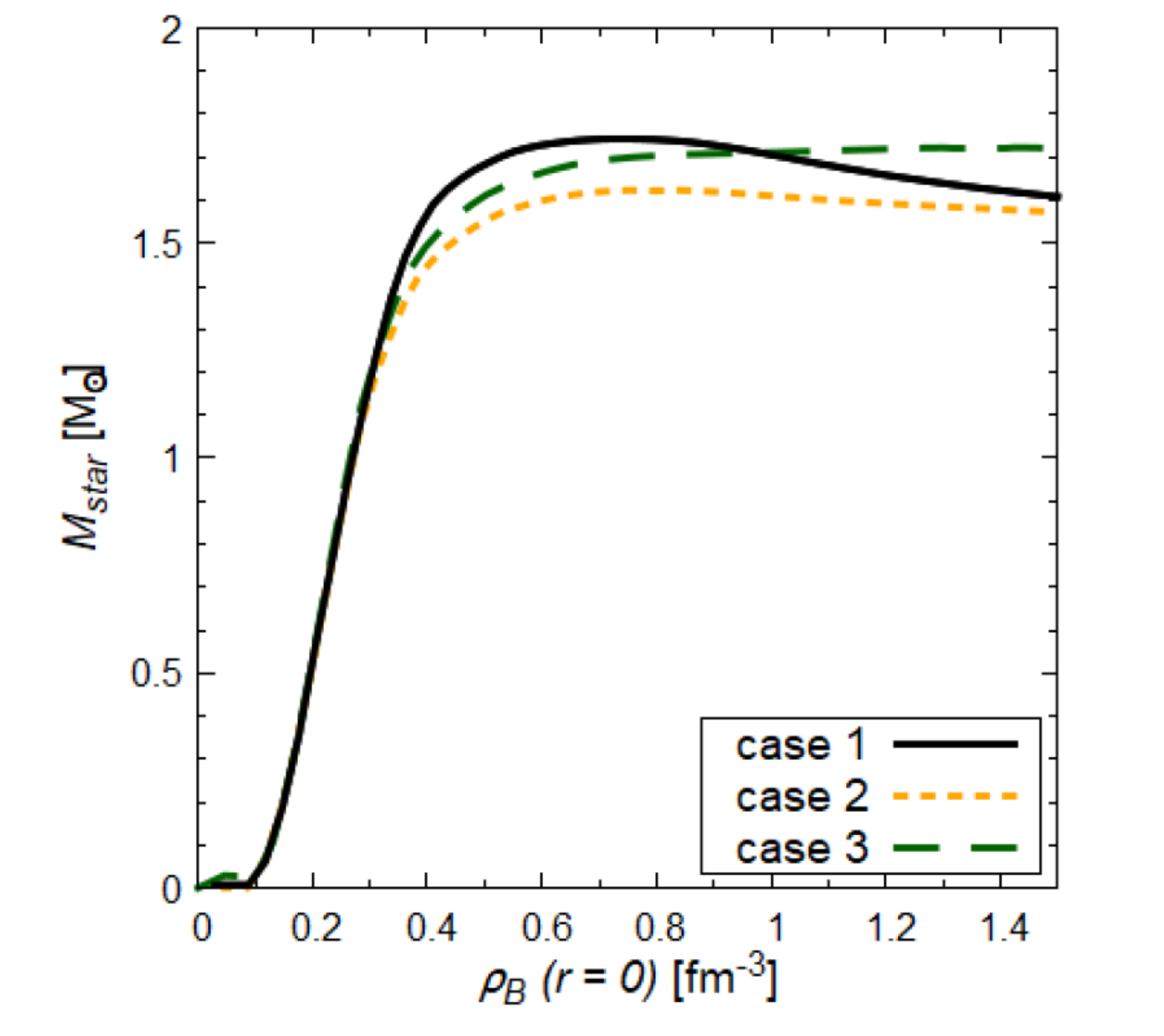}
\caption{(Color online) The pressure in neutron star matter as function of the baryon density ({\it top panel}) and
the resulting neutron star masses as function of the central baryon density ({\it bottom panel})
for the three cases listed in Tab.~\ref{tab:sixeight}. Case 1 is identical to the ``full result''
for the 4-fermi interaction case, shown by the solid lines in Fig.~\ref{fig:Press_4fermi}.} 
\label{fig:Press_468fermi}
\end{figure}
%===============================================================================

The top panel of Fig.~\ref{fig:Press_468fermi} shows the results for the pressure in neutron star matter
for the three cases listed in Tab.~\ref{tab:sixeight} as functions of the baryon density, and the bottom panel shows the resulting neutron star masses as functions of the central baryon density. Although the 6-fermi interaction (case 2) leads only to a slight decrease
of the pressure in the region of $\rho_B = 0.35 \sim 0.8$ fm$^{-3}$, the resulting decrease in neutron star masses is quite significant.\footnote{As in other works, for example Ref.~\cite{Weissenborn:2011ut}, we find that 
small changes of the pressure in this region of baryon densities can lead to appreciable
changes in the star masses.} On the other hand, the $(vv)$-type 8-fermi interaction (last term in Eq.~(\ref{eq:eightfermilag})),
with a very moderate coupling constant and counterbalanced by the $(ss)$-type interaction so as not to change the
saturation properties of symmetric nuclear matter,
gives a strongly increasing pressure for $\rho_B > 0.7$ fm$^{-3}$ and stabilizes the neutron stars against collapse
for central densities larger than $0.7$ fm$^{-3}$. 

Taken together, the 6-fermi and 8-fermi interactions do not
lead to noticeable changes of the maximum star masses, but rather work towards stabilization of the stars
for high central densities. The resulting star masses for case 3 in the range
of central densities between $0.7$ and $1.5$ fm$^{-3}$ are all around $1.7 \, M_{\odot}$, with
radii decreasing from $11.5$ km to $9.5$ km. 
We finally give the values of the maximum central baryon densities which gives stable stars, 
the maximum star masses, 
and the radii of the stars with maximum mass for the three cases shown in Fig.~\ref{fig:Press_468fermi}: 
\begin{align}
(\rho_{B}^{\rm max}(r=0), \, M_{\rm star}^{\rm max}, \, R) 
= (0.72 \, {\rm fm}^{-3}, \,\, 1.73 \, M_{\odot}, \,\,  12.3 \, {\rm km})  
\nonumber
\end{align}
for the case 1;
\begin{align}
(\rho_{B}^{\rm max}(r=0), \, M_{\rm star}^{\rm max}, \, R) 
= ( 0.8 \, {\rm fm}^{-3}, \,\, 1.62 \, M_{\odot}, \,\, 11.9 \, {\rm km}) 
\nonumber
\end{align}
for the case 2; and
\begin{align}
(\rho_{B}^{\rm max}(r=0), \, M_{\rm star}^{\rm max}, \, R) 
= ( 1.4 \, {\rm fm}^{-3},  \,\, 1.72 \, M_{\odot}, \,\,  9.8 \, {\rm km})
\nonumber
\end{align}
for the case 3.

\section{SUMMARY\label{sec:summary}}

In this paper we used the 3-flavor NJL model as an effective quark theory of QCD to describe the octet baryons as quark-diquark
bound states, and the equations of state of nuclear and neutron star matter in the relativistic mean field approximation
based on quark degrees of freedom. One of our basic concepts was to preserve the flavor and chiral 
symmetries of the interaction Lagrangian, i.e., to allow explicit symmetry breakings only by the current quark
masses and not by ad-hoc changes of model parameters. 
In Sec.~\ref{sec:Intro} we stated the four main purposes of our work, so let us now summarize our
results in this order. 

First, the internal quark structure of baryons leads to density dependent meson-baryon coupling constants and meson
masses which strongly reduce the attractive parts of the interactions in nuclear matter. The main reason for this effect is the nonlinear behavior 
of the hadron masses as functions of the constituent quark masses. In particular, we found that the attraction experienced by the $\Sigma$ 
baryon immersed in nuclear matter is reduced more strongly than that for the $\Lambda$ baryon, and we could verify that the
mass difference between the $\Sigma$ and the $\Lambda$ baryons immersed in the nuclear medium increases with increasing
density. However, we found that this effect, which is based on the different quark-diquark structures of those two
baryons, is not sufficient to make the $\Sigma$ unbound in the region of normal nuclear matter density.

Second, we used concepts of the relativistic Fermi liquid theory to derive the effective meson exchange interaction
between octet baryons in the nuclear medium, and the analogue of the Landau relation between the
energies of the baryons and the interactions between them. We also used the same concepts to
discuss the renormalization of currents carried by baryons, as well as the effects of nucleon density variations 
on the energies of hyperons immersed in the nuclear medium. To the best of our knowledge, some of these relations
cannot be found in the literature, and we hope that our results will be useful for further
investigations.

Third, we designed our mean field approximation so that it reflects the basic symmetries of the model and their dynamical breakings,
regardless of possible disagreements with observations. 
To appreciate this point, let us suppose for the moment that we had explicitly broken the 
flavor and chiral symmetries, as specified below Eq.~(\ref{eq:lagrangian}), by choosing a different coupling constant (say $G_{\rho}$) for the isovector term in the second line of Eq.~(\ref{eq:lagrangian}):
$G_{\rho} \, \left[ \left(\overline{q} \lambda_i \, \gamma^{\mu} \,q \right)^2 +  
\left(\overline{q} \lambda_i \, \gamma^{\mu} \gamma_5 \,q \right)^2\right]$, where $i=1,2,3$. 
By choosing $G_{\rho} \simeq 3 G_v \simeq 18$ GeV$^{-2}$, we
could reproduce the empirical symmetry energy $a_s = 32$ MeV (see Sec.~\ref{sec:fermiliquid1}), the shallow bound state
of pure neutron matter in Fig.~\ref{fig:energies_snm} would disappear, neutron stars made of nucleons and leptons would become heavier, 
and the onset of the $\Sigma^-$ baryon would move to higher densities or disappear, because its energy 
gets a positive shift from the vector isovector potential, twice as large as for the neutron (see upper panel of Fig.~\ref{fig:chemdens_nsm}, and Eq.(\ref{eq:baryonenergy1})). 
This would delay the onset of the decrease of $M_s$ in neutron star matter
(see Fig.~\ref{fig:effq_nsm}) and thereby hinder the succession of further hyperons (see lower panel of Fig.~\ref{fig:chemdens_nsm}), leading again to larger star masses.
While this ad-hoc modification may still have some phenomenological justification, one may think of more
drastic changes, like for example enhancing the coupling constant in
the vector potential acting on the $s$-quark in Eq.~(\ref{eq:fields}), or 
introducing a phenomenological repulsive function into the energy density which grows
asymptotically for large densities. In these or other ways one could ``improve'' the results, 
%but not learn anything new.  
but only little can be learned from it.

Fourth, we found that the so called hyperon puzzle persists in the NJL model for composite octet baryons in the mean field
approximation, and 6-fermi and 8-fermi interactions - with coupling constants chosen so as not to spoil the saturation
properties of normal nuclear matter - do not solve the problem. On the positive side, we have shown that a special kind of 8-fermi
interaction, characterized by a product of four quark current operators, is able to support stable stars up to 
$1.7$ solar masses over a large region of central densities. In view of the extremely large baryons densities involved in the investigation
of neutron stars, we believe that any solution to the hyperon
puzzle must involve quark degrees of freedom, not only quarks in individual hadrons but also quarks
which belong to two or more hadrons, or to the whole system. An investigation along these lines would naturally lead to an 
examination of various patterns of phase transitions to 3-flavor quark matter, including pairing and condensation phenomena.

\begin{acknowledgments}
K. N.  wishes to thank the staff and students of the Department of Physics at Tokai University for their discussions and advice. 
W.B. acknowledges very helpful advice from Prof. H. Tamura and Prof. F. Weber. The work of I.C. was supported by the U.S. Department of Energy, Office of Science, Office of Nuclear Physics, contract no. DE-AC02-06CH11357.
This work was supported partially through USJHPE (U.S. - Japan Hadronic Physics Exchange Program for Studies
of Hadron Structure and QCD) by the US Department of Energy under grant DE-SC0006758.
\end{acknowledgments}

\appendix

\section{BARYONS AS QUARK - DIQUARK BOUND STATES\label{app:baryons}}

The quark - diquark model, based on the static approximation to the Faddeev equation, for octet baryons in the limit of  isospin symmetry ($M_u = M_d$) has been described in 
Refs.~\cite{Carrillo-Serrano:2014zta,Carrillo-Serrano:2016igi}. As explained in the main text, in our present work
we still assume isospin symmetry in the vacuum, and therefore equal current quark masses ($m_u=m_d\equiv m$) and equal constituent
quark masses in the vacuum ($M_{u0} = M_{d0} \equiv M_0$). However, a consistent description of
isospin asymmetric systems, like neutron star matter, in the framework of an effective quark theory requires to consider the spontaneous breaking of isospin 
symmetry due to the presence of the medium, i.e., $M_u \neq M_d$. 
In this appendix, we therefore briefly explain the main points of 
our model for the octet baryons, treating the masses $M_u, M_d, M_s$ as independent quantities.

The chiral invariant interaction Lagrangian in the $qq$ channel is given by~\cite{Vogl:1991qt}
\begin{align}
{\cal L}_I^{(qq)} &= G_S \left[ \left( \widebar{q} \gamma_5 \, C \, \lambda_a \, \lambda^{(C)}_A \, \widebar{q}^T \right) \,
\left( {q}^T C^{-1} \gamma_5 \, \lambda_a \, \lambda^{(C)}_A \, q \right)  \right.   \nonumber \\
& \left. - \left( \widebar{q}  \, C \, \lambda_a \, \lambda^{(C)}_A \, \widebar{q}^T \right) \,  
\left( {q}^T C^{-1}  \,  \lambda_a \, \lambda^{(C)}_A \, q \right)  \right]   \nonumber \\
&+ G_A \left[ \left( \widebar{q} \gamma_{\mu} \, C \, \lambda_s \, \lambda^{(C)}_A \, \widebar{q}^T \right) \,
\left( {q}^T C^{-1} \gamma_{\mu} \, \lambda_s \, \lambda^{(C)}_A \, q \right) \right.    \nonumber \\
& \left. + \left( \widebar{q}  \gamma_{\mu} \gamma_5 \, C \, \lambda_a \, \lambda^{(C)}_A \, \widebar{q}^T \right) \,  
\left( {q}^T C^{-1}  \, \gamma^{\mu} \gamma_5 \, \lambda_a \, \lambda^{(C)}_A \, q \right)  \right] \,.
\label{eq:lagqq}
\end{align}
Here $\lambda_a$ ($a=2,5,7)$ are the antisymmetric Gell-Mann flavor matrices, $\lambda_s$ ($s = 0,1,3,4,6,8$)
are the symmetric ones, and the antisymmetric Gell-Mann color matrices $\lambda^{(C)}_A$ ($A = 2,5,7$)  
project to color $\widebar{3}$ diquark states. (There are also interaction terms in the color $6$ diquark channels,
which are not shown here because they do not contribute to colorless baryon states.) The charge conjugation Dirac matrix
is $C = i \gamma_2 \gamma_0$. The first line in~(\ref{eq:lagqq}) is the interaction in the scalar diquark ($0^+$) channel,
the second line shows the pseudoscalar diquark ($0^-$) channel, the third line the axial vector diquark ($1^+$) channel,
and the fourth line the vector diquark ($1^-$) channel. Following previous works~\cite{Carrillo-Serrano:2014zta,Carrillo-Serrano:2016igi}, we will include only the scalar and the
axial vector diquark channels, which are expected to be dominant from the nonrelativistic analogy.

By simple manipulations in flavor space, we can identically rewrite the 2 terms relevant for our calculation as follows:
\begin{align}
{\cal L}^{(qq)} &= 
G_S \left( \widebar{q} \gamma_5 \, C \, t_a \, \lambda^{(C)}_A \, \widebar{q}^T \right) \,
\left( {q}^T C^{-1} \gamma_5 \, t_a^{\dagger} \, \lambda^{(C)}_A \, q \right)    \nonumber \\
& + G_A \left( \widebar{q} \gamma_{\mu} \, C \, t_s \, \lambda^{(C)}_A \, \widebar{q}^T \right) \,
\left( {q}^T C^{-1} \gamma_{\mu} \, t_s^{\dagger} \, \lambda^{(C)}_A \, q \right) \,, 
\label{eq:lagqq1}
\end{align}   
where we introduced the three antisymmetric and anti-Hermitian $3 \times 3$ flavor matrices 
$t_a \equiv \left(t_{[ud]}, t_{[us]}, t_{[ds]} \right)$, and the six symmetric and Hermitian 
$3 \times 3$ flavor matrices $t_s \equiv 
\left(t_{\{ud\}}, t_{\{us\}}, t_{\{ds\}}, t_{\{uu\}}, t_{\{dd\}}, t_{\{ss\}} \right)$.  
For example, $t_{[ud]}$ is given by
$\displaystyle{ t_{[ud]} = \left(  \begin{array}{ccc}
0 & 1 & 0 \\
-1 & 0 & 0 \\
0 & 0 & 0    \end{array}   \right)}$, and corresponds to the antisymmetric flavor combination expressed by $[ud]$.
The matrices $t_{[us]}$, $t_{[ds]}$ are defined similarly, and correspond to the antisymmetric flavor combinations
$[us]$ and $[ds]$. The symmetric matrix 
$t_{\{ud\}}$ has the same structure as $t_{[ud]}$, but with the $-1$ replaced by $+1$ in the (2,1) component, and
corresponds to the symmetric flavor combination expressed by $\{ud\}$. The matrices $t_{\{us\}}$ and $t_{\{ds\}}$
are defined in a similar way for the symmetric flavor combinations $\{us\}$ and $\{ds\}$. 
Finally, the matrices $t_{\{uu\}}$, $t_{\{dd\}}$ and $t_{\{ss\}}$ have a $\sqrt{2}$ as the (1,1) component, the (2,2)
component, and the (3,3) component, respectively, with all other components equal to zero.

\begin{figure}
\centering\includegraphics[width=\columnwidth]{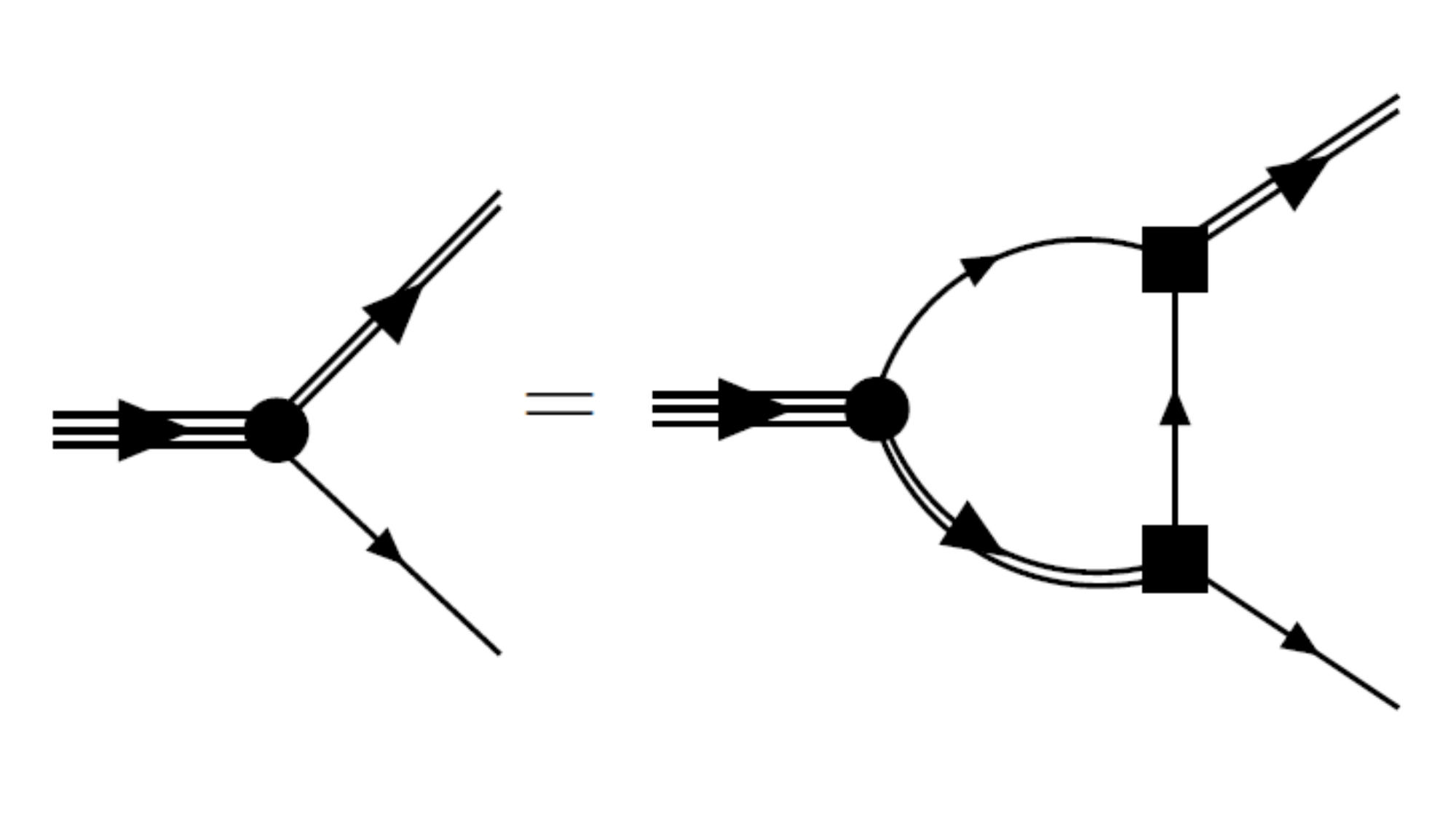}
%\vspace{-0.5cm}  
\caption{Graphical representation of the Faddeev equation~(\ref{eq:fad}).
The black dot represents the vertex function $X$, the square the Dirac-flavor vertex functions
$\Lambda \, t$ in~(\ref{eq:z}), the single lines the quark propagator $S$, and the
double line the diquark propagator $\tau$. External baryon, diquark and quark lines are amputated.} 
\label{fig:Faddeev}
\end{figure}

We express the Faddeev vertex functions for a given baryon by $X^a_{i}$, where 
$a$ denotes the diquark channels explained above ($a = [ud] , \dots ,\{ss\}$), and $i$ is the flavor of the third quark.
For example, for the proton the diquark-quark channels are labeled by $[ud]u$, $\{ud\}u$, and $\{uu\}d$.
The Faddeev equations for the vertex functions $X^a_i(p,q)$, describing a baryon of momentum $p$ as a
bound state of a quark (momentum $q$) and a diquark (momentum $p-q$),  are 
\begin{align}
X^a_i(p,q) = \int \frac{{\rm d}^4 k}{(2 \pi)^4} \, Z^{ab}_{ij} \, S_j(k) \,\tau^{bc}_{(ki)}(p-k) \,X^c_j(p,k) \,, 
\label{eq:fad}
\end{align}
which is shown graphically in Fig.~\ref{fig:Faddeev}. Here the quark exchange kernel is given by
\begin{align}
Z^{ab}_{ij} = -3 \Lambda^b \, \left(t^b \,  S(k+q-p)  t^{a\dagger} \right)_{ij} \, \Lambda^a \,,
\label{eq:z}
\end{align}
where the factor $-3$ comes from projection to color singlet states, and we used the identity
$C S^T(k) C^{-1} = S(-k)$ to process the charge conjugation matrices. 
The Dirac matrices $\Lambda$ are given by $\Lambda^a = \gamma_5$ for the scalar diquark channels 
(flavor index $a =[ud], [us], [ds]$) and $\Lambda^a = \gamma^{\mu} \gamma_5$ for the axial vector diquark channels
(flavor index $a = \{ud\}, \{us\}, \{ds\}, \{uu\}, \{dd\}, \{ss\}$). The quantities $\tau^{bc}_{(ki)}$ in Eq.~(\ref{eq:fad})
are diagonal in the diquark flavor indices, $\tau^{bc}_{(ki)} = \delta_{b c} \tau^b_{(ki)}$, where
$\tau^b_{(ki)}$ is the reduced $t$-matrix in the diquark channel $b$ with interacting quark flavors $k, i$.
Therefore, $\tau^b_{(ki)} \equiv \tau_{[ki]}$ in the scalar diquark channels ($[ki] =[ud], [us], [ds]$), and
$\tau^b_{(ki)} \equiv \tau^{\mu \nu}_{\{ki\}}$ in the axial vector diquark channels 
($\{ki\} = \{ud\}, \{us\}, \{ds\}, \{uu\}, \{dd\}, \{ss\}$). The explicit forms of the reduced diquark $t$-matrices
$\tau_{[ki]}$ and $\tau^{\mu \nu}_{\{ki\}}$ are given in Ref.~\cite{Carrillo-Serrano:2014zta,Carrillo-Serrano:2016igi}. 
In the last factor $\left( \dots \right)$ of~(\ref{eq:z}), the quark propagator is considered as a $3 \times 3$ diagonal
matrix with diagonal elements $S_k = (S_u, S_d, S_s)$.

%===============================================================================
%===============================================================================

The formalism described so far is the Faddeev framework in the NJL model, where the only 
assumptions are the ladder approximation for the 2-body $t$-matrices and the restriction
to the scalar and axial vector diquark channels.    
The quark-diquark model used in the calculations of the main text is obtained by the replacement 
$S \rightarrow - \frac{1}{M}$ in the
quark exchange kernel~(\ref{eq:z}) for each quark flavor $k$, i.e., this approximation neglects the momentum dependence of the quark exchange kernel and is called
the ``static approximation'' of the Faddeev kernel~\cite{Hellstern:1997nv}.
In this approximation, the vertex functions $X^a_i$ of~(\ref{eq:fad}) depend only on the total momentum $p$, and the
integral in Eq.~(\ref{eq:fad}) extends only over the product $S_j(k) \, \tau^{bc}_{(ki)}(p-k)$, which we regularize according to
the proper time scheme (see App.~\ref{app:regularization}), which avoids unphysical thresholds for the decay into quarks. The Dirac structure of the vertex
function can also be determined analytically. Let us again take the proton as an example. Arranging the three interacting
channels mentioned above into a vector, the vertex function can be expressed in the form
\footnote{The coupling of the time component of the axial vector diquark and the
quark to the total spin $(\frac{1}{2}, S_N)$ gives rise to the structure $\frac{p^{\mu}}{M_N} \gamma_5 u(p, S_N)$, and the coupling of the 3-vector components of the axial vector diquark to the quark gives the structure 
\begin{align}
\sum_{\lambda, s} \left(1 \frac{1}{2}, \lambda s | \frac{1}{2} S_N \right) \varepsilon^{\mu}(p, \lambda) \, u(p, s) 
= \frac{-1}{\sqrt{3}} \left( \frac{p^{\mu}}{M_N} + \gamma^{\mu} \right) \gamma_5 \, u(p, S_N) \,, 
\nonumber
\end{align}
where $\varepsilon^{\mu}(p,\lambda)$ is the Lorentz 4-vector for spin 1 with mass $M_N$, and $u(p, S_N)$ is the Dirac spinor with mass $M_N$.}
\begin{align}
|p \rangle = \left( \begin{array}{c}
X^{[ud]}_u (p) \\ X^{\{ud\}}_u (p) \\  X^{\{uu\}}_d (p)
\end{array}  \right)  
\equiv \left( \begin{array}{c}
\alpha_1 \, [ud] u \\
\left(\alpha_2 \frac{p^{\mu}}{M_p} + \alpha_3 \gamma^{\mu}\right) \gamma_5  \, \{ud \} u \\
\left(\alpha_4 \frac{p^{\mu}}{M_p} + \alpha_5 \gamma^{\mu}\right) \gamma_5  \, \{uu \} d   
\end{array}  \right) \, u_p(p) \,,
\label{eq:proton}
\end{align}
where $u_p(p)$ is the Dirac spinor with the mass of the proton ($M_p$). Inserting~(\ref{eq:proton})
into Eq.~(\ref{eq:fad}) then gives homogeneous equations for the coefficients $\alpha_i$, and the characteristic
equation gives the proton mass $M_p = M_p(M_u, M_d)$.

The vertex functions of the other members of the octet with two identical quark flavors
are similar to~(\ref{eq:proton}), with obvious replacements of quark flavors. For the
$\Sigma^0$, the flavor structure can be obtained by acting with the isospin lowering operator ($T_-$) on
$|\Sigma^+ \rangle$, which generates 
\begin{align}
|\Sigma^0 \rangle = \left( \begin{array}{c}
X^{[us]}_d (p) \\ X^{[ds]}_u (p) \\  X^{\{us\}}_d (p)  \\  X^{\{ds\}}_u (p)  \\  X^{\{ud\}}_s (p)
\end{array}  \right)  
\equiv \left( \begin{array}{c}
\alpha_1 \, [us] d \\
\alpha_2  \, [ds] u  \\
\left(\alpha_3 \frac{p^{\mu}}{M_{\Sigma^0}} + \alpha_4 \gamma^{\mu}\right) \gamma_5  \, \{us \} d \\
\left(\alpha_5 \frac{p^{\mu}}{M_{\Sigma^0}} + \alpha_6 \gamma^{\mu}\right) \gamma_5  \, \{ds \} u  \\
\left(\alpha_7 \frac{p^{\mu}}{M_{\Sigma^0}} + \alpha_8 \gamma^{\mu}\right) \gamma_5  \, \{ud \} s  
\end{array}  \right) \, u_{\Sigma^0}(p) \,,
\label{eq:sigma}
\end{align}
where $u_{\Sigma^0}(p)$ is the Dirac spinor with the mass $M_{\Sigma^0}$. 
Note that there is no component with the flavor structure $[ud]s$ in the $\Sigma^0$,
and, of course, also no components where the 2 light quarks form a scalar diquark in 
$\Sigma^{\pm}$ because those vanish identically ($[uu]=[dd]=0$).

For the $\Lambda$, we first construct a state $U_+ | \Xi^0 \rangle$, where the raising
$U$-spin operator converts an $s$-quark into a $d$-quark,  
and orthogonalize this state to $|\Sigma^0 \rangle$. This gives
\begin{align}
|\Lambda \rangle = \left( \begin{array}{c}
X^{[ud]}_s (p) \\ X^{[us]}_d (p) \\  X^{[ds]}_u (p)  \\  X^{\{us\}}_d (p)  \\  X^{\{ds\}}_u (p)
\end{array}  \right)  
\equiv \left( \begin{array}{c}
\alpha_1 \, [ud] s \\
\alpha_2  \, [us] d  \\
\alpha_3  \, [ds] u  \\
\left(\alpha_4 \frac{p^{\mu}}{M_{\Lambda}} + \alpha_5 \gamma^{\mu}\right) \gamma_5  \, \{us \} d \\
\left(\alpha_6 \frac{p^{\mu}}{M_{\Lambda}} + \alpha_7 \gamma^{\mu}\right) \gamma_5  \, \{ds \} u  
\end{array}  \right) \, u_{\Lambda}(p) \,,
\label{eq:lambda}
\end{align}
where $u_{\Lambda}(p)$ is the Dirac spinor with the mass $M_{\Lambda}$.
Note that there is no component with the flavor structure $\{ud\}s$ in the $\Lambda$.

In the calculations of the main text we only need the masses of the octet baryons as functions of the
constituent quark masses. In isospin asymmetric systems like neutron star matter, the isospin symmetry is
obviously broken, but the charge symmetry is intact if we simultaneously reverse the signs of the isospin $z$-components
of the baryons and the constituent quarks in the baryon. 
%see the related comments at the end of Sec.\ref{sec:mf}.  
We therefore have the following five independent
functions of $M_u, M_d$ (omitting the obvious dependence on $M_s$ for the hyperons to simplify the notations):
\begin{align}
M_p &= M_p(M_u, M_d) \,, \,\,\,\,\,\,\,  
M_{\Sigma^+} = M_{\Sigma^+}(M_u) \,, \,\,\,\,\,\, 
M_{\Xi^+} = M_{\Xi^+}(M_u) \,,  \nonumber \\
M_{\Sigma^0} &= M_{\Sigma^0}(M_u, M_d) = M_{\Sigma^0}(M_d, M_u) \,,  \nonumber \\
M_{\Lambda} &= M_{\Lambda}(M_u, M_d) = M_{\Lambda}(M_d, M_u) \,.
\nonumber
\end{align}
The masses of the remaining baryons can then be expressed by
\begin{align}
M_n &= M_p(M_d, M_u) \,, &  
M_{\Sigma^-} &= M_{\Sigma^+}(M_d) \,, \nonumber \\
M_{\Xi^0}  &=  M_{\Xi^+}(M_d)  \,.
\label{eq:cs}
\end{align}

The vertex functions and masses of the decuplet baryons are calculated in a similar way. The calculation is simplified by the fact that
here only the axial vector diquark channels (symmetric combinations of quark flavors $\{q_1 q_2 \}$) contribute, which leaves only one possible 
Dirac-Lorentz structure for all components of the baryon vertex, namely the Rarita-Schwinger spinor $u^{\mu}(p,S_b)$.

In the present work, we determined the coupling constants $G_S$, $G_A$ of the Lagrangian~(\ref{eq:lagqq}) so as to reproduce the
observed masses of the nucleon ($M_N=0.94$ GeV) and the Delta baryon ($M_{\Delta} = 1.232$ GeV) in the vacuum ($M_u=M_d=0.4$ GeV).
We also determined our vacuum value of the strange quark mass so as to reproduce the observed mass of the $\Omega$ baryon 
($M_{\Omega} = 1.67$ GeV). In this way we obtain 
\begin{align}
G_S &= 8.76 \,{\rm GeV}^{-2} \,, &  G_A &= 7.36 \,{\rm GeV}^{-2}\,, &  M_s &= 0.562 \,{\rm GeV} \,.
\label{eq:paramb}
\end{align}  
The resulting masses of octet baryons are given in Tab.~\ref{tab:octetmasses} of the main text. The masses of the
decuplet baryons - except for the $\Delta$ and the $\Omega$ which were fitted - 
are also well reproduced in this calculation; we obtain $M_{\Sigma^*} = 1.38$ GeV, and $M_{\Xi^*} = 1.53$ GeV. The mass of the kaon, however, is underestimated ($0.43$ GeV for the case of 4-fermi couplings);
in order to reproduce its observed mass we would need a larger value of $M_s$. Because the focus of our
present work is on the baryons, we made no attempt to reproduce the meson masses well.
   
The masses of the diquarks (poles of the quantities $\tau$ in Eq.~(\ref{eq:fad})) are
$M^{[\ell \ell']} = 0.768$ GeV and $M^{[\ell s]} = 0.902$ GeV for the scalar diquarks with 
$\ell, \ell' = u,d$,
and $M^{\{\ell \ell' \}} = 0.929$ GeV, $M^{\{\ell s\}} = 1.04$ GeV,  $M^{\{s s\}} = 1.15$ GeV for the axial vector diquarks. 

We finally mention that the values of $G_S$ and $G_A$ given in~(\ref{eq:paramb}) are different from those used in a previous
work on the flavor $SU(2)$ case~\cite{Tanimoto:2019tsl}. There $G_S$ and $G_A$ were fitted to the mass of the free nucleon and its 
axial vector coupling constant ($g_A=1.26$). The values of $G_S$ ($G_A$) obtained in that way were larger (smaller) than the
values given in~(\ref{eq:paramb}), which indicates that the dominance of the scalar diquark channel,
which increases in-medium because of the decreasing scalar diquark mass, was more pronounced in Ref.~\cite{Tanimoto:2019tsl}
than in the present work. This stronger attraction in-medium, however, was eventually canceled by a stronger
repulsion in the vector-isovector $q \widebar{q}$ channel, because in the flavor $SU(2)$ case it was possible
to reproduce the symmetry energy in the mean field approximation without violating the chiral symmetry of the
interaction Lagrangian. As a result, the pressure in neutron star matter and the star
masses calculated in Ref.~\cite{Tanimoto:2019tsl} were almost identical to the results shown by the dashed lines
in Fig.~\ref{fig:Press_4fermi} of the main text. 
 
In our present work, we found it more essential to reproduce the $N - \Delta$ mass difference, because one
of our motivations was to see how the $\Sigma - \Lambda$ mass difference evolves with density if the spin dependent
diquark correlations are constrained to the vacuum value of the $N - \Delta$ mass difference from the outset.
The value of $g_A$, obtained with our present coupling constants (\ref{eq:paramb}), is larger than the observed value by
about 10$\%$. In future investigations, we wish to see whether the inclusion of additional diquark channels
allows one to find a set of coupling constants which reproduces the nucleon mass, the Delta baryon mass, and 
$g_A$ simultaneously.

\section{Meson exchange in symmetric nuclear matter\label{app:mesonexchange}}

%===============================================================================
\begin{figure}
  \centering\includegraphics[width=\columnwidth]{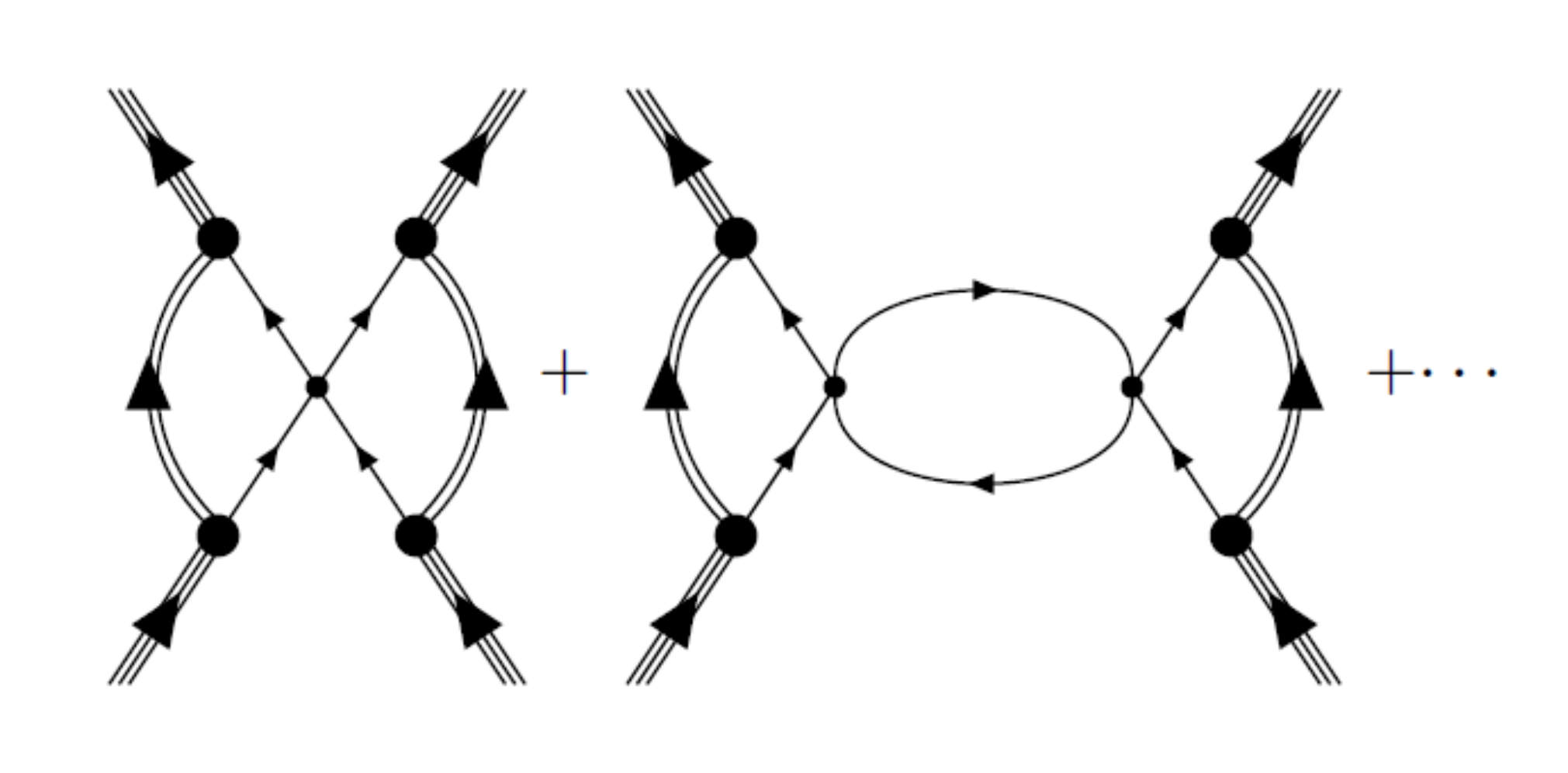}
%\vspace{-0.5cm}  
\caption{Graphical representation of a meson exchange interaction in the quark-diquark model for the baryons.
Only the quark loop contributions are shown, and the dots indicate the higher orders in the
RPA-type series of $\overline{q} q$ bubble graphs. The nucleon loop contributions in the denominators
of Eqs.~(\ref{eq:isoscalar}) and (\ref{eq:isovector}) are not shown here for simplicity.
The small dots represent the 4-fermi interaction in the $\overline{q}q$ channel, and the other
symbols are explained in the caption to Fig.~\ref{fig:Faddeev}.} 
\label{fig:MEX}
\end{figure}
%===============================================================================

As mentioned in Sec.~\ref{sec:effbb}, in order to express the effective baryon-nucleon interactions~(\ref{eq:isoscalar}) and (\ref{eq:isovector}) in terms of meson exchange processes of the type shown 
in Fig.~\ref{fig:MEX}, one should multiply the numerator and denominator functions
in the first two lines of those expressions by the quark-meson couplings  
\begin{align}
g_{\sigma}^{(q)2} = g_{\delta}^{(q)2} = \frac{-1}{\Pi'_{s}(q^2=0)} \,,   
\label{eq:mesonquark1}
\end{align}
and similarly in the third lines by
\begin{align}
g_{\omega}^{(q)2} = g_{\rho}^{(q)2} = \frac{-1}{\Pi'_{v}(q^2=0)} \,.   
\label{eq:mesonquark2}
\end{align}
Here the $\overline{q} q$ bubble graphs in the scalar and vector channels are given by
\begin{align}
\Pi_s(q^2) &= 12 i \int \frac{{\rm d}^4 k}{(2\pi)^4} \nonumber \\ 
&  \hspace{-1cm} \times \left[ \frac{-2}{k^2 - M^2} + \left(q^2 - 4 M^2\right) \int_0^1 \, {\rm d}x
\frac{1}{\left(k^2 - M^2 + q^2 x(1-x) \right)^2} \right] \,,   \label{eq:pis}  \\
\Pi_v(q^2) &=  48 i q^2 \, \int \frac{{\rm d}^4 k}{(2\pi)^4} \,
\int_0^1 \, {\rm d}x
\frac{ x(1-x)}{\left(k^2 - M^2 + q^2 x(1-x) \right)^2} \,,    \label{eq:piv}
\end{align}
and the primes in~(\ref{eq:mesonquark1}) and~(\ref{eq:mesonquark2}) mean differentiation
w.r.t. $q^2$.

%===============================================================================
\begin{table}
\addtolength{\extrarowheight}{2.2pt}
\centering
\caption{Effective coupling constants and masses of $\sigma$, $\omega$, $\delta$ and $\rho$ mesons for four values of the baryon density in symmetric nuclear matter. Coupling constants are dimensionless, and masses are given in units of GeV. 
For definitions, see Eqs.~(\ref{f0mex}),~(\ref{f0pmex}), and text.}
\begin{tabular*}{\columnwidth}{@{\extracolsep{\fill}}c|c|cccc|c}
\hline\hline
density  & $g_{\sigma}^{(q)}$  &  $g_{\sigma}^{(N)}$  &  $g_{\sigma}^{(\Lambda)}$  &  $g_{\sigma}^{(\Sigma)}$  &  
$g_{\sigma}^{(\Xi)}$ &  $M_{\sigma}$ \\
\hline
0    & 6.63   & 18.21 &  12.13  & 10.28    & 5.70    & 1.25  \\
0.15 & 4.85   & 10.06 &  7.05   & 5.59     & 3.15    & 0.96  \\
0.3  & 4.20   & 6.25  &  4.80   & 3.59     & 2.06    & 0.97  \\
0.5  & 3.85   & 4.01  &  3.45   & 2.44     & 1.43    & 1.05  \\  
\hline\hline
density  & $g_{\omega}^{(q)}$  &  $g_{\omega}^{(N)}$  &  $g_{\omega}^{(\Lambda)}$  &  $g_{\omega}^{(\Sigma)}$  &  
$g_{\omega}^{(\Xi)}$ &  $M_{\omega}$ \\
\hline
0    & 5.27   & 15.80 &  10.54  & 10.54    & 5.27    & 1.52  \\
0.15 & 4.51   & 13.53 &  9.02   & 9.02     & 4.51    & 1.30  \\
0.3  & 4.18   & 12.53 &  8.35   & 8.35     & 4.18    & 1.20   \\
0.5  & 3.99   & 11.96 &  7.97   & 7.97     & 3.99    & 1.15  \\    
\hline\hline
density  & $g_{\delta}^{(q)}$  &  $g_{\delta}^{(p)}$  &  $g_{\delta}^{(\Lambda)}$  &  $g_{\delta}^{(\Sigma^+)}$  &  
$g_{\delta}^{(\Xi^0)}$ &  $M_{\delta}$ \\
\hline
0    & 6.63   & 4.64  & 0  & 10.28    & 5.70   & 1.25  \\
0.15 & 4.85   & 2.38  & 0  & 5.59     & 3.15   & 0.99  \\
0.3  & 4.20   & 1.38  & 0  & 3.59     & 2.06   & 1.00   \\
0.5  & 3.85   & 0.81  & 0  & 2.44     & 1.43   & 1.10   \\  
\hline\hline
density  & $g_{\rho}^{(q)}$  &  $g_{\rho}^{(p)}$  &  $g_{\rho}^{(\Lambda)}$  &  $g_{\rho}^{(\Sigma^+)}$  &  
$g_{\rho}^{(\Xi^0)}$ &  $M_{\rho}$ \\
\hline
0    & 5.27   & 5.27  &  0  &   10.54    & 5.27   & 1.52  \\
0.15 & 4.51   & 4.51  &  0  &   9.02     & 4.51   & 1.30  \\
0.3  & 4.18   & 4.18  &  0  &   8.35     & 4.18   & 1.20   \\
0.5  & 3.99   & 3.99  &  0  &   7.97     & 3.99   & 1.15  \\         
\hline\hline
\end{tabular*}
\label{tab:MEX}
\end{table}
%===============================================================================

For simplicity we consider only the $\ell=0$ terms in~(\ref{eq:isoscalar}) and~(\ref{eq:isovector}).
They can be expressed in the following form
\begin{align}
f_{0, bN} &= - \frac{M_b}{E_b} \frac{M_N}{E_N} \, \frac{ g_{\sigma}^{(b)} g_{\sigma}^{(N)}} {M_{\sigma}^2} 
+ \frac{ g_{\omega}^{(b)} g_{\omega}^{(N)}} {M_{\omega}^2}  \,,   
\label{f0mex}  \\ 
f'_{0, bN} &= - \frac{M_b}{E_b} \frac{M_N}{E_N} \, \frac{ g_{\delta}^{(b)} g_{\delta}^{(p)}} {M_{\delta}^2} 
+ \frac{ g_{\rho}^{(b)} g_{\rho}^{(p)}} {M_{\rho}^2} \,.  
\label{f0pmex}
\end{align}
Here all meson-baryon coupling constants and meson masses are defined at zero momentum, and are different
from the values at the meson poles. The resulting values for the effective coupling constants and
masses are summarized for three values of the baryon density in Tab.~\ref{tab:MEX}. 
In relation to our discussions in Sec.~\ref{sec:numericalnm}, we note that $g_{\sigma}^{(\Lambda)} > g_{\sigma}^{(\Sigma)}$,
which reflects the different internal quark-diquark structure of the $\Lambda$ and the $\Sigma$ baryons.

We finally add a few comments on the definition of effective coupling constants and meson masses used here:
First, the multiplication of the density dependent scaling factors~(\ref{eq:mesonquark1}) and~(\ref{eq:mesonquark2}) 
to the numerators and denominators of~(\ref{eq:isoscalar}) and~(\ref{eq:isovector}) obscures the simplicity
of those basic expressions, and for better orientation the values listed in Tab.~\ref{tab:factors} of the main text is more useful. 
Nevertheless, it is necessary for a proper definition of coupling constants and meson masses at
zero momentum of the mesons. For the coupling constants, this is immediately clear from Fig.~\ref{fig:MEX}. 
For the meson masses, consider for example the case of the $\sigma$ meson. The reduced
$t$-matrix in the $0^+$ $\overline{q} q$ channel is given by 
\begin{align}
\tau_{\sigma}(q^2) = \frac{ - 2 G_{\pi}}{1 + 2 G_{\pi} \Pi_{s}(q^2) + 2 G_{\pi} \, \delta M_{\sigma}^2} \,,
\label{eq:tausigma}
\end{align}
where the nucleon loop contributions, approximated by their forms at $q=0$, are denoted by $\delta M_{\sigma}^2$.
Expanding~(\ref{eq:tausigma}) around $q^2 = 0$ gives the approximate Yukawa-like form
\begin{align}
\tau_{\sigma}(q^2 ) = \frac{g_{\sigma}^{(q)2}}{q^2 - M_{\sigma}^2} \,,  
\nonumber
\end{align}
where $g_{\sigma}^{(q)}$ is defined by (\ref{eq:mesonquark1}), and 
\begin{align}
M_{\sigma}^2 = g_{\sigma}^{(q)2} \left( \frac{1}{2 G_{\pi}} + \Pi_{s}(q^2=0) + \delta M_{\sigma}^2 \right) \,. 
\label{eq:sigmamass2}
\end{align}
The terms $\left( \dots \right)$ in (\ref{eq:sigmamass2}) agree with the denominator in the second line of 
Eq.~(\ref{eq:isoscalar}) because of the relation $\Pi_s(q^2=0) = 2 g(M)$, where $g(M)$ is given by Eq.~(\ref{eq:g}).

\section{Regularization method\label{app:regularization}}

To evaluate 4-dimensional integrals, we introduce Feynman parameters and perform shifts of the
loop momentum so that the integrand depends only on $k^2$, where $k$ is the loop momentum, besides
other fixed variables.
We then perform a Wick rotation and use 4-dimensional spherical polar coordinates to obtain
\begin{align}
\int \dd^4 k\, f(k^2) =  2\pi^{2}i\int^{\infty}_{0}\dd k_{E}\,k_{E}^3\,f(-k_{E}^2) \,,
\nonumber
\end{align}
where $k_{E}=\sqrt{k_{0}^2+\vect{k}^2}$ is the Euclidean length. 
Next, we consider the following identities:
\begin{align}
\ln \frac{D}{D_0} &= - \int_0^{\infty} \frac{{\rm d} \tau}{\tau} \left( e^{-\tau D} - e^{-\tau D_0} \right) \,, 
\label{eq:reg1} \\
\frac{1}{D^{n}} &= \frac{1}{(n-1)!} \int^{\infty}_{0} \, {\rm d} \tau \, \tau^{n-1}e^{-\tau D}
\,\,\,\,\,\,\,\,\,\,\,\,(n \geq 0) \,,   
\label{eq:reg2}
\end{align}
where $D$ is a function of $k_E^2$ and other fixed variables. 
In the proper time regularization scheme, the infrared cutoff ($\Lambda_{\rm IR}$) is introduced
by replacing the upper integration limits in~(\ref{eq:reg1}),~(\ref{eq:reg2}) by
$1/ \Lambda_{\rm IR}^2$, and the ultraviolet cutoff ($\Lambda_{\rm UV}$) by replacing the
lower integration limits by $1/ \Lambda_{\rm UV}^2$.
After these replacements, one performs the integration over $k_E$.
The ultraviolet cutoff makes the integrals finite, while the
infrared cutoff eliminates unphysical thresholds (imaginary parts) for the decay of hadrons into
quarks, thus simulating the role of confinement.

%*************************REVISION(begin)**********************************************************

\section{Sizes of quark cores in the nuclear medium\label{app:sizes}}

The rms radius of the baryon density distribution of the quark core of the nucleons in the medium is related to the isoscalar combination of the corresponding electric charge radii of protons and neutrons by
\begin{align}
r_N(\rho_B) = \sqrt{\langle r_{Ep}^2 \rangle(\rho_B) + \langle r_{En}^2 \rangle(\rho_B)} \,.
\label{radius}
\end{align}
In the language of Feynman diagrams used in Ref.\cite{Cloet:2014rja}, the corresponding isoscalar 
baryon form factor is obtained by
the operator insertion $\frac{1}{3} \gamma^0$ on each quark line. 
For free nucleons (zero density) the result of the NJL model calculations of Ref.~\cite{Cloet:2014rja}, using the same parameters as in the present paper, is $r_N(0) = 0.475$ fm.
%\footnote{
%This value is obtained by replacing the dressed quark form factors in Sect.~VI of Ref. ~\cite{Cloet:2014rja}
%by their bare values ($F_{1U} = \frac{2}{3}, F_{1D} = - \frac{1}{3}, F_{2U}=F_{2D}=0$).
Note that this is the value for the quark core without meson cloud corrections, obtained 
by replacing the dressed quark form factors in Sect.~VI of Ref. ~\cite{Cloet:2014rja}
by their bare values ($F_{1U} = \frac{2}{3}, F_{1D} = - \frac{1}{3}, F_{2U}=F_{2D}=0$).
The pion cloud contributions to the isoscalar quantity $r_N$ are very small. 
A simple estimate of $\omega$ meson cloud effects, using our present value of $G_v$,  
%leads to a multiplicative factor of $1/(1 + 2 G_v \pi_v(q^2))$ to the baryon form factor, which 
gives only a small correction, but a more realistic treatment, following the lines of the vector meson dominance
model with the observed $\omega$ meson pole,  
increases the isoscalar baryon radius to $0.78$ fm (see Table VI of Ref.~\cite{Cloet:2014rja}), which is close to the experimental value.
As mentioned in the main text, however, the quantity which seems more relevant for the role of the Pauli principle 
is the baryon radius of the quark core without meson cloud effects. This is simply because the mesons are bosons,
and the overlap of the meson clouds just corresponds to the meson exchange interactions. Therefore the results shown in
Sec.~\ref{sec:sizes} of the main text refer to this quantity.

Based on naive geometric intuition, the volume fraction occupied by the quark cores 
can be defined as
\begin{align}
f(\rho_B) = \rho_B \, v_N(\rho_B)\,, 
\label{fractions}
\end{align}
where $v_N(\rho_B) = \frac{4 \pi}{3} r_N^3(\rho_B)$. Estimates based on this expression are also given
in the main text.

%**********************************REVISION(end)**************************************************

\bibliographystyle{apsrev4-1}
\bibliography{bib}

\begin{thebibliography}{103}%
\makeatletter
\providecommand \@ifxundefined [1]{%
 \@ifx{#1\undefined}
}%
\providecommand \@ifnum [1]{%
 \ifnum #1\expandafter \@firstoftwo
 \else \expandafter \@secondoftwo
 \fi
}%
\providecommand \@ifx [1]{%
 \ifx #1\expandafter \@firstoftwo
 \else \expandafter \@secondoftwo
 \fi
}%
\providecommand \natexlab [1]{#1}%
\providecommand \enquote  [1]{``#1''}%
\providecommand \bibnamefont  [1]{#1}%
\providecommand \bibfnamefont [1]{#1}%
\providecommand \citenamefont [1]{#1}%
\providecommand \href@noop [0]{\@secondoftwo}%
\providecommand \href [0]{\begingroup \@sanitize@url \@href}%
\providecommand \@href[1]{\@@startlink{#1}\@@href}%
\providecommand \@@href[1]{\endgroup#1\@@endlink}%
\providecommand \@sanitize@url [0]{\catcode `\\12\catcode `\$12\catcode `\&12\catcode `\#12\catcode `\^12\catcode `\_12\catcode `\%12\relax}%
\providecommand \@@startlink[1]{}%
\providecommand \@@endlink[0]{}%
\providecommand \url  [0]{\begingroup\@sanitize@url \@url }%
\providecommand \@url [1]{\endgroup\@href {#1}{\urlprefix }}%
\providecommand \urlprefix  [0]{URL }%
\providecommand \Eprint [0]{\href }%
\providecommand \doibase [0]{http://dx.doi.org/}%
\providecommand \selectlanguage [0]{\@gobble}%
\providecommand \bibinfo  [0]{\@secondoftwo}%
\providecommand \bibfield  [0]{\@secondoftwo}%
\providecommand \translation [1]{[#1]}%
\providecommand \BibitemOpen [0]{}%
\providecommand \bibitemStop [0]{}%
\providecommand \bibitemNoStop [0]{.\EOS\space}%
\providecommand \EOS [0]{\spacefactor3000\relax}%
\providecommand \BibitemShut  [1]{\csname bibitem#1\endcsname}%
\let\auto@bib@innerbib\@empty
%</preamble>
\bibitem [{\citenamefont {Haidenbauer}\ and\ \citenamefont {Meissner}(2005)}]{Haidenbauer:2005zh}%
  \BibitemOpen
  \bibfield  {author} {\bibinfo {author} {\bibfnamefont {J.}~\bibnamefont {Haidenbauer}}\ and\ \bibinfo {author} {\bibfnamefont {U.-G.}\ \bibnamefont {Meissner}},\ }\href {\doibase 10.1103/PhysRevC.72.044005} {\bibfield  {journal} {\bibinfo  {journal} {Phys. Rev. C}\ }\textbf {\bibinfo {volume} {72}},\ \bibinfo {pages} {044005} (\bibinfo {year} {2005})},\ \Eprint {http://arxiv.org/abs/nucl-th/0506019} {arXiv:nucl-th/0506019} \BibitemShut {NoStop}%
\bibitem [{\citenamefont {Rijken}\ and\ \citenamefont {Yamamoto}(2006)}]{Rijken:2006ep}%
  \BibitemOpen
  \bibfield  {author} {\bibinfo {author} {\bibfnamefont {T.~A.}\ \bibnamefont {Rijken}}\ and\ \bibinfo {author} {\bibfnamefont {Y.}~\bibnamefont {Yamamoto}},\ }\href {\doibase 10.1103/PhysRevC.73.044008} {\bibfield  {journal} {\bibinfo  {journal} {Phys. Rev. C}\ }\textbf {\bibinfo {volume} {73}},\ \bibinfo {pages} {044008} (\bibinfo {year} {2006})},\ \Eprint {http://arxiv.org/abs/nucl-th/0603042} {arXiv:nucl-th/0603042} \BibitemShut {NoStop}%
\bibitem [{\citenamefont {Bogner}\ \emph {et~al.}(2010)\citenamefont {Bogner}, \citenamefont {Furnstahl},\ and\ \citenamefont {Schwenk}}]{Bogner:2009bt}%
  \BibitemOpen
  \bibfield  {author} {\bibinfo {author} {\bibfnamefont {S.~K.}\ \bibnamefont {Bogner}}, \bibinfo {author} {\bibfnamefont {R.~J.}\ \bibnamefont {Furnstahl}}, \ and\ \bibinfo {author} {\bibfnamefont {A.}~\bibnamefont {Schwenk}},\ }\href {\doibase 10.1016/j.ppnp.2010.03.001} {\bibfield  {journal} {\bibinfo  {journal} {Prog. Part. Nucl. Phys.}\ }\textbf {\bibinfo {volume} {65}},\ \bibinfo {pages} {94} (\bibinfo {year} {2010})},\ \Eprint {http://arxiv.org/abs/0912.3688} {arXiv:0912.3688 [nucl-th]} \BibitemShut {NoStop}%
\bibitem [{\citenamefont {Petschauer}\ \emph {et~al.}(2020)\citenamefont {Petschauer}, \citenamefont {Haidenbauer}, \citenamefont {Kaiser}, \citenamefont {Mei\ss{}ner},\ and\ \citenamefont {Weise}}]{Petschauer:2020urh}%
  \BibitemOpen
  \bibfield  {author} {\bibinfo {author} {\bibfnamefont {S.}~\bibnamefont {Petschauer}}, \bibinfo {author} {\bibfnamefont {J.}~\bibnamefont {Haidenbauer}}, \bibinfo {author} {\bibfnamefont {N.}~\bibnamefont {Kaiser}}, \bibinfo {author} {\bibfnamefont {U.-G.}\ \bibnamefont {Mei\ss{}ner}}, \ and\ \bibinfo {author} {\bibfnamefont {W.}~\bibnamefont {Weise}},\ }\href {\doibase 10.3389/fphy.2020.00012} {\bibfield  {journal} {\bibinfo  {journal} {Front. in Phys.}\ }\textbf {\bibinfo {volume} {8}},\ \bibinfo {pages} {12} (\bibinfo {year} {2020})},\ \Eprint {http://arxiv.org/abs/2002.00424} {arXiv:2002.00424 [nucl-th]} \BibitemShut {NoStop}%
\bibitem [{\citenamefont {Lonardoni}\ \emph {et~al.}(2014)\citenamefont {Lonardoni}, \citenamefont {Pederiva},\ and\ \citenamefont {Gandolfi}}]{Lonardoni:2013gta}%
  \BibitemOpen
  \bibfield  {author} {\bibinfo {author} {\bibfnamefont {D.}~\bibnamefont {Lonardoni}}, \bibinfo {author} {\bibfnamefont {F.}~\bibnamefont {Pederiva}}, \ and\ \bibinfo {author} {\bibfnamefont {S.}~\bibnamefont {Gandolfi}},\ }\href {\doibase 10.1103/PhysRevC.89.014314} {\bibfield  {journal} {\bibinfo  {journal} {Phys. Rev. C}\ }\textbf {\bibinfo {volume} {89}},\ \bibinfo {pages} {014314} (\bibinfo {year} {2014})},\ \Eprint {http://arxiv.org/abs/1312.3844} {arXiv:1312.3844 [nucl-th]} \BibitemShut {NoStop}%
\bibitem [{\citenamefont {Faessler}\ \emph {et~al.}(1982)\citenamefont {Faessler}, \citenamefont {Fernandez}, \citenamefont {Lubeck},\ and\ \citenamefont {Shimizu}}]{Faessler:1982ik}%
  \BibitemOpen
  \bibfield  {author} {\bibinfo {author} {\bibfnamefont {A.}~\bibnamefont {Faessler}}, \bibinfo {author} {\bibfnamefont {F.}~\bibnamefont {Fernandez}}, \bibinfo {author} {\bibfnamefont {G.}~\bibnamefont {Lubeck}}, \ and\ \bibinfo {author} {\bibfnamefont {K.}~\bibnamefont {Shimizu}},\ }\href {\doibase 10.1016/0370-2693(82)90961-3} {\bibfield  {journal} {\bibinfo  {journal} {Phys. Lett. B}\ }\textbf {\bibinfo {volume} {112}},\ \bibinfo {pages} {201} (\bibinfo {year} {1982})}\BibitemShut {NoStop}%
\bibitem [{\citenamefont {Oka}\ \emph {et~al.}(1987)\citenamefont {Oka}, \citenamefont {Shimizu},\ and\ \citenamefont {Yazaki}}]{Oka:1986fr}%
  \BibitemOpen
  \bibfield  {author} {\bibinfo {author} {\bibfnamefont {M.}~\bibnamefont {Oka}}, \bibinfo {author} {\bibfnamefont {K.}~\bibnamefont {Shimizu}}, \ and\ \bibinfo {author} {\bibfnamefont {K.}~\bibnamefont {Yazaki}},\ }\href {\doibase 10.1016/0375-9474(87)90371-X} {\bibfield  {journal} {\bibinfo  {journal} {Nucl. Phys. A}\ }\textbf {\bibinfo {volume} {464}},\ \bibinfo {pages} {700} (\bibinfo {year} {1987})}\BibitemShut {NoStop}%
\bibitem [{\citenamefont {Fujiwara}\ \emph {et~al.}(2007)\citenamefont {Fujiwara}, \citenamefont {Suzuki},\ and\ \citenamefont {Nakamoto}}]{Fujiwara:2006yh}%
  \BibitemOpen
  \bibfield  {author} {\bibinfo {author} {\bibfnamefont {Y.}~\bibnamefont {Fujiwara}}, \bibinfo {author} {\bibfnamefont {Y.}~\bibnamefont {Suzuki}}, \ and\ \bibinfo {author} {\bibfnamefont {C.}~\bibnamefont {Nakamoto}},\ }\href {\doibase 10.1016/j.ppnp.2006.08.001} {\bibfield  {journal} {\bibinfo  {journal} {Prog. Part. Nucl. Phys.}\ }\textbf {\bibinfo {volume} {58}},\ \bibinfo {pages} {439} (\bibinfo {year} {2007})},\ \Eprint {http://arxiv.org/abs/nucl-th/0607013} {arXiv:nucl-th/0607013} \BibitemShut {NoStop}%
\bibitem [{\citenamefont {Oka}(2023)}]{Oka:2023hdc}%
  \BibitemOpen
  \bibfield  {author} {\bibinfo {author} {\bibfnamefont {M.}~\bibnamefont {Oka}},\ }\href@noop {} {\  (\bibinfo {year} {2023})},\ \Eprint {http://arxiv.org/abs/2301.06026} {arXiv:2301.06026 [nucl-th]} \BibitemShut {NoStop}%
\bibitem [{\citenamefont {Inoue}\ \emph {et~al.}(2010)\citenamefont {Inoue}, \citenamefont {Ishii}, \citenamefont {Aoki}, \citenamefont {Doi}, \citenamefont {Hatsuda}, \citenamefont {Ikeda}, \citenamefont {Murano}, \citenamefont {Nemura},\ and\ \citenamefont {Sasaki}}]{Inoue:2010hs}%
  \BibitemOpen
  \bibfield  {author} {\bibinfo {author} {\bibfnamefont {T.}~\bibnamefont {Inoue}}, \bibinfo {author} {\bibfnamefont {N.}~\bibnamefont {Ishii}}, \bibinfo {author} {\bibfnamefont {S.}~\bibnamefont {Aoki}}, \bibinfo {author} {\bibfnamefont {T.}~\bibnamefont {Doi}}, \bibinfo {author} {\bibfnamefont {T.}~\bibnamefont {Hatsuda}}, \bibinfo {author} {\bibfnamefont {Y.}~\bibnamefont {Ikeda}}, \bibinfo {author} {\bibfnamefont {K.}~\bibnamefont {Murano}}, \bibinfo {author} {\bibfnamefont {H.}~\bibnamefont {Nemura}}, \ and\ \bibinfo {author} {\bibfnamefont {K.}~\bibnamefont {Sasaki}} (\bibinfo {collaboration} {HAL QCD}),\ }\href {\doibase 10.1143/PTP.124.591} {\bibfield  {journal} {\bibinfo  {journal} {Prog. Theor. Phys.}\ }\textbf {\bibinfo {volume} {124}},\ \bibinfo {pages} {591} (\bibinfo {year} {2010})},\ \Eprint {http://arxiv.org/abs/1007.3559} {arXiv:1007.3559 [hep-lat]} \BibitemShut {NoStop}%
\bibitem [{\citenamefont {Hyodo}\ and\ \citenamefont {Niiyama}(2021)}]{Hyodo:2020czb}%
  \BibitemOpen
  \bibfield  {author} {\bibinfo {author} {\bibfnamefont {T.}~\bibnamefont {Hyodo}}\ and\ \bibinfo {author} {\bibfnamefont {M.}~\bibnamefont {Niiyama}},\ }\href {\doibase 10.1016/j.ppnp.2021.103868} {\bibfield  {journal} {\bibinfo  {journal} {Prog. Part. Nucl. Phys.}\ }\textbf {\bibinfo {volume} {120}},\ \bibinfo {pages} {103868} (\bibinfo {year} {2021})},\ \Eprint {http://arxiv.org/abs/2010.07592} {arXiv:2010.07592 [hep-ph]} \BibitemShut {NoStop}%
\bibitem [{\citenamefont {Gibson}\ \emph {et~al.}(2010)\citenamefont {Gibson}, \citenamefont {Imai}, \citenamefont {Motoba}, \citenamefont {Nagae},\ and\ \citenamefont {Ohnishi}}]{Gibson:2010zz}%
  \BibitemOpen
  \bibinfo {editor} {\bibfnamefont {B.~F.}\ \bibnamefont {Gibson}}, \bibinfo {editor} {\bibfnamefont {K.}~\bibnamefont {Imai}}, \bibinfo {editor} {\bibfnamefont {T.}~\bibnamefont {Motoba}}, \bibinfo {editor} {\bibfnamefont {T.}~\bibnamefont {Nagae}}, \ and\ \bibinfo {editor} {\bibfnamefont {A.}~\bibnamefont {Ohnishi}},\ eds.,\ \href@noop {} {\emph {\bibinfo {title} {{Proceedings, 10th International Conference on Hypernuclear and strange particle physics (HYP 2009)}: {Tokai, Japan, September 14-18, 2009}}}},\ Vol.\ \bibinfo {volume} {835}\ (\bibinfo {year} {2010})\BibitemShut {NoStop}%
\bibitem [{\citenamefont {Gal}\ \emph {et~al.}(2016)\citenamefont {Gal}, \citenamefont {Hungerford},\ and\ \citenamefont {Millener}}]{Gal:2016boi}%
  \BibitemOpen
  \bibfield  {author} {\bibinfo {author} {\bibfnamefont {A.}~\bibnamefont {Gal}}, \bibinfo {author} {\bibfnamefont {E.~V.}\ \bibnamefont {Hungerford}}, \ and\ \bibinfo {author} {\bibfnamefont {D.~J.}\ \bibnamefont {Millener}},\ }\href {\doibase 10.1103/RevModPhys.88.035004} {\bibfield  {journal} {\bibinfo  {journal} {Rev. Mod. Phys.}\ }\textbf {\bibinfo {volume} {88}},\ \bibinfo {pages} {035004} (\bibinfo {year} {2016})},\ \Eprint {http://arxiv.org/abs/1605.00557} {arXiv:1605.00557 [nucl-th]} \BibitemShut {NoStop}%
\bibitem [{\citenamefont {Hiyama}\ and\ \citenamefont {Nakazawa}(2018)}]{Hiyama:2018lgs}%
  \BibitemOpen
  \bibfield  {author} {\bibinfo {author} {\bibfnamefont {E.}~\bibnamefont {Hiyama}}\ and\ \bibinfo {author} {\bibfnamefont {K.}~\bibnamefont {Nakazawa}},\ }\href {\doibase 10.1146/annurev-nucl-101917-021108} {\bibfield  {journal} {\bibinfo  {journal} {Ann. Rev. Nucl. Part. Sci.}\ }\textbf {\bibinfo {volume} {68}},\ \bibinfo {pages} {131} (\bibinfo {year} {2018})}\BibitemShut {NoStop}%
\bibitem [{\citenamefont {Tamura}(2022)}]{Tamura:2022jet}%
  \BibitemOpen
  \bibfield  {author} {\bibinfo {author} {\bibfnamefont {H.}~\bibnamefont {Tamura}},\ }\href {\doibase 10.1051/epjconf/202227112001} {\bibfield  {journal} {\bibinfo  {journal} {EPJ Web Conf.}\ }\textbf {\bibinfo {volume} {271}},\ \bibinfo {pages} {12001} (\bibinfo {year} {2022})}\BibitemShut {NoStop}%
\bibitem [{\citenamefont {Miwa}\ \emph {et~al.}(2022)\citenamefont {Miwa} \emph {et~al.}}]{Miwa:2022coz}%
  \BibitemOpen
  \bibfield  {author} {\bibinfo {author} {\bibfnamefont {K.}~\bibnamefont {Miwa}} \emph {et~al.},\ }\href {\doibase 10.1051/epjconf/202227104001} {\bibfield  {journal} {\bibinfo  {journal} {EPJ Web Conf.}\ }\textbf {\bibinfo {volume} {271}},\ \bibinfo {pages} {04001} (\bibinfo {year} {2022})}\BibitemShut {NoStop}%
\bibitem [{\citenamefont {Nanamura}\ \emph {et~al.}(2022)\citenamefont {Nanamura} \emph {et~al.}}]{J-PARCE40:2022nvq}%
  \BibitemOpen
  \bibfield  {author} {\bibinfo {author} {\bibfnamefont {T.}~\bibnamefont {Nanamura}} \emph {et~al.} (\bibinfo {collaboration} {J-PARC E40}),\ }\href {\doibase 10.1093/ptep/ptac101} {\bibfield  {journal} {\bibinfo  {journal} {PTEP}\ }\textbf {\bibinfo {volume} {2022}},\ \bibinfo {pages} {093D01} (\bibinfo {year} {2022})},\ \Eprint {http://arxiv.org/abs/2203.08393} {arXiv:2203.08393 [nucl-ex]} \BibitemShut {NoStop}%
\bibitem [{\citenamefont {Demorest}\ \emph {et~al.}(2010)\citenamefont {Demorest}, \citenamefont {Pennucci}, \citenamefont {Ransom}, \citenamefont {Roberts},\ and\ \citenamefont {Hessels}}]{Demorest:2010bx}%
  \BibitemOpen
  \bibfield  {author} {\bibinfo {author} {\bibfnamefont {P.}~\bibnamefont {Demorest}}, \bibinfo {author} {\bibfnamefont {T.}~\bibnamefont {Pennucci}}, \bibinfo {author} {\bibfnamefont {S.}~\bibnamefont {Ransom}}, \bibinfo {author} {\bibfnamefont {M.}~\bibnamefont {Roberts}}, \ and\ \bibinfo {author} {\bibfnamefont {J.}~\bibnamefont {Hessels}},\ }\href {\doibase 10.1038/nature09466} {\bibfield  {journal} {\bibinfo  {journal} {Nature}\ }\textbf {\bibinfo {volume} {467}},\ \bibinfo {pages} {1081} (\bibinfo {year} {2010})},\ \Eprint {http://arxiv.org/abs/1010.5788} {arXiv:1010.5788 [astro-ph.HE]} \BibitemShut {NoStop}%
%%CITATION = ARXIV:1010.5788;%%
\bibitem [{\citenamefont {Antoniadis}\ \emph {et~al.}(2013)\citenamefont {Antoniadis} \emph {et~al.}}]{Antoniadis:2013pzd}%
  \BibitemOpen
  \bibfield  {author} {\bibinfo {author} {\bibfnamefont {J.}~\bibnamefont {Antoniadis}} \emph {et~al.},\ }\href {\doibase 10.1126/science.1233232} {\bibfield  {journal} {\bibinfo  {journal} {Science}\ }\textbf {\bibinfo {volume} {340}},\ \bibinfo {pages} {6131} (\bibinfo {year} {2013})},\ \Eprint {http://arxiv.org/abs/1304.6875} {arXiv:1304.6875 [astro-ph.HE]} \BibitemShut {NoStop}%
%%CITATION = ARXIV:1304.6875;%%
\bibitem [{\citenamefont {Riley}\ \emph {et~al.}(2021)\citenamefont {Riley} \emph {et~al.}}]{Riley:2021pdl}%
  \BibitemOpen
  \bibfield  {author} {\bibinfo {author} {\bibfnamefont {T.~E.}\ \bibnamefont {Riley}} \emph {et~al.},\ }\href {\doibase 10.3847/2041-8213/ac0a81} {\bibfield  {journal} {\bibinfo  {journal} {Astrophys. J. Lett.}\ }\textbf {\bibinfo {volume} {918}},\ \bibinfo {pages} {L27} (\bibinfo {year} {2021})},\ \Eprint {http://arxiv.org/abs/2105.06980} {arXiv:2105.06980 [astro-ph.HE]} \BibitemShut {NoStop}%
\bibitem [{\citenamefont {Fonseca}\ \emph {et~al.}(2021)\citenamefont {Fonseca} \emph {et~al.}}]{Fonseca:2021wxt}%
  \BibitemOpen
  \bibfield  {author} {\bibinfo {author} {\bibfnamefont {E.}~\bibnamefont {Fonseca}} \emph {et~al.},\ }\href {\doibase 10.3847/2041-8213/ac03b8} {\bibfield  {journal} {\bibinfo  {journal} {Astrophys. J. Lett.}\ }\textbf {\bibinfo {volume} {915}},\ \bibinfo {pages} {L12} (\bibinfo {year} {2021})},\ \Eprint {http://arxiv.org/abs/2104.00880} {arXiv:2104.00880 [astro-ph.HE]} \BibitemShut {NoStop}%
\bibitem [{\citenamefont {Glendenning}(1997)}]{Glendenning:1997wn}%
  \BibitemOpen
  \bibfield  {author} {\bibinfo {author} {\bibfnamefont {N.~K.}\ \bibnamefont {Glendenning}},\ }\href@noop {} {\emph {\bibinfo {title} {{Compact stars: Nuclear physics, particle physics, and general relativity}}}}\ (\bibinfo {year} {1997})\BibitemShut {NoStop}%
%%CITATION = INSPIRE-456851;%%
\bibitem [{\citenamefont {Bombaci}(2017)}]{Bombaci:2016xzl}%
  \BibitemOpen
  \bibfield  {author} {\bibinfo {author} {\bibfnamefont {I.}~\bibnamefont {Bombaci}},\ }\href {\doibase 10.7566/JPSCP.17.101002} {\bibfield  {journal} {\bibinfo  {journal} {JPS Conf. Proc.}\ }\textbf {\bibinfo {volume} {17}},\ \bibinfo {pages} {101002} (\bibinfo {year} {2017})},\ \Eprint {http://arxiv.org/abs/1601.05339} {arXiv:1601.05339 [nucl-th]} \BibitemShut {NoStop}%
\bibitem [{\citenamefont {Chatterjee}\ and\ \citenamefont {Vida\~na}(2016)}]{Chatterjee:2015pua}%
  \BibitemOpen
  \bibfield  {author} {\bibinfo {author} {\bibfnamefont {D.}~\bibnamefont {Chatterjee}}\ and\ \bibinfo {author} {\bibfnamefont {I.}~\bibnamefont {Vida\~na}},\ }\href {\doibase 10.1140/epja/i2016-16029-x} {\bibfield  {journal} {\bibinfo  {journal} {Eur. Phys. J. A}\ }\textbf {\bibinfo {volume} {52}},\ \bibinfo {pages} {29} (\bibinfo {year} {2016})},\ \Eprint {http://arxiv.org/abs/1510.06306} {arXiv:1510.06306 [nucl-th]} \BibitemShut {NoStop}%
\bibitem [{\citenamefont {Burgio}\ \emph {et~al.}(2021)\citenamefont {Burgio}, \citenamefont {Schulze}, \citenamefont {Vidana},\ and\ \citenamefont {Wei}}]{Burgio:2021vgk}%
  \BibitemOpen
  \bibfield  {author} {\bibinfo {author} {\bibfnamefont {G.~F.}\ \bibnamefont {Burgio}}, \bibinfo {author} {\bibfnamefont {H.~J.}\ \bibnamefont {Schulze}}, \bibinfo {author} {\bibfnamefont {I.}~\bibnamefont {Vidana}}, \ and\ \bibinfo {author} {\bibfnamefont {J.~B.}\ \bibnamefont {Wei}},\ }\href {\doibase 10.1016/j.ppnp.2021.103879} {\bibfield  {journal} {\bibinfo  {journal} {Prog. Part. Nucl. Phys.}\ }\textbf {\bibinfo {volume} {120}},\ \bibinfo {pages} {103879} (\bibinfo {year} {2021})},\ \Eprint {http://arxiv.org/abs/2105.03747} {arXiv:2105.03747 [nucl-th]} \BibitemShut {NoStop}%
\bibitem [{\citenamefont {Weissenborn}\ \emph {et~al.}(2012)\citenamefont {Weissenborn}, \citenamefont {Chatterjee},\ and\ \citenamefont {Schaffner-Bielich}}]{Weissenborn:2011ut}%
  \BibitemOpen
  \bibfield  {author} {\bibinfo {author} {\bibfnamefont {S.}~\bibnamefont {Weissenborn}}, \bibinfo {author} {\bibfnamefont {D.}~\bibnamefont {Chatterjee}}, \ and\ \bibinfo {author} {\bibfnamefont {J.}~\bibnamefont {Schaffner-Bielich}},\ }\href {\doibase 10.1103/PhysRevC.85.065802} {\bibfield  {journal} {\bibinfo  {journal} {Phys. Rev. C}\ }\textbf {\bibinfo {volume} {85}},\ \bibinfo {pages} {065802} (\bibinfo {year} {2012})},\ \bibinfo {note} {[Erratum: Phys.Rev.C 90, 019904 (2014)]},\ \Eprint {http://arxiv.org/abs/1112.0234} {arXiv:1112.0234 [astro-ph.HE]} \BibitemShut {NoStop}%
\bibitem [{\citenamefont {Spinella}\ and\ \citenamefont {Weber}(2019)}]{Spinella:2018dab}%
  \BibitemOpen
  \bibfield  {author} {\bibinfo {author} {\bibfnamefont {W.~M.}\ \bibnamefont {Spinella}}\ and\ \bibinfo {author} {\bibfnamefont {F.}~\bibnamefont {Weber}},\ }\href {\doibase 10.1002/asna.201913579} {\bibfield  {journal} {\bibinfo  {journal} {Astron. Nachr.}\ }\textbf {\bibinfo {volume} {340}},\ \bibinfo {pages} {145} (\bibinfo {year} {2019})},\ \Eprint {http://arxiv.org/abs/1812.03600} {arXiv:1812.03600 [nucl-th]} \BibitemShut {NoStop}%
\bibitem [{\citenamefont {Yamamoto}\ \emph {et~al.}(2013)\citenamefont {Yamamoto}, \citenamefont {Furumoto}, \citenamefont {Yasutake},\ and\ \citenamefont {Rijken}}]{Yamamoto:2013ada}%
  \BibitemOpen
  \bibfield  {author} {\bibinfo {author} {\bibfnamefont {Y.}~\bibnamefont {Yamamoto}}, \bibinfo {author} {\bibfnamefont {T.}~\bibnamefont {Furumoto}}, \bibinfo {author} {\bibfnamefont {N.}~\bibnamefont {Yasutake}}, \ and\ \bibinfo {author} {\bibfnamefont {T.~A.}\ \bibnamefont {Rijken}},\ }\href {\doibase 10.1103/PhysRevC.88.022801} {\bibfield  {journal} {\bibinfo  {journal} {Phys. Rev. C}\ }\textbf {\bibinfo {volume} {88}},\ \bibinfo {pages} {022801} (\bibinfo {year} {2013})},\ \Eprint {http://arxiv.org/abs/1308.2130} {arXiv:1308.2130 [nucl-th]} \BibitemShut {NoStop}%
\bibitem [{\citenamefont {Haidenbauer}\ \emph {et~al.}(2017)\citenamefont {Haidenbauer}, \citenamefont {Mei\ss{}ner}, \citenamefont {Kaiser},\ and\ \citenamefont {Weise}}]{Haidenbauer:2016vfq}%
  \BibitemOpen
  \bibfield  {author} {\bibinfo {author} {\bibfnamefont {J.}~\bibnamefont {Haidenbauer}}, \bibinfo {author} {\bibfnamefont {U.~G.}\ \bibnamefont {Mei\ss{}ner}}, \bibinfo {author} {\bibfnamefont {N.}~\bibnamefont {Kaiser}}, \ and\ \bibinfo {author} {\bibfnamefont {W.}~\bibnamefont {Weise}},\ }\href {\doibase 10.1140/epja/i2017-12316-4} {\bibfield  {journal} {\bibinfo  {journal} {Eur. Phys. J. A}\ }\textbf {\bibinfo {volume} {53}},\ \bibinfo {pages} {121} (\bibinfo {year} {2017})},\ \Eprint {http://arxiv.org/abs/1612.03758} {arXiv:1612.03758 [nucl-th]} \BibitemShut {NoStop}%
\bibitem [{\citenamefont {Kohno}(2018)}]{Kohno:2018gby}%
  \BibitemOpen
  \bibfield  {author} {\bibinfo {author} {\bibfnamefont {M.}~\bibnamefont {Kohno}},\ }\href {\doibase 10.1103/PhysRevC.97.035206} {\bibfield  {journal} {\bibinfo  {journal} {Phys. Rev. C}\ }\textbf {\bibinfo {volume} {97}},\ \bibinfo {pages} {035206} (\bibinfo {year} {2018})},\ \Eprint {http://arxiv.org/abs/1802.05388} {arXiv:1802.05388 [nucl-th]} \BibitemShut {NoStop}%
\bibitem [{\citenamefont {Logoteta}\ \emph {et~al.}(2019)\citenamefont {Logoteta}, \citenamefont {Vidana},\ and\ \citenamefont {Bombaci}}]{Logoteta:2019utx}%
  \BibitemOpen
  \bibfield  {author} {\bibinfo {author} {\bibfnamefont {D.}~\bibnamefont {Logoteta}}, \bibinfo {author} {\bibfnamefont {I.}~\bibnamefont {Vidana}}, \ and\ \bibinfo {author} {\bibfnamefont {I.}~\bibnamefont {Bombaci}},\ }\href {\doibase 10.1140/epja/i2019-12909-9} {\bibfield  {journal} {\bibinfo  {journal} {Eur. Phys. J. A}\ }\textbf {\bibinfo {volume} {55}},\ \bibinfo {pages} {207} (\bibinfo {year} {2019})},\ \Eprint {http://arxiv.org/abs/1906.11722} {arXiv:1906.11722 [nucl-th]} \BibitemShut {NoStop}%
\bibitem [{\citenamefont {Gerstung}\ \emph {et~al.}(2020)\citenamefont {Gerstung}, \citenamefont {Kaiser},\ and\ \citenamefont {Weise}}]{Gerstung:2020ktv}%
  \BibitemOpen
  \bibfield  {author} {\bibinfo {author} {\bibfnamefont {D.}~\bibnamefont {Gerstung}}, \bibinfo {author} {\bibfnamefont {N.}~\bibnamefont {Kaiser}}, \ and\ \bibinfo {author} {\bibfnamefont {W.}~\bibnamefont {Weise}},\ }\href {\doibase 10.1140/epja/s10050-020-00180-2} {\bibfield  {journal} {\bibinfo  {journal} {Eur. Phys. J. A}\ }\textbf {\bibinfo {volume} {56}},\ \bibinfo {pages} {175} (\bibinfo {year} {2020})},\ \Eprint {http://arxiv.org/abs/2001.10563} {arXiv:2001.10563 [nucl-th]} \BibitemShut {NoStop}%
\bibitem [{\citenamefont {Contrera}\ \emph {et~al.}(2022)\citenamefont {Contrera}, \citenamefont {Blaschke}, \citenamefont {Carlomagno}, \citenamefont {Grunfeld},\ and\ \citenamefont {Liebing}}]{Contrera:2022tqh}%
  \BibitemOpen
  \bibfield  {author} {\bibinfo {author} {\bibfnamefont {G.~A.}\ \bibnamefont {Contrera}}, \bibinfo {author} {\bibfnamefont {D.}~\bibnamefont {Blaschke}}, \bibinfo {author} {\bibfnamefont {J.~P.}\ \bibnamefont {Carlomagno}}, \bibinfo {author} {\bibfnamefont {A.~G.}\ \bibnamefont {Grunfeld}}, \ and\ \bibinfo {author} {\bibfnamefont {S.}~\bibnamefont {Liebing}},\ }\href {\doibase 10.1103/PhysRevC.105.045808} {\bibfield  {journal} {\bibinfo  {journal} {Phys. Rev. C}\ }\textbf {\bibinfo {volume} {105}},\ \bibinfo {pages} {045808} (\bibinfo {year} {2022})},\ \Eprint {http://arxiv.org/abs/2201.00477} {arXiv:2201.00477 [nucl-th]} \BibitemShut {NoStop}%
\bibitem [{\citenamefont {Buballa}(2005)}]{Buballa:2003qv}%
  \BibitemOpen
  \bibfield  {author} {\bibinfo {author} {\bibfnamefont {M.}~\bibnamefont {Buballa}},\ }\href {\doibase 10.1016/j.physrep.2004.11.004} {\bibfield  {journal} {\bibinfo  {journal} {Phys. Rept.}\ }\textbf {\bibinfo {volume} {407}},\ \bibinfo {pages} {205} (\bibinfo {year} {2005})},\ \Eprint {http://arxiv.org/abs/hep-ph/0402234} {arXiv:hep-ph/0402234 [hep-ph]} \BibitemShut {NoStop}%
%%CITATION = HEP-PH/0402234;%%
\bibitem [{\citenamefont {Alford}\ \emph {et~al.}(2008)\citenamefont {Alford}, \citenamefont {Schmitt}, \citenamefont {Rajagopal},\ and\ \citenamefont {Sch\"afer}}]{Alford:2007xm}%
  \BibitemOpen
  \bibfield  {author} {\bibinfo {author} {\bibfnamefont {M.~G.}\ \bibnamefont {Alford}}, \bibinfo {author} {\bibfnamefont {A.}~\bibnamefont {Schmitt}}, \bibinfo {author} {\bibfnamefont {K.}~\bibnamefont {Rajagopal}}, \ and\ \bibinfo {author} {\bibfnamefont {T.}~\bibnamefont {Sch\"afer}},\ }\href {\doibase 10.1103/RevModPhys.80.1455} {\bibfield  {journal} {\bibinfo  {journal} {Rev. Mod. Phys.}\ }\textbf {\bibinfo {volume} {80}},\ \bibinfo {pages} {1455} (\bibinfo {year} {2008})},\ \Eprint {http://arxiv.org/abs/0709.4635} {arXiv:0709.4635 [hep-ph]} \BibitemShut {NoStop}%
\bibitem [{\citenamefont {Guichon}(1988)}]{Guichon:1987jp}%
  \BibitemOpen
  \bibfield  {author} {\bibinfo {author} {\bibfnamefont {P.~A.~M.}\ \bibnamefont {Guichon}},\ }\href {\doibase 10.1016/0370-2693(88)90762-9} {\bibfield  {journal} {\bibinfo  {journal} {Phys. Lett. B}\ }\textbf {\bibinfo {volume} {200}},\ \bibinfo {pages} {235} (\bibinfo {year} {1988})}\BibitemShut {NoStop}%
\bibitem [{\citenamefont {Chodos}\ \emph {et~al.}(1974)\citenamefont {Chodos}, \citenamefont {Jaffe}, \citenamefont {Johnson}, \citenamefont {Thorn},\ and\ \citenamefont {Weisskopf}}]{Chodos:1974je}%
  \BibitemOpen
  \bibfield  {author} {\bibinfo {author} {\bibfnamefont {A.}~\bibnamefont {Chodos}}, \bibinfo {author} {\bibfnamefont {R.~L.}\ \bibnamefont {Jaffe}}, \bibinfo {author} {\bibfnamefont {K.}~\bibnamefont {Johnson}}, \bibinfo {author} {\bibfnamefont {C.~B.}\ \bibnamefont {Thorn}}, \ and\ \bibinfo {author} {\bibfnamefont {V.~F.}\ \bibnamefont {Weisskopf}},\ }\href {\doibase 10.1103/PhysRevD.9.3471} {\bibfield  {journal} {\bibinfo  {journal} {Phys. Rev. D}\ }\textbf {\bibinfo {volume} {9}},\ \bibinfo {pages} {3471} (\bibinfo {year} {1974})}\BibitemShut {NoStop}%
\bibitem [{\citenamefont {Nambu}\ and\ \citenamefont {Jona-Lasinio}(1961{\natexlab{a}})}]{Nambu:1961tp}%
  \BibitemOpen
  \bibfield  {author} {\bibinfo {author} {\bibfnamefont {Y.}~\bibnamefont {Nambu}}\ and\ \bibinfo {author} {\bibfnamefont {G.}~\bibnamefont {Jona-Lasinio}},\ }\href {\doibase 10.1103/PhysRev.122.345} {\bibfield  {journal} {\bibinfo  {journal} {Phys. Rev.}\ }\textbf {\bibinfo {volume} {122}},\ \bibinfo {pages} {345} (\bibinfo {year} {1961}{\natexlab{a}})},\ \bibinfo {note} {[,127(1961)]}\BibitemShut {NoStop}%
%%CITATION = PHRVA,122,345;%%
\bibitem [{\citenamefont {Nambu}\ and\ \citenamefont {Jona-Lasinio}(1961{\natexlab{b}})}]{Nambu:1961fr}%
  \BibitemOpen
  \bibfield  {author} {\bibinfo {author} {\bibfnamefont {Y.}~\bibnamefont {Nambu}}\ and\ \bibinfo {author} {\bibfnamefont {G.}~\bibnamefont {Jona-Lasinio}},\ }\href {\doibase 10.1103/PhysRev.124.246} {\bibfield  {journal} {\bibinfo  {journal} {Phys. Rev.}\ }\textbf {\bibinfo {volume} {124}},\ \bibinfo {pages} {246} (\bibinfo {year} {1961}{\natexlab{b}})},\ \bibinfo {note} {[,141(1961)]}\BibitemShut {NoStop}%
%%CITATION = PHRVA,124,246;%%
\bibitem [{\citenamefont {Vogl}\ and\ \citenamefont {Weise}(1991)}]{Vogl:1991qt}%
  \BibitemOpen
  \bibfield  {author} {\bibinfo {author} {\bibfnamefont {U.}~\bibnamefont {Vogl}}\ and\ \bibinfo {author} {\bibfnamefont {W.}~\bibnamefont {Weise}},\ }\href {\doibase 10.1016/0146-6410(91)90005-9} {\bibfield  {journal} {\bibinfo  {journal} {Prog. Part. Nucl. Phys.}\ }\textbf {\bibinfo {volume} {27}},\ \bibinfo {pages} {195} (\bibinfo {year} {1991})}\BibitemShut {NoStop}%
%%CITATION = PPNPD,27,195;%%
\bibitem [{\citenamefont {Hatsuda}\ and\ \citenamefont {Kunihiro}(1994)}]{Hatsuda:1994pi}%
  \BibitemOpen
  \bibfield  {author} {\bibinfo {author} {\bibfnamefont {T.}~\bibnamefont {Hatsuda}}\ and\ \bibinfo {author} {\bibfnamefont {T.}~\bibnamefont {Kunihiro}},\ }\href {\doibase 10.1016/0370-1573(94)90022-1} {\bibfield  {journal} {\bibinfo  {journal} {Phys. Rept.}\ }\textbf {\bibinfo {volume} {247}},\ \bibinfo {pages} {221} (\bibinfo {year} {1994})},\ \Eprint {http://arxiv.org/abs/hep-ph/9401310} {arXiv:hep-ph/9401310 [hep-ph]} \BibitemShut {NoStop}%
%%CITATION = HEP-PH/9401310;%%
\bibitem [{\citenamefont {Ishii}\ \emph {et~al.}(1995)\citenamefont {Ishii}, \citenamefont {Bentz},\ and\ \citenamefont {Yazaki}}]{Ishii:1995bu}%
  \BibitemOpen
  \bibfield  {author} {\bibinfo {author} {\bibfnamefont {N.}~\bibnamefont {Ishii}}, \bibinfo {author} {\bibfnamefont {W.}~\bibnamefont {Bentz}}, \ and\ \bibinfo {author} {\bibfnamefont {K.}~\bibnamefont {Yazaki}},\ }\href {\doibase 10.1016/0375-9474(95)00032-V} {\bibfield  {journal} {\bibinfo  {journal} {Nucl. Phys. A}\ }\textbf {\bibinfo {volume} {587}},\ \bibinfo {pages} {617} (\bibinfo {year} {1995})}\BibitemShut {NoStop}%
\bibitem [{\citenamefont {Bentz}\ and\ \citenamefont {Thomas}(2001)}]{Bentz:2001vc}%
  \BibitemOpen
  \bibfield  {author} {\bibinfo {author} {\bibfnamefont {W.}~\bibnamefont {Bentz}}\ and\ \bibinfo {author} {\bibfnamefont {A.~W.}\ \bibnamefont {Thomas}},\ }\href {\doibase 10.1016/S0375-9474(01)01119-8} {\bibfield  {journal} {\bibinfo  {journal} {Nucl. Phys.}\ }\textbf {\bibinfo {volume} {A696}},\ \bibinfo {pages} {138} (\bibinfo {year} {2001})},\ \Eprint {http://arxiv.org/abs/nucl-th/0105022} {arXiv:nucl-th/0105022 [nucl-th]} \BibitemShut {NoStop}%
%%CITATION = NUCL-TH/0105022;%%
\bibitem [{\citenamefont {Geesaman}\ \emph {et~al.}(1995)\citenamefont {Geesaman}, \citenamefont {Saito},\ and\ \citenamefont {Thomas}}]{Geesaman:1995yd}%
  \BibitemOpen
  \bibfield  {author} {\bibinfo {author} {\bibfnamefont {D.~F.}\ \bibnamefont {Geesaman}}, \bibinfo {author} {\bibfnamefont {K.}~\bibnamefont {Saito}}, \ and\ \bibinfo {author} {\bibfnamefont {A.~W.}\ \bibnamefont {Thomas}},\ }\href {\doibase 10.1146/annurev.ns.45.120195.002005} {\bibfield  {journal} {\bibinfo  {journal} {Ann. Rev. Nucl. Part. Sci.}\ }\textbf {\bibinfo {volume} {45}},\ \bibinfo {pages} {337} (\bibinfo {year} {1995})}\BibitemShut {NoStop}%
\bibitem [{\citenamefont {Lu}\ \emph {et~al.}(1998)\citenamefont {Lu}, \citenamefont {Thomas}, \citenamefont {Tsushima}, \citenamefont {Williams},\ and\ \citenamefont {Saito}}]{Lu:1997mu}%
  \BibitemOpen
  \bibfield  {author} {\bibinfo {author} {\bibfnamefont {D.-H.}\ \bibnamefont {Lu}}, \bibinfo {author} {\bibfnamefont {A.~W.}\ \bibnamefont {Thomas}}, \bibinfo {author} {\bibfnamefont {K.}~\bibnamefont {Tsushima}}, \bibinfo {author} {\bibfnamefont {A.~G.}\ \bibnamefont {Williams}}, \ and\ \bibinfo {author} {\bibfnamefont {K.}~\bibnamefont {Saito}},\ }\href {\doibase 10.1016/S0370-2693(97)01385-3} {\bibfield  {journal} {\bibinfo  {journal} {Phys. Lett. B}\ }\textbf {\bibinfo {volume} {417}},\ \bibinfo {pages} {217} (\bibinfo {year} {1998})},\ \Eprint {http://arxiv.org/abs/nucl-th/9706043} {arXiv:nucl-th/9706043} \BibitemShut {NoStop}%
\bibitem [{\citenamefont {Stone}\ \emph {et~al.}(2016)\citenamefont {Stone}, \citenamefont {Guichon}, \citenamefont {Reinhard},\ and\ \citenamefont {Thomas}}]{Stone:2016qmi}%
  \BibitemOpen
  \bibfield  {author} {\bibinfo {author} {\bibfnamefont {J.~R.}\ \bibnamefont {Stone}}, \bibinfo {author} {\bibfnamefont {P.~A.~M.}\ \bibnamefont {Guichon}}, \bibinfo {author} {\bibfnamefont {P.~G.}\ \bibnamefont {Reinhard}}, \ and\ \bibinfo {author} {\bibfnamefont {A.~W.}\ \bibnamefont {Thomas}},\ }\href {\doibase 10.1103/PhysRevLett.116.092501} {\bibfield  {journal} {\bibinfo  {journal} {Phys. Rev. Lett.}\ }\textbf {\bibinfo {volume} {116}},\ \bibinfo {pages} {092501} (\bibinfo {year} {2016})},\ \Eprint {http://arxiv.org/abs/1601.08131} {arXiv:1601.08131 [nucl-th]} \BibitemShut {NoStop}%
\bibitem [{\citenamefont {Cloët}\ \emph {et~al.}(2005)\citenamefont {Cloët}, \citenamefont {Bentz},\ and\ \citenamefont {Thomas}}]{Cloet:2005rt}%
  \BibitemOpen
  \bibfield  {author} {\bibinfo {author} {\bibfnamefont {I.~C.}\ \bibnamefont {Cloët}}, \bibinfo {author} {\bibfnamefont {W.}~\bibnamefont {Bentz}}, \ and\ \bibinfo {author} {\bibfnamefont {A.~W.}\ \bibnamefont {Thomas}},\ }\href {\doibase 10.1103/PhysRevLett.95.052302} {\bibfield  {journal} {\bibinfo  {journal} {Phys. Rev. Lett.}\ }\textbf {\bibinfo {volume} {95}},\ \bibinfo {pages} {052302} (\bibinfo {year} {2005})},\ \Eprint {http://arxiv.org/abs/nucl-th/0504019} {arXiv:nucl-th/0504019 [nucl-th]} \BibitemShut {NoStop}%
%%CITATION = NUCL-TH/0504019;%%
\bibitem [{\citenamefont {Cloet}\ \emph {et~al.}(2006)\citenamefont {Cloet}, \citenamefont {Bentz},\ and\ \citenamefont {Thomas}}]{Cloet:2006bq}%
  \BibitemOpen
  \bibfield  {author} {\bibinfo {author} {\bibfnamefont {I.~C.}\ \bibnamefont {Cloet}}, \bibinfo {author} {\bibfnamefont {W.}~\bibnamefont {Bentz}}, \ and\ \bibinfo {author} {\bibfnamefont {A.~W.}\ \bibnamefont {Thomas}},\ }\href {\doibase 10.1016/j.physletb.2006.08.076} {\bibfield  {journal} {\bibinfo  {journal} {Phys. Lett. B}\ }\textbf {\bibinfo {volume} {642}},\ \bibinfo {pages} {210} (\bibinfo {year} {2006})},\ \Eprint {http://arxiv.org/abs/nucl-th/0605061} {arXiv:nucl-th/0605061} \BibitemShut {NoStop}%
\bibitem [{\citenamefont {Clo\"et}\ \emph {et~al.}(2016)\citenamefont {Clo\"et}, \citenamefont {Bentz},\ and\ \citenamefont {Thomas}}]{Cloet:2015tha}%
  \BibitemOpen
  \bibfield  {author} {\bibinfo {author} {\bibfnamefont {I.~C.}\ \bibnamefont {Clo\"et}}, \bibinfo {author} {\bibfnamefont {W.}~\bibnamefont {Bentz}}, \ and\ \bibinfo {author} {\bibfnamefont {A.~W.}\ \bibnamefont {Thomas}},\ }\href {\doibase 10.1103/PhysRevLett.116.032701} {\bibfield  {journal} {\bibinfo  {journal} {Phys. Rev. Lett.}\ }\textbf {\bibinfo {volume} {116}},\ \bibinfo {pages} {032701} (\bibinfo {year} {2016})},\ \Eprint {http://arxiv.org/abs/1506.05875} {arXiv:1506.05875 [nucl-th]} \BibitemShut {NoStop}%
\bibitem [{\citenamefont {Saito}\ \emph {et~al.}(2007)\citenamefont {Saito}, \citenamefont {Tsushima},\ and\ \citenamefont {Thomas}}]{Saito:2005rv}%
  \BibitemOpen
  \bibfield  {author} {\bibinfo {author} {\bibfnamefont {K.}~\bibnamefont {Saito}}, \bibinfo {author} {\bibfnamefont {K.}~\bibnamefont {Tsushima}}, \ and\ \bibinfo {author} {\bibfnamefont {A.~W.}\ \bibnamefont {Thomas}},\ }\href {\doibase 10.1016/j.ppnp.2005.07.003} {\bibfield  {journal} {\bibinfo  {journal} {Prog. Part. Nucl. Phys.}\ }\textbf {\bibinfo {volume} {58}},\ \bibinfo {pages} {1} (\bibinfo {year} {2007})},\ \Eprint {http://arxiv.org/abs/hep-ph/0506314} {arXiv:hep-ph/0506314} \BibitemShut {NoStop}%
\bibitem [{\citenamefont {Sammarruca}(2008)}]{Sammarruca:2008hy}%
  \BibitemOpen
  \bibfield  {author} {\bibinfo {author} {\bibfnamefont {F.}~\bibnamefont {Sammarruca}},\ }\href@noop {} {\  (\bibinfo {year} {2008})},\ \Eprint {http://arxiv.org/abs/0801.0879} {arXiv:0801.0879 [nucl-th]} \BibitemShut {NoStop}%
\bibitem [{\citenamefont {Guichon}\ \emph {et~al.}(2008)\citenamefont {Guichon}, \citenamefont {Thomas},\ and\ \citenamefont {Tsushima}}]{Guichon:2008zz}%
  \BibitemOpen
  \bibfield  {author} {\bibinfo {author} {\bibfnamefont {P.~A.~M.}\ \bibnamefont {Guichon}}, \bibinfo {author} {\bibfnamefont {A.~W.}\ \bibnamefont {Thomas}}, \ and\ \bibinfo {author} {\bibfnamefont {K.}~\bibnamefont {Tsushima}},\ }\href {\doibase 10.1016/j.nuclphysa.2008.10.001} {\bibfield  {journal} {\bibinfo  {journal} {Nucl. Phys. A}\ }\textbf {\bibinfo {volume} {814}},\ \bibinfo {pages} {66} (\bibinfo {year} {2008})},\ \Eprint {http://arxiv.org/abs/0712.1925} {arXiv:0712.1925 [nucl-th]} \BibitemShut {NoStop}%
\bibitem [{\citenamefont {Close}(1979)}]{Close:1979bt}%
  \BibitemOpen
  \bibfield  {author} {\bibinfo {author} {\bibfnamefont {F.~E.}\ \bibnamefont {Close}},\ }\href@noop {} {\emph {\bibinfo {title} {{An Introduction to Quarks and Partons}}}}\ (\bibinfo {year} {1979})\BibitemShut {NoStop}%
\bibitem [{\citenamefont {Carrillo-Serrano}\ \emph {et~al.}(2016)\citenamefont {Carrillo-Serrano}, \citenamefont {Bentz}, \citenamefont {Cloët},\ and\ \citenamefont {Thomas}}]{Carrillo-Serrano:2016igi}%
  \BibitemOpen
  \bibfield  {author} {\bibinfo {author} {\bibfnamefont {M.~E.}\ \bibnamefont {Carrillo-Serrano}}, \bibinfo {author} {\bibfnamefont {W.}~\bibnamefont {Bentz}}, \bibinfo {author} {\bibfnamefont {I.~C.}\ \bibnamefont {Cloët}}, \ and\ \bibinfo {author} {\bibfnamefont {A.~W.}\ \bibnamefont {Thomas}},\ }\href {\doibase 10.1016/j.physletb.2016.05.065} {\bibfield  {journal} {\bibinfo  {journal} {Phys. Lett.}\ }\textbf {\bibinfo {volume} {B759}},\ \bibinfo {pages} {178} (\bibinfo {year} {2016})},\ \Eprint {http://arxiv.org/abs/1603.02741} {arXiv:1603.02741 [nucl-th]} \BibitemShut {NoStop}%
%%CITATION = ARXIV:1603.02741;%%
\bibitem [{\citenamefont {Landau}(1956)}]{Landau:1956aa}%
  \BibitemOpen
  \bibfield  {author} {\bibinfo {author} {\bibfnamefont {L.~D.}\ \bibnamefont {Landau}},\ }\href {http://www.jetp.ac.ru/cgi-bin/e/index/e/3/6/p920?a=list} {\bibfield  {journal} {\bibinfo  {journal} {Sov. Phys. JETP}\ }\textbf {\bibinfo {volume} {3}},\ \bibinfo {pages} {920} (\bibinfo {year} {1956})}\BibitemShut {NoStop}%
\bibitem [{\citenamefont {Landau}(1957)}]{Landau:1957aa}%
  \BibitemOpen
  \bibfield  {author} {\bibinfo {author} {\bibfnamefont {L.~D.}\ \bibnamefont {Landau}},\ }\href {http://www.jetp.ac.ru/cgi-bin/e/index/e/5/1/p101?a=list} {\bibfield  {journal} {\bibinfo  {journal} {Sov. Phys. JETP}\ }\textbf {\bibinfo {volume} {5}},\ \bibinfo {pages} {101} (\bibinfo {year} {1957})}\BibitemShut {NoStop}%
\bibitem [{\citenamefont {Landau}(1959)}]{Landau:1959aa}%
  \BibitemOpen
  \bibfield  {author} {\bibinfo {author} {\bibfnamefont {L.~D.}\ \bibnamefont {Landau}},\ }\href {http://www.jetp.ac.ru/cgi-bin/e/index/e/8/1/p70?a=list} {\bibfield  {journal} {\bibinfo  {journal} {Sov. Phys. JETP}\ }\textbf {\bibinfo {volume} {8}},\ \bibinfo {pages} {70} (\bibinfo {year} {1959})}\BibitemShut {NoStop}%
\bibitem [{\citenamefont {Migdal}(1967)}]{Migdal:1967aa}%
  \BibitemOpen
  \bibfield  {author} {\bibinfo {author} {\bibfnamefont {A.~B.}\ \bibnamefont {Migdal}},\ }\href@noop {} {\emph {\bibinfo {title} {{Theory of finite Fermi systems and applications to atomic nuclei}}}}\ (\bibinfo  {publisher} {New York, Wiley},\ \bibinfo {year} {1967})\BibitemShut {NoStop}%
%%CITATION = INSPIRE-559956;%%
\bibitem [{\citenamefont {Migdal}\ \emph {et~al.}(1990)\citenamefont {Migdal}, \citenamefont {Saperstein}, \citenamefont {Troitsky},\ and\ \citenamefont {Voskresensky}}]{Migdal:1990vm}%
  \BibitemOpen
  \bibfield  {author} {\bibinfo {author} {\bibfnamefont {A.~B.}\ \bibnamefont {Migdal}}, \bibinfo {author} {\bibfnamefont {E.~E.}\ \bibnamefont {Saperstein}}, \bibinfo {author} {\bibfnamefont {M.~A.}\ \bibnamefont {Troitsky}}, \ and\ \bibinfo {author} {\bibfnamefont {D.~N.}\ \bibnamefont {Voskresensky}},\ }\href {\doibase 10.1016/0370-1573(90)90132-L} {\bibfield  {journal} {\bibinfo  {journal} {Phys. Rept.}\ }\textbf {\bibinfo {volume} {192}},\ \bibinfo {pages} {179} (\bibinfo {year} {1990})}\BibitemShut {NoStop}%
%%CITATION = PRPLC,192,179;%%
\bibitem [{\citenamefont {Baym}\ and\ \citenamefont {Chin}(1976)}]{Baym:1975va}%
  \BibitemOpen
  \bibfield  {author} {\bibinfo {author} {\bibfnamefont {G.}~\bibnamefont {Baym}}\ and\ \bibinfo {author} {\bibfnamefont {S.~A.}\ \bibnamefont {Chin}},\ }\href {\doibase 10.1016/0375-9474(76)90513-3} {\bibfield  {journal} {\bibinfo  {journal} {Nucl. Phys.}\ }\textbf {\bibinfo {volume} {A262}},\ \bibinfo {pages} {527} (\bibinfo {year} {1976})}\BibitemShut {NoStop}%
%%CITATION = NUPHA,A262,527;%%
\bibitem [{\citenamefont {Nozi{\`e}res}(1964)}]{Nozieres:1964aa}%
  \BibitemOpen
  \bibfield  {author} {\bibinfo {author} {\bibfnamefont {P.}~\bibnamefont {Nozi{\`e}res}},\ }\href@noop {} {\emph {\bibinfo {title} {Theory of interacting Fermi systems}}}\ (\bibinfo  {publisher} {W.A. Benjamin},\ \bibinfo {address} {New York},\ \bibinfo {year} {1964})\BibitemShut {NoStop}%
\bibitem [{\citenamefont {Negele}\ and\ \citenamefont {Orland}(1998)}]{Negele:1988aa}%
  \BibitemOpen
  \bibfield  {author} {\bibinfo {author} {\bibfnamefont {J.}~\bibnamefont {Negele}}\ and\ \bibinfo {author} {\bibfnamefont {H.}~\bibnamefont {Orland}},\ }\href {http://www.worldcat.org/isbn/0738200522} {\emph {\bibinfo {title} {Quantum Many-particle Systems}}}\ (\bibinfo  {publisher} {Westview Press},\ \bibinfo {year} {1998})\BibitemShut {NoStop}%
\bibitem [{\citenamefont {Shankar}(1994)}]{Shankar:1993pf}%
  \BibitemOpen
  \bibfield  {author} {\bibinfo {author} {\bibfnamefont {R.}~\bibnamefont {Shankar}},\ }\href {\doibase 10.1103/RevModPhys.66.129} {\bibfield  {journal} {\bibinfo  {journal} {Rev. Mod. Phys.}\ }\textbf {\bibinfo {volume} {66}},\ \bibinfo {pages} {129} (\bibinfo {year} {1994})},\ \Eprint {http://arxiv.org/abs/cond-mat/9307009} {arXiv:cond-mat/9307009} \BibitemShut {NoStop}%
\bibitem [{\citenamefont {'t~Hooft}(1976)}]{tHooft:1976snw}%
  \BibitemOpen
  \bibfield  {author} {\bibinfo {author} {\bibfnamefont {G.}~\bibnamefont {'t~Hooft}},\ }\href {\doibase 10.1103/PhysRevD.14.3432} {\bibfield  {journal} {\bibinfo  {journal} {Phys. Rev. D}\ }\textbf {\bibinfo {volume} {14}},\ \bibinfo {pages} {3432} (\bibinfo {year} {1976})},\ \bibinfo {note} {[Erratum: Phys.Rev.D 18, 2199 (1978)]}\BibitemShut {NoStop}%
\bibitem [{\citenamefont {Osipov}\ \emph {et~al.}(2007)\citenamefont {Osipov}, \citenamefont {Hiller}, \citenamefont {Blin},\ and\ \citenamefont {da~Providencia}}]{Osipov:2006ns}%
  \BibitemOpen
  \bibfield  {author} {\bibinfo {author} {\bibfnamefont {A.~A.}\ \bibnamefont {Osipov}}, \bibinfo {author} {\bibfnamefont {B.}~\bibnamefont {Hiller}}, \bibinfo {author} {\bibfnamefont {A.~H.}\ \bibnamefont {Blin}}, \ and\ \bibinfo {author} {\bibfnamefont {J.}~\bibnamefont {da~Providencia}},\ }\href {\doibase 10.1016/j.aop.2006.08.004} {\bibfield  {journal} {\bibinfo  {journal} {Annals Phys.}\ }\textbf {\bibinfo {volume} {322}},\ \bibinfo {pages} {2021} (\bibinfo {year} {2007})},\ \Eprint {http://arxiv.org/abs/hep-ph/0607066} {arXiv:hep-ph/0607066} \BibitemShut {NoStop}%
\bibitem [{\citenamefont {Tanimoto}\ \emph {et~al.}(2020)\citenamefont {Tanimoto}, \citenamefont {Bentz},\ and\ \citenamefont {Clo\"et}}]{Tanimoto:2019tsl}%
  \BibitemOpen
  \bibfield  {author} {\bibinfo {author} {\bibfnamefont {T.}~\bibnamefont {Tanimoto}}, \bibinfo {author} {\bibfnamefont {W.}~\bibnamefont {Bentz}}, \ and\ \bibinfo {author} {\bibfnamefont {I.~C.}\ \bibnamefont {Clo\"et}},\ }\href {\doibase 10.1103/PhysRevC.101.055204} {\bibfield  {journal} {\bibinfo  {journal} {Phys. Rev. C}\ }\textbf {\bibinfo {volume} {101}},\ \bibinfo {pages} {055204} (\bibinfo {year} {2020})},\ \Eprint {http://arxiv.org/abs/1903.06851} {arXiv:1903.06851 [nucl-th]} \BibitemShut {NoStop}%
\bibitem [{\citenamefont {Carrillo-Serrano}\ \emph {et~al.}(2014)\citenamefont {Carrillo-Serrano}, \citenamefont {Clo\"et},\ and\ \citenamefont {Thomas}}]{Carrillo-Serrano:2014zta}%
  \BibitemOpen
  \bibfield  {author} {\bibinfo {author} {\bibfnamefont {M.~E.}\ \bibnamefont {Carrillo-Serrano}}, \bibinfo {author} {\bibfnamefont {I.~C.}\ \bibnamefont {Clo\"et}}, \ and\ \bibinfo {author} {\bibfnamefont {A.~W.}\ \bibnamefont {Thomas}},\ }\href {\doibase 10.1103/PhysRevC.90.064316} {\bibfield  {journal} {\bibinfo  {journal} {Phys. Rev. C}\ }\textbf {\bibinfo {volume} {90}},\ \bibinfo {pages} {064316} (\bibinfo {year} {2014})},\ \Eprint {http://arxiv.org/abs/1409.1653} {arXiv:1409.1653 [nucl-th]} \BibitemShut {NoStop}%
\bibitem [{\citenamefont {Saito}\ and\ \citenamefont {Thomas}(1994)}]{Saito:1994ki}%
  \BibitemOpen
  \bibfield  {author} {\bibinfo {author} {\bibfnamefont {K.}~\bibnamefont {Saito}}\ and\ \bibinfo {author} {\bibfnamefont {A.~W.}\ \bibnamefont {Thomas}},\ }\href {\doibase 10.1016/0370-2693(94)91520-2} {\bibfield  {journal} {\bibinfo  {journal} {Phys. Lett. B}\ }\textbf {\bibinfo {volume} {327}},\ \bibinfo {pages} {9} (\bibinfo {year} {1994})},\ \Eprint {http://arxiv.org/abs/nucl-th/9403015} {arXiv:nucl-th/9403015} \BibitemShut {NoStop}%
\bibitem [{\citenamefont {Birse}(1995)}]{Birse:1994eg}%
  \BibitemOpen
  \bibfield  {author} {\bibinfo {author} {\bibfnamefont {M.~C.}\ \bibnamefont {Birse}},\ }\href {\doibase 10.1103/PhysRevC.51.R1083} {\bibfield  {journal} {\bibinfo  {journal} {Phys. Rev. C}\ }\textbf {\bibinfo {volume} {51}},\ \bibinfo {pages} {R1083} (\bibinfo {year} {1995})},\ \Eprint {http://arxiv.org/abs/nucl-th/9412001} {arXiv:nucl-th/9412001} \BibitemShut {NoStop}%
\bibitem [{\citenamefont {Saito}\ and\ \citenamefont {Thomas}(1995)}]{Saito:1995up}%
  \BibitemOpen
  \bibfield  {author} {\bibinfo {author} {\bibfnamefont {K.}~\bibnamefont {Saito}}\ and\ \bibinfo {author} {\bibfnamefont {A.~W.}\ \bibnamefont {Thomas}},\ }\href {\doibase 10.1103/PhysRevC.52.2789} {\bibfield  {journal} {\bibinfo  {journal} {Phys. Rev. C}\ }\textbf {\bibinfo {volume} {52}},\ \bibinfo {pages} {2789} (\bibinfo {year} {1995})},\ \Eprint {http://arxiv.org/abs/nucl-th/9506003} {arXiv:nucl-th/9506003} \BibitemShut {NoStop}%
\bibitem [{\citenamefont {Wallace}\ \emph {et~al.}(1995)\citenamefont {Wallace}, \citenamefont {Gross},\ and\ \citenamefont {Tjon}}]{Wallace:1994ux}%
  \BibitemOpen
  \bibfield  {author} {\bibinfo {author} {\bibfnamefont {S.~J.}\ \bibnamefont {Wallace}}, \bibinfo {author} {\bibfnamefont {F.}~\bibnamefont {Gross}}, \ and\ \bibinfo {author} {\bibfnamefont {J.~A.}\ \bibnamefont {Tjon}},\ }\href {\doibase 10.1103/PhysRevLett.74.228} {\bibfield  {journal} {\bibinfo  {journal} {Phys. Rev. Lett.}\ }\textbf {\bibinfo {volume} {74}},\ \bibinfo {pages} {228} (\bibinfo {year} {1995})},\ \Eprint {http://arxiv.org/abs/nucl-th/9407040} {arXiv:nucl-th/9407040} \BibitemShut {NoStop}%
\bibitem [{\citenamefont {Brown}\ and\ \citenamefont {Weise}(1987)}]{Brown:1985gt}%
  \BibitemOpen
  \bibfield  {author} {\bibinfo {author} {\bibfnamefont {G.~E.}\ \bibnamefont {Brown}}\ and\ \bibinfo {author} {\bibfnamefont {W.}~\bibnamefont {Weise}},\ }\href {\doibase 10.1088/0031-8949/36/2/003} {\bibfield  {journal} {\bibinfo  {journal} {Comments Nucl. Part. Phys.}\ }\textbf {\bibinfo {volume} {17}},\ \bibinfo {pages} {39} (\bibinfo {year} {1987})}\BibitemShut {NoStop}%
\bibitem [{\citenamefont {Wallace}(1998)}]{Wallace:1998mx}%
  \BibitemOpen
  \bibfield  {author} {\bibinfo {author} {\bibfnamefont {S.~J.}\ \bibnamefont {Wallace}},\ }\href {\doibase 10.1016/S0375-9474(98)00020-7} {\bibfield  {journal} {\bibinfo  {journal} {Nucl. Phys. A}\ }\textbf {\bibinfo {volume} {631}},\ \bibinfo {pages} {137C} (\bibinfo {year} {1998})}\BibitemShut {NoStop}%
\bibitem [{\citenamefont {Bentz}\ \emph {et~al.}(1985)\citenamefont {Bentz}, \citenamefont {Arima}, \citenamefont {Hyuga}, \citenamefont {Shimizu},\ and\ \citenamefont {Yazaki}}]{Bentz:1985qh}%
  \BibitemOpen
  \bibfield  {author} {\bibinfo {author} {\bibfnamefont {W.}~\bibnamefont {Bentz}}, \bibinfo {author} {\bibfnamefont {A.}~\bibnamefont {Arima}}, \bibinfo {author} {\bibfnamefont {H.}~\bibnamefont {Hyuga}}, \bibinfo {author} {\bibfnamefont {K.}~\bibnamefont {Shimizu}}, \ and\ \bibinfo {author} {\bibfnamefont {K.}~\bibnamefont {Yazaki}},\ }\href {\doibase 10.1016/0375-9474(85)90550-0} {\bibfield  {journal} {\bibinfo  {journal} {Nucl. Phys. A}\ }\textbf {\bibinfo {volume} {436}},\ \bibinfo {pages} {593} (\bibinfo {year} {1985})}\BibitemShut {NoStop}%
\bibitem [{\citenamefont {Schwinger}(1951)}]{Schwinger:1951nm}%
  \BibitemOpen
  \bibfield  {author} {\bibinfo {author} {\bibfnamefont {J.~S.}\ \bibnamefont {Schwinger}},\ }\href {\doibase 10.1103/PhysRev.82.664} {\bibfield  {journal} {\bibinfo  {journal} {Phys. Rev.}\ }\textbf {\bibinfo {volume} {82}},\ \bibinfo {pages} {664} (\bibinfo {year} {1951})}\BibitemShut {NoStop}%
\bibitem [{\citenamefont {Hellstern}\ \emph {et~al.}(1997)\citenamefont {Hellstern}, \citenamefont {Alkofer},\ and\ \citenamefont {Reinhardt}}]{Hellstern:1997nv}%
  \BibitemOpen
  \bibfield  {author} {\bibinfo {author} {\bibfnamefont {G.}~\bibnamefont {Hellstern}}, \bibinfo {author} {\bibfnamefont {R.}~\bibnamefont {Alkofer}}, \ and\ \bibinfo {author} {\bibfnamefont {H.}~\bibnamefont {Reinhardt}},\ }\href {\doibase 10.1016/S0375-9474(97)00412-0} {\bibfield  {journal} {\bibinfo  {journal} {Nucl. Phys.}\ }\textbf {\bibinfo {volume} {A625}},\ \bibinfo {pages} {697} (\bibinfo {year} {1997})},\ \Eprint {http://arxiv.org/abs/hep-ph/9706551} {arXiv:hep-ph/9706551 [hep-ph]} \BibitemShut {NoStop}%
%%CITATION = HEP-PH/9706551;%%
\bibitem [{\citenamefont {Gell-Mann}(1962)}]{Gell-Mann:1962yej}%
  \BibitemOpen
  \bibfield  {author} {\bibinfo {author} {\bibfnamefont {M.}~\bibnamefont {Gell-Mann}},\ }\href {\doibase 10.1103/PhysRev.125.1067} {\bibfield  {journal} {\bibinfo  {journal} {Phys. Rev.}\ }\textbf {\bibinfo {volume} {125}},\ \bibinfo {pages} {1067} (\bibinfo {year} {1962})}\BibitemShut {NoStop}%
\bibitem [{\citenamefont {Okubo}(1962)}]{Okubo:1961jc}%
  \BibitemOpen
  \bibfield  {author} {\bibinfo {author} {\bibfnamefont {S.}~\bibnamefont {Okubo}},\ }\href {\doibase 10.1143/PTP.27.949} {\bibfield  {journal} {\bibinfo  {journal} {Prog. Theor. Phys.}\ }\textbf {\bibinfo {volume} {27}},\ \bibinfo {pages} {949} (\bibinfo {year} {1962})}\BibitemShut {NoStop}%
\bibitem [{\citenamefont {Bentz}\ \emph {et~al.}(2003)\citenamefont {Bentz}, \citenamefont {Horikawa}, \citenamefont {Ishii},\ and\ \citenamefont {Thomas}}]{Bentz:2002um}%
  \BibitemOpen
  \bibfield  {author} {\bibinfo {author} {\bibfnamefont {W.}~\bibnamefont {Bentz}}, \bibinfo {author} {\bibfnamefont {T.}~\bibnamefont {Horikawa}}, \bibinfo {author} {\bibfnamefont {N.}~\bibnamefont {Ishii}}, \ and\ \bibinfo {author} {\bibfnamefont {A.~W.}\ \bibnamefont {Thomas}},\ }\href {\doibase 10.1016/S0375-9474(03)00635-3} {\bibfield  {journal} {\bibinfo  {journal} {Nucl. Phys.}\ }\textbf {\bibinfo {volume} {A720}},\ \bibinfo {pages} {95} (\bibinfo {year} {2003})},\ \Eprint {http://arxiv.org/abs/nucl-th/0210067} {arXiv:nucl-th/0210067 [nucl-th]} \BibitemShut {NoStop}%
%%CITATION = NUCL-TH/0210067;%%
\bibitem [{\citenamefont {Nagata}\ and\ \citenamefont {Hosaka}(2004)}]{Nagata:2003gg}%
  \BibitemOpen
  \bibfield  {author} {\bibinfo {author} {\bibfnamefont {K.}~\bibnamefont {Nagata}}\ and\ \bibinfo {author} {\bibfnamefont {A.}~\bibnamefont {Hosaka}},\ }\href {\doibase 10.1143/PTP.111.857} {\bibfield  {journal} {\bibinfo  {journal} {Prog. Theor. Phys.}\ }\textbf {\bibinfo {volume} {111}},\ \bibinfo {pages} {857} (\bibinfo {year} {2004})},\ \Eprint {http://arxiv.org/abs/hep-ph/0312161} {arXiv:hep-ph/0312161} \BibitemShut {NoStop}%
\bibitem [{\citenamefont {Ulrych}\ and\ \citenamefont {Muther}(1997)}]{Ulrych:1997es}%
  \BibitemOpen
  \bibfield  {author} {\bibinfo {author} {\bibfnamefont {S.}~\bibnamefont {Ulrych}}\ and\ \bibinfo {author} {\bibfnamefont {H.}~\bibnamefont {Muther}},\ }\href {\doibase 10.1103/PhysRevC.56.1788} {\bibfield  {journal} {\bibinfo  {journal} {Phys. Rev. C}\ }\textbf {\bibinfo {volume} {56}},\ \bibinfo {pages} {1788} (\bibinfo {year} {1997})},\ \Eprint {http://arxiv.org/abs/nucl-th/9706030} {arXiv:nucl-th/9706030} \BibitemShut {NoStop}%
\bibitem [{\citenamefont {Bentz}\ and\ \citenamefont {Clo\"et}(2022)}]{Bentz:2020mdk}%
  \BibitemOpen
  \bibfield  {author} {\bibinfo {author} {\bibfnamefont {W.}~\bibnamefont {Bentz}}\ and\ \bibinfo {author} {\bibfnamefont {I.~C.}\ \bibnamefont {Clo\"et}},\ }\href {\doibase 10.1103/PhysRevC.105.014320} {\bibfield  {journal} {\bibinfo  {journal} {Phys. Rev. C}\ }\textbf {\bibinfo {volume} {105}},\ \bibinfo {pages} {014320} (\bibinfo {year} {2022})},\ \Eprint {http://arxiv.org/abs/2004.11605} {arXiv:2004.11605 [nucl-th]} \BibitemShut {NoStop}%
\bibitem [{\citenamefont {McNeil}\ \emph {et~al.}(1986)\citenamefont {McNeil}, \citenamefont {Amado}, \citenamefont {Horowitz}, \citenamefont {Oka}, \citenamefont {Shepard},\ and\ \citenamefont {Sparrow}}]{McNeil:1986zz}%
  \BibitemOpen
  \bibfield  {author} {\bibinfo {author} {\bibfnamefont {J.~A.}\ \bibnamefont {McNeil}}, \bibinfo {author} {\bibfnamefont {R.~D.}\ \bibnamefont {Amado}}, \bibinfo {author} {\bibfnamefont {C.~J.}\ \bibnamefont {Horowitz}}, \bibinfo {author} {\bibfnamefont {M.}~\bibnamefont {Oka}}, \bibinfo {author} {\bibfnamefont {J.~R.}\ \bibnamefont {Shepard}}, \ and\ \bibinfo {author} {\bibfnamefont {D.~A.}\ \bibnamefont {Sparrow}},\ }\href {\doibase 10.1103/PhysRevC.34.746} {\bibfield  {journal} {\bibinfo  {journal} {Phys. Rev. C}\ }\textbf {\bibinfo {volume} {34}},\ \bibinfo {pages} {746} (\bibinfo {year} {1986})}\BibitemShut {NoStop}%
\bibitem [{\citenamefont {Ichii}\ \emph {et~al.}(1988)\citenamefont {Ichii}, \citenamefont {Bentz}, \citenamefont {Arima},\ and\ \citenamefont {Suzuki}}]{Ichii:1988jn}%
  \BibitemOpen
  \bibfield  {author} {\bibinfo {author} {\bibfnamefont {S.}~\bibnamefont {Ichii}}, \bibinfo {author} {\bibfnamefont {W.}~\bibnamefont {Bentz}}, \bibinfo {author} {\bibfnamefont {A.}~\bibnamefont {Arima}}, \ and\ \bibinfo {author} {\bibfnamefont {T.}~\bibnamefont {Suzuki}},\ }\href {\doibase 10.1016/0375-9474(88)90026-7} {\bibfield  {journal} {\bibinfo  {journal} {Nucl. Phys. A}\ }\textbf {\bibinfo {volume} {487}},\ \bibinfo {pages} {493} (\bibinfo {year} {1988})}\BibitemShut {NoStop}%
\bibitem [{\citenamefont {Cohen}\ and\ \citenamefont {Furnstahl}(1987)}]{Cohen:1986tt}%
  \BibitemOpen
  \bibfield  {author} {\bibinfo {author} {\bibfnamefont {J.}~\bibnamefont {Cohen}}\ and\ \bibinfo {author} {\bibfnamefont {R.~J.}\ \bibnamefont {Furnstahl}},\ }\href {\doibase 10.1103/PhysRevC.35.2231} {\bibfield  {journal} {\bibinfo  {journal} {Phys. Rev. C}\ }\textbf {\bibinfo {volume} {35}},\ \bibinfo {pages} {2231} (\bibinfo {year} {1987})}\BibitemShut {NoStop}%
\bibitem [{\citenamefont {Cohen}(1993)}]{Cohen:1993zr}%
  \BibitemOpen
  \bibfield  {author} {\bibinfo {author} {\bibfnamefont {J.}~\bibnamefont {Cohen}},\ }\href {\doibase 10.1103/PhysRevC.48.1346} {\bibfield  {journal} {\bibinfo  {journal} {Phys. Rev. C}\ }\textbf {\bibinfo {volume} {48}},\ \bibinfo {pages} {1346} (\bibinfo {year} {1993})}\BibitemShut {NoStop}%
\bibitem [{\citenamefont {Arima}\ \emph {et~al.}(1987)\citenamefont {Arima}, \citenamefont {Shimizu}, \citenamefont {Bentz},\ and\ \citenamefont {Hyuga}}]{Arima:1987hib}%
  \BibitemOpen
  \bibfield  {author} {\bibinfo {author} {\bibfnamefont {A.}~\bibnamefont {Arima}}, \bibinfo {author} {\bibfnamefont {K.}~\bibnamefont {Shimizu}}, \bibinfo {author} {\bibfnamefont {W.}~\bibnamefont {Bentz}}, \ and\ \bibinfo {author} {\bibfnamefont {H.}~\bibnamefont {Hyuga}},\ }\href@noop {} {\bibfield  {journal} {\bibinfo  {journal} {Adv. Nucl. Phys.}\ }\textbf {\bibinfo {volume} {18}},\ \bibinfo {pages} {1} (\bibinfo {year} {1987})}\BibitemShut {NoStop}%
\bibitem [{\citenamefont {Furnstahl}\ and\ \citenamefont {Serot}(1987)}]{Furnstahl:1987rd}%
  \BibitemOpen
  \bibfield  {author} {\bibinfo {author} {\bibfnamefont {R.~J.}\ \bibnamefont {Furnstahl}}\ and\ \bibinfo {author} {\bibfnamefont {B.~D.}\ \bibnamefont {Serot}},\ }\href {\doibase 10.1016/0375-9474(87)90182-5} {\bibfield  {journal} {\bibinfo  {journal} {Nucl. Phys. A}\ }\textbf {\bibinfo {volume} {468}},\ \bibinfo {pages} {539} (\bibinfo {year} {1987})}\BibitemShut {NoStop}%
\bibitem [{\citenamefont {Panda}\ \emph {et~al.}(2002)\citenamefont {Panda}, \citenamefont {Bracco}, \citenamefont {Chiapparini}, \citenamefont {Conte},\ and\ \citenamefont {Krein}}]{Panda:2002iu}%
  \BibitemOpen
  \bibfield  {author} {\bibinfo {author} {\bibfnamefont {P.~K.}\ \bibnamefont {Panda}}, \bibinfo {author} {\bibfnamefont {M.~E.}\ \bibnamefont {Bracco}}, \bibinfo {author} {\bibfnamefont {M.}~\bibnamefont {Chiapparini}}, \bibinfo {author} {\bibfnamefont {E.}~\bibnamefont {Conte}}, \ and\ \bibinfo {author} {\bibfnamefont {G.}~\bibnamefont {Krein}},\ }\href {\doibase 10.1103/PhysRevC.65.065206} {\bibfield  {journal} {\bibinfo  {journal} {Phys. Rev. C}\ }\textbf {\bibinfo {volume} {65}},\ \bibinfo {pages} {065206} (\bibinfo {year} {2002})},\ \Eprint {http://arxiv.org/abs/nucl-th/0205051} {arXiv:nucl-th/0205051} \BibitemShut {NoStop}%
\bibitem [{\citenamefont {Leong}\ \emph {et~al.}(2023{\natexlab{a}})\citenamefont {Leong}, \citenamefont {Thomas},\ and\ \citenamefont {Guichon}}]{Leong:2023lmw}%
  \BibitemOpen
  \bibfield  {author} {\bibinfo {author} {\bibfnamefont {J.}~\bibnamefont {Leong}}, \bibinfo {author} {\bibfnamefont {A.~W.}\ \bibnamefont {Thomas}}, \ and\ \bibinfo {author} {\bibfnamefont {P.~A.~M.}\ \bibnamefont {Guichon}},\ }\href@noop {} {\  (\bibinfo {year} {2023}{\natexlab{a}})},\ \Eprint {http://arxiv.org/abs/2308.08987} {arXiv:2308.08987 [nucl-th]} \BibitemShut {NoStop}%
\bibitem [{\citenamefont {Okubo}(1963)}]{Okubo:1963fa}%
  \BibitemOpen
  \bibfield  {author} {\bibinfo {author} {\bibfnamefont {S.}~\bibnamefont {Okubo}},\ }\href {\doibase 10.1016/S0375-9601(63)92548-9} {\bibfield  {journal} {\bibinfo  {journal} {Phys. Lett.}\ }\textbf {\bibinfo {volume} {5}},\ \bibinfo {pages} {165} (\bibinfo {year} {1963})}\BibitemShut {NoStop}%
\bibitem [{\citenamefont {Zweig}(1964)}]{Zweig:1964jf}%
  \BibitemOpen
  \bibfield  {author} {\bibinfo {author} {\bibfnamefont {G.}~\bibnamefont {Zweig}},\ }\enquote {\bibinfo {title} {{An SU(3) model for strong interaction symmetry and its breaking. Version 2}},}\ in\ \href@noop {} {\emph {\bibinfo {booktitle} {{DEVELOPMENTS IN THE QUARK THEORY OF HADRONS. VOL. 1. 1964 - 1978}}}},\ \bibinfo {editor} {edited by\ \bibinfo {editor} {\bibfnamefont {D.~B.}\ \bibnamefont {Lichtenberg}}\ and\ \bibinfo {editor} {\bibfnamefont {S.~P.}\ \bibnamefont {Rosen}}}\ (\bibinfo {year} {1964})\ pp.\ \bibinfo {pages} {22--101}\BibitemShut {NoStop}%
\bibitem [{\citenamefont {Iizuka}(1966)}]{Iizuka:1966fk}%
  \BibitemOpen
  \bibfield  {author} {\bibinfo {author} {\bibfnamefont {J.}~\bibnamefont {Iizuka}},\ }\href {\doibase 10.1143/PTPS.37.21} {\bibfield  {journal} {\bibinfo  {journal} {Prog. Theor. Phys. Suppl.}\ }\textbf {\bibinfo {volume} {37}},\ \bibinfo {pages} {21} (\bibinfo {year} {1966})}\BibitemShut {NoStop}%
\bibitem [{\citenamefont {Heiselberg}\ and\ \citenamefont {Hjorth-Jensen}(2000)}]{Heiselberg:1999mq}%
  \BibitemOpen
  \bibfield  {author} {\bibinfo {author} {\bibfnamefont {H.}~\bibnamefont {Heiselberg}}\ and\ \bibinfo {author} {\bibfnamefont {M.}~\bibnamefont {Hjorth-Jensen}},\ }\href {\doibase 10.1016/S0370-1573(99)00110-6} {\bibfield  {journal} {\bibinfo  {journal} {Phys. Rept.}\ }\textbf {\bibinfo {volume} {328}},\ \bibinfo {pages} {237} (\bibinfo {year} {2000})},\ \Eprint {http://arxiv.org/abs/nucl-th/9902033} {arXiv:nucl-th/9902033} \BibitemShut {NoStop}%
\bibitem [{\citenamefont {Stone}\ \emph {et~al.}(2021)\citenamefont {Stone}, \citenamefont {Dexheimer}, \citenamefont {Guichon}, \citenamefont {Thomas},\ and\ \citenamefont {Typel}}]{Stone:2019blq}%
  \BibitemOpen
  \bibfield  {author} {\bibinfo {author} {\bibfnamefont {J.~R.}\ \bibnamefont {Stone}}, \bibinfo {author} {\bibfnamefont {V.}~\bibnamefont {Dexheimer}}, \bibinfo {author} {\bibfnamefont {P.~A.~M.}\ \bibnamefont {Guichon}}, \bibinfo {author} {\bibfnamefont {A.~W.}\ \bibnamefont {Thomas}}, \ and\ \bibinfo {author} {\bibfnamefont {S.}~\bibnamefont {Typel}},\ }\href {\doibase 10.1093/mnras/staa4006} {\bibfield  {journal} {\bibinfo  {journal} {Mon. Not. Roy. Astron. Soc.}\ }\textbf {\bibinfo {volume} {502}},\ \bibinfo {pages} {3476} (\bibinfo {year} {2021})},\ \Eprint {http://arxiv.org/abs/1906.11100} {arXiv:1906.11100 [nucl-th]} \BibitemShut {NoStop}%
\bibitem [{\citenamefont {Motta}\ and\ \citenamefont {Thomas}(2022)}]{Motta:2022nlj}%
  \BibitemOpen
  \bibfield  {author} {\bibinfo {author} {\bibfnamefont {T.~F.}\ \bibnamefont {Motta}}\ and\ \bibinfo {author} {\bibfnamefont {A.~W.}\ \bibnamefont {Thomas}},\ }\href {\doibase 10.1142/S0217732322300014} {\bibfield  {journal} {\bibinfo  {journal} {Mod. Phys. Lett. A}\ }\textbf {\bibinfo {volume} {37}},\ \bibinfo {pages} {2230001} (\bibinfo {year} {2022})},\ \Eprint {http://arxiv.org/abs/2201.11549} {arXiv:2201.11549 [nucl-th]} \BibitemShut {NoStop}%
\bibitem [{\citenamefont {Leong}\ \emph {et~al.}(2023{\natexlab{b}})\citenamefont {Leong}, \citenamefont {Motta}, \citenamefont {Thomas},\ and\ \citenamefont {Guichon}}]{Leong:2023yma}%
  \BibitemOpen
  \bibfield  {author} {\bibinfo {author} {\bibfnamefont {J.}~\bibnamefont {Leong}}, \bibinfo {author} {\bibfnamefont {T.~F.}\ \bibnamefont {Motta}}, \bibinfo {author} {\bibfnamefont {A.~W.}\ \bibnamefont {Thomas}}, \ and\ \bibinfo {author} {\bibfnamefont {P.~A.~M.}\ \bibnamefont {Guichon}},\ }\href {\doibase 10.1103/PhysRevC.108.015804} {\bibfield  {journal} {\bibinfo  {journal} {Phys. Rev. C}\ }\textbf {\bibinfo {volume} {108}},\ \bibinfo {pages} {015804} (\bibinfo {year} {2023}{\natexlab{b}})}\BibitemShut {NoStop}%
\bibitem [{\citenamefont {Tolman}(1939)}]{Tolman:1939jz}%
  \BibitemOpen
  \bibfield  {author} {\bibinfo {author} {\bibfnamefont {R.~C.}\ \bibnamefont {Tolman}},\ }\href {\doibase 10.1103/PhysRev.55.364} {\bibfield  {journal} {\bibinfo  {journal} {Phys. Rev.}\ }\textbf {\bibinfo {volume} {55}},\ \bibinfo {pages} {364} (\bibinfo {year} {1939})}\BibitemShut {NoStop}%
%%CITATION = PHRVA,55,364;%%
\bibitem [{\citenamefont {Oppenheimer}\ and\ \citenamefont {Volkoff}(1939)}]{Oppenheimer:1939ne}%
  \BibitemOpen
  \bibfield  {author} {\bibinfo {author} {\bibfnamefont {J.~R.}\ \bibnamefont {Oppenheimer}}\ and\ \bibinfo {author} {\bibfnamefont {G.~M.}\ \bibnamefont {Volkoff}},\ }\href {\doibase 10.1103/PhysRev.55.374} {\bibfield  {journal} {\bibinfo  {journal} {Phys. Rev.}\ }\textbf {\bibinfo {volume} {55}},\ \bibinfo {pages} {374} (\bibinfo {year} {1939})}\BibitemShut {NoStop}%
%%CITATION = PHRVA,55,374;%%
\bibitem [{\citenamefont {Hatsuda}\ and\ \citenamefont {Kunihiro}(1991)}]{Hatsuda:1991jn}%
  \BibitemOpen
  \bibfield  {author} {\bibinfo {author} {\bibfnamefont {T.}~\bibnamefont {Hatsuda}}\ and\ \bibinfo {author} {\bibfnamefont {T.}~\bibnamefont {Kunihiro}},\ }\href {\doibase 10.1007/BF01579559} {\bibfield  {journal} {\bibinfo  {journal} {Z. Phys. C}\ }\textbf {\bibinfo {volume} {51}},\ \bibinfo {pages} {49} (\bibinfo {year} {1991})}\BibitemShut {NoStop}%
\bibitem [{\citenamefont {Rehberg}\ \emph {et~al.}(1996)\citenamefont {Rehberg}, \citenamefont {Klevansky},\ and\ \citenamefont {Hufner}}]{Rehberg:1995kh}%
  \BibitemOpen
  \bibfield  {author} {\bibinfo {author} {\bibfnamefont {P.}~\bibnamefont {Rehberg}}, \bibinfo {author} {\bibfnamefont {S.~P.}\ \bibnamefont {Klevansky}}, \ and\ \bibinfo {author} {\bibfnamefont {J.}~\bibnamefont {Hufner}},\ }\href {\doibase 10.1103/PhysRevC.53.410} {\bibfield  {journal} {\bibinfo  {journal} {Phys. Rev. C}\ }\textbf {\bibinfo {volume} {53}},\ \bibinfo {pages} {410} (\bibinfo {year} {1996})},\ \Eprint {http://arxiv.org/abs/hep-ph/9506436} {arXiv:hep-ph/9506436} \BibitemShut {NoStop}%
\bibitem [{\citenamefont {Kato}\ \emph {et~al.}(1993)\citenamefont {Kato}, \citenamefont {Bentz}, \citenamefont {Yazaki},\ and\ \citenamefont {Tanaka}}]{Kato:1993np}%
  \BibitemOpen
  \bibfield  {author} {\bibinfo {author} {\bibfnamefont {M.}~\bibnamefont {Kato}}, \bibinfo {author} {\bibfnamefont {W.}~\bibnamefont {Bentz}}, \bibinfo {author} {\bibfnamefont {K.}~\bibnamefont {Yazaki}}, \ and\ \bibinfo {author} {\bibfnamefont {K.}~\bibnamefont {Tanaka}},\ }\href {\doibase 10.1016/0375-9474(93)90266-Z} {\bibfield  {journal} {\bibinfo  {journal} {Nucl. Phys. A}\ }\textbf {\bibinfo {volume} {551}},\ \bibinfo {pages} {541} (\bibinfo {year} {1993})}\BibitemShut {NoStop}%
\bibitem [{\citenamefont {Cloët}\ \emph {et~al.}(2014)\citenamefont {Cloët}, \citenamefont {Bentz},\ and\ \citenamefont {Thomas}}]{Cloet:2014rja}%
  \BibitemOpen
  \bibfield  {author} {\bibinfo {author} {\bibfnamefont {I.~C.}\ \bibnamefont {Cloët}}, \bibinfo {author} {\bibfnamefont {W.}~\bibnamefont {Bentz}}, \ and\ \bibinfo {author} {\bibfnamefont {A.~W.}\ \bibnamefont {Thomas}},\ }\href {\doibase 10.1103/PhysRevC.90.045202} {\bibfield  {journal} {\bibinfo  {journal} {Phys. Rev.}\ }\textbf {\bibinfo {volume} {C90}},\ \bibinfo {pages} {045202} (\bibinfo {year} {2014})},\ \Eprint {http://arxiv.org/abs/1405.5542} {arXiv:1405.5542 [nucl-th]} \BibitemShut {NoStop}%
%%CITATION = ARXIV:1405.5542;%%
\end{thebibliography}%

\end{document}